\documentclass[graybox, secnum,usenames,dvipsnames]{svmult}

\usepackage{amsfonts}

\usepackage{mathptmx}       
\usepackage{helvet}         
\usepackage{courier}        
\usepackage{type1cm}        
\usepackage{makeidx}         
\usepackage{graphicx}        
\usepackage{multicol}        
\usepackage[bottom]{footmisc}
\usepackage{hyperref}        
\usepackage{soul}            
\hypersetup{colorlinks=true,urlcolor=blue}
\usepackage[square,numbers]{natbib}
\makeindex             
                       
\def\be{\begin{equation}}
\def\ee{\end{equation}}
\def\bea{\begin{eqnarray}}
\def\eea{\end{eqnarray}}

\usepackage{tikz}
\usetikzlibrary{patterns}

\relax 

\usepackage{wrapfig}
\usepackage{amsmath}
\usepackage{slashed}
 
\usepackage{bm}

\usepackage{xcolor,colortbl}

\usepackage{pgfplots}

\definecolor{ashgrey}{rgb}{0.7, 0.75, 0.71}

\definecolor{mygray}{gray}{0.9}

\definecolor{medgray}{gray}{0.8}

\definecolor{ovgreen}{RGB}{108, 193, 10}

\usetikzlibrary{shapes.geometric}

\begin{document}
\title*{ D-branes}
 
\author{Constantin Bachas}
\institute{Constantin Bachas \at Laboratoire de Physique de l'Ecole Normale Sup\'erieure, 24 rue Lhomond, 75005 Paris\\ \email{costas.bachas@ens.fr}}
%
%
\maketitle
\abstract{This is an   introduction to the 
non-perturbative
 excitations  of  string theory known as D-branes. Topics covered include their definition and main properties, their role in   dualities and their dynamics. 
Based on lectures given at LACES 2019, and at the Amsterdam-Brussels-Geneva-Paris Solvay Doctoral School. 
}

\section*{Keywords} 
D-branes, string theory, dualities, solitons, supergravity,  gauge theories


\section{Introduction}

 The discovery of D-branes \cite{Polchinski:1995mt} revolutionized string theory. Although their existence was anticipated in  earlier
works, see e.g. \cite{Dai:1989ua,Horava:1989ga,Shenker:1990uf,Green:1991et,Bianchi:1991eu}, it was the realisation that they are
exact solitonic excitations of closed-string theory that unleashed a series of developments in a variety of subjects  including  supersymmetric gauge theories,  dualities, 
string compactifications, quantum black holes,  and the gauge/gravity or AdS/CFT correspondence. These developments are covered elsewhere in  the present  volume,
so  I will  only touch upon  them occasionally and briefly. 
The  purpose of this contribution is  to provide the  background material that is necessary  for  the study of these more advanced topics. 

 The   notes are   based on the early review ref.\,\cite{Bachas:1998rg}, and  on lectures given during  several editions of the Amsterdam-Brussels-Geneva-Paris Doctoral School 
 and in  the 2019  LACES school  in Florence. Most of the material is standard and  can be found in many  textbooks, in particular
 \cite{Polchinski:1998rq,Johnson:2003glb,Zwiebach:2004tj,Recknagel:2013uja,Kiritsis:2019npv}.  The presentation is   biased by the expertise and taste (or lack thereof\,!) of the author.
To make the lectures  self-contained and to define  the notation and conventions,   I include  a very brief  introduction to string theory. This and
some other parts of the present
chapter will overlap  with  other contributions to this volume.


\section{{The free bosonic string}}\label{sec:2}

\subsection{\textit{Polyakov and Nambu-Goto actions}}\label{subsec:21}
 The starting point is the Nambu-Goto action which is proportional to the (pseudo)area of the 
 worldsheet  $X^\mu(\sigma^\alpha)$ of the string, 
\be
S_{\rm NG} =  -T_{\rm F}  \int d^2\sigma\,  [-{\rm det} (  \partial_\alpha X^\mu \partial_\beta X^\nu \eta_{\mu\nu} )]^{1/2}\ . 
\label{NG}
\ee
Here $T_{\rm F}  \equiv  (2\pi\alpha^\prime)^{-1}$ is the tension of the string, $\mu,\nu= 0, 1 \cdots d-1$ and $\alpha, \beta = 0,1$. 
 The string moves in flat Minkowski spacetime and its worldsheet is parametrized by $\sigma^\alpha$.
   The signature of the metric is $(-+ \cdots +)$.

The above action  is {classically} equivalent to  that  of $d$  free massless scalar fields 
coupling minimally 
 to an auxiliary    two-dimensional  metric $g_{\alpha\beta}$ \cite{Polyakov:1981rd},\,\footnote{This action is called  
 the Polyakov action, demonstrating  in the words of Polyakov himself  \cite{Polyakov:2008zx} 
  Arnold's theorem 
 that ``things are never called after their true inventors.''   
The trick was indeed implicit in the study of minimal surfaces by
 J. Douglas and others, and it was  used explicitly for the worldsheet action of the superstring \cite{Brink:1976sc}. But it is in Polyakov's treatment of the quantum string 
  that it becomes truly  instrumental. I thank Bengt Nilsson for a discussion on this point.}
\be 
S_{\rm Polyakov} =  - {T_{\rm F}  \over 2} \int d^2\sigma\,  \sqrt{-g} \, g^{\alpha\beta}   \partial_\alpha X^\mu \partial_\beta X^\nu \eta_{\mu\nu}  \ . 
\label{Polyakov}
\ee
Both actions are invariant under reparametrizations
of the worldsheet, $\sigma^\alpha \to \tilde\sigma^\alpha(\sigma^\beta)$. 
One may choose   conformal coordinates
so  that $g_{\alpha\beta} =e^\phi  \eta_{\alpha\beta}$. The Liouville field $\phi$ 
 drops out of the 
classical action (\ref{Polyakov}), leading to  the
field equations ($\sigma^\pm = \sigma^0 \pm \sigma^1$)
 \be
 \partial_+ \partial_-  X^\mu = 0\, , \qquad { \partial_\pm X^\mu \partial_\pm X^\nu \eta_{\mu\nu} = 0\ .  }  
\label{classicaleq}
\ee
 The $d$ equations on the left follow by varying $X^\mu$,  while the two right-hand-side equations 
 follow from variations of  the auxiliary metric, 
\be
{2\over \sqrt{-g}}\, {\delta S_{\rm Polyakov}\over \delta g^{\alpha\beta}} \equiv T_{\alpha\beta}\ 
 \propto\  \partial_\alpha X^\mu
\partial_\beta X_\mu - {1\over 2} g_{\alpha\beta} (\partial_\gamma X^\mu \partial^\gamma X_\mu) = 0\ . 
\ee
These  equations impose that the energy-momentum tensor $T_{\alpha\beta}$ of   `matter' fields is zero. 
Because the trace  vanishes identically there are only two non-trivial equations. 
Furthermore,  the conservation $\partial^\alpha T_{\alpha\beta} = 0$ means  that  they  need
only to be imposed at some initial time. 
 They   are  therefore  {{phase-space constraints}},  familiar from the Hamiltonian formulation of gravity. In  string theory  they are called the  {Virasoro conditions. }

Had we started instead with  the Nambu-Goto action (\ref{NG}),  we could  again choose conformal coordinates  such that the {induced} 
metric $ \partial_\alpha X^\mu \partial_\beta X^\nu \eta_{\mu\nu} \propto \eta_{\alpha\beta}$. The classical equations are then again  
(\ref{classicaleq})  with  the Virasoro constraints now arising as   gauge-fixing conditions. 
The two actions are thus classically  equivalent.
\smallskip

Note that one may add to the Polyakov action
an  Einstein term $-\Phi_0  
\int {d^2\sigma\over 4\pi} \,  \sqrt{-g} \, R$, but   $\sqrt{g} \,R =  - \nabla^2 \phi $  is a total derivative  that  does not affect   the Virasoro conditions. 
This term plays nevertheless  an important  role because 
  \bea\label{Euler}
  \chi_\Sigma \equiv {1\over 4\pi}  \int_\Sigma  \sqrt{g} \,R \  = \  2-2h_\Sigma -b_\Sigma \ , 
 \eea 
where $h_\Sigma$  is the number of handles and $b_\Sigma$  the number of boundaries  of the    Riemann surface $\Sigma$. Different  worldsheet topologies  
are thus weighted with different powers of  the string coupling $    g_s\equiv \exp(\Phi_0)$. 
The topological invariant  $\chi_\Sigma$ is called the  Euler characteristic.  It  equals  2 for the sphere, 1 for the disk and 0 for the torus.


 
\subsection{\textit{Classical motion of  open and closed  strings}}\label{sec:2.2}

The  wave equation  $\partial_+ \partial_-  X = 0$ is solved by a sum of left- and
right-moving waves, $X = X_R(\sigma^-) + X_L(\sigma^+) $. For closed strings $\sigma^1\equiv \sigma^1 + 2\pi$ is periodically identified 
 and the most general solution reads
 \be\label{modesclosed}
 {\textcolor{darkgray}{\bf closed}}: \quad 
 X^\mu(\sigma) =  x^\mu +   \alpha^\prime P^\mu \sigma^0 +  i\sqrt{\alpha^\prime\over 2} \sum_{n\not= 0}  {1\over n} ( a_n^\mu e^{- in\sigma^-} + \tilde a_n^\mu e^{- in\sigma^+})
 \ee
with   $n$ running  over non-zero  integers.
The normalizations  have been  fixed  so that  
$P^\mu = T_F \int_0^{2\pi} d\sigma^1 \partial_0 X^\mu$ is the center-of-mass momentum
  of the string,  and the canonical Poisson brackets imply  $\{ a_n^\mu , a_m^\nu \} = \{ \tilde a_n^\mu , \tilde a_m^\nu \} = 
   in\,  \delta_{n+m,0} \delta^{\mu\nu}$. Reality requires $(a_n^\mu)^*  = a_{-n}^\mu$,  and
 likewise for the tilde variables.   
\smallskip

  The  Virasoro constraints can be  solved explicitly in
   light-cone gauge, 
\be\label{LCgauge}
 X^{+} = \alpha^\prime P^+ \sigma^0 \quad
 {\rm and}\quad   \partial_\pm X^- = {2\over \alpha^\prime P^+} \sum_{j=2}^{d-1}\partial_\pm X^j \partial_\pm X^j\ 
\ee
with $X^\pm = X^0 \pm X^{1}$. Note that by
 choosing    $ X^+_L =  {1\over 2} \alpha^\prime P^+\sigma^+ $ and  
$ X^{+}_R=  {1\over 2} \alpha^\prime P^+\sigma^-$ we removed  the residual freedom under coordinate transformations
that preserve  the conformal gauge, 
  $\sigma^+ \to f(\sigma^+)$ and $\sigma^- \to  \tilde f (\sigma^-)$.
The  phase space of a closed string is  therefore  parametrized  by  the center-of-mass positions and momenta,  
and by  the  amplitudes of oscillation  in the  transverse dimensions.
  These
 are   only  subject   to the {mass-shell} 
and  {level-matching}\,conditions which are obtained by integrating 
 eqs.(\ref{LCgauge}) around the string, 
\be
 {\textcolor{darkgray}{\rm\bf closed}}: \quad 
  M^2 =  - P^\mu P_\mu =   {2\over \alpha^\prime} \sum_{j=2}^{d-1}\,
  \sum_{n\not=0}  a_{-n}^j  a^j_{n}\  = \ 
{2 \over \alpha^\prime} \sum_{j=2}^{d-1}\, \sum_{n\not=0}   \tilde a_{-n}^j \tilde a^j_{n} \ \ . 
\label{Virclosed}
\ee

 The reader may wonder why not go to the center-of-mass frame  and 
choose  the more intuitive temporal gauge $X^0 = \sigma^0$.  The  reason is that in this gauge the Virasoro conditions are quadratic 
in all  fields,  
\bea\vert\partial_+ \vec X\vert^2 = \vert\partial_- \vec X\vert^2
= {1\over 4}\quad {\rm where}\quad \vec X = (X^1, \cdots X^{d-1})\ , 
\label{temp} 
\eea
making it hard to eliminate  the redundant oscillation amplitudes. Furthermore, as we will see in the next subsection,  
some states of the quantized string are massless. 
 Gauge (\ref{temp})  is nevertheless  useful for proving that cusps, which are strong  emitters of gravitational waves \cite{Damour:2000wa}, 
 would be a generic feature of   cosmic strings if these exist  \cite{Turok:1984cn}. 
 Cusps are points that move at the speed of light, which is  equivalent to  $\partial_+ \vec X  = \partial_- \vec X $. 
But $\partial_\pm  \vec X(\sigma^\pm)$  trace  curves on the round sphere as their arguments  vary, so
they generically intersect  at some   $(\sigma^+, \sigma^-)$ {\small Q.E.D.}

 \smallskip
  After this amusing digression we turn now  to open strings. 
   For worldsheets $\Sigma$ with  {boundary} ${\partial\Sigma}$
   the  variation of the  Polyakov action in conformal gauge  gives
\bea
\delta S_{\rm Polyakov} \propto&&  \int_\Sigma  d^2\sigma\,  \partial^\alpha (\delta X_\mu) \, \partial_\alpha X^\mu  
\nonumber \\
&& =   \int_{\partial\Sigma} d\sigma^\alpha \epsilon_{\alpha\beta} \partial^\beta X^\mu \, \delta X_\mu  -
\int_\Sigma d^2\sigma\,    \delta X_\mu  \  \partial^\alpha \partial_\alpha X^\mu  = 0\ . 
\eea
Free endpoints, i.e. those for which $\delta X^\mu$ is arbitrary, must therefore obey  
 {{Neumann}} conditions, 
$\partial_\perp X^\mu = 0$ where $\perp$ stands for  normal to the boundary. 
The most  general solution with  {{Neumann}} conditions at both ends
  reads\,\footnote{The  choice of  parametrization is such   that
the  canonical Poisson brackets are  the same  for both open- and   closed-string  scillation amplitudes.}
 \be\label{modesNN}
  {\textcolor{darkgray}{\rm\bf  open\  (NN)}} : 
  \quad    X^\mu =  x^\mu +   2\alpha^\prime P^\mu \sigma^0 +
     i\sqrt{\alpha^\prime\over 2} \sum_{n \not=0}  {1\over n}  a_n^\mu\,  ( e^{- in\sigma^-}  +   e^{- in\sigma^+})\ ,  
  \ee
where  $n\in  \mathbb{Z}$ and 
 $\sigma^1 \in [0,\pi]$.
 String  excitations  are reflected at the endpoints and form  standing waves, which is why
 left- and right-moving amplitudes  are identified. 
 Note also that when  $\partial_1 X^\mu =0$ the Virasoro conditions reduce to  $\partial_0 X^\mu \partial_0 X^\nu \eta_{\mu\nu} = 0$,  
which shows that   free   endpoints
 travel at the speed of light.
 
 Another natural boundary condition    is the  {Dirichlet} condition which  fixes the  position of the string endpoints. 
A relativistic violin string,  for example,  would have ${\bf X}(\sigma^1=0)= {\bf x}$  and ${\bf X}(\sigma^1=\pi)= {\bf x}+ \Delta {\bf x}$. 
The mode expansion would  in this case read 
  \be\label{modesDDa}
 {\textcolor{darkgray}{\rm\bf  open (DD)}}: \quad 
    X^j  =   x^j + {{\sigma^1\over \pi} \, \Delta x^j}+ 
     i\sqrt{\alpha^\prime\over 2} \sum_{n\not= 0}  {1\over n}  a_n^j\,  ( e^{- in\sigma^-}  -   e^{- in\sigma^+})\ .   
  \ee
Comparing  with eq.\,(\ref{modesNN})  we see that the  center-of-mass 
position and momentum is here replaced by ${\bf x}$ and $\Delta {\bf x}$. 
 Both are fixed,  so they are not part of the open-string phase space.

 If  the $X^{\mu}$ 
 obey  Neumann conditions at both ends for $\mu = 0, 1, \cdots, p$ 
and   Dirichlet conditions for  $\mu  = p+1, \cdots , d-1$, then the  string endpoints 
 are forced to move   on static parallel $p$-dimensional
hyperplanes,  as  illustrated in   figure {\textcolor{red} 1} below.  
 One can  think of these hyperplanes  as  
  {extended $p$-dimensional objects}. They  will   turn out to be   dynamical,  non-perturbative  excitations of string theory  dubbed
   {{D$p$ branes}}. Freely moving open strings live on D$(d-1)$ branes, whereas our  relativistic violin string is attached to two D0 branes, alias D-particles.

 \vskip 6mm

\begin{center}
\hskip 1.1cm
 \begin{tikzpicture}[scale=1.1]
 \draw[pattern=north west lines, pattern color=ovgreen] (0,-1) -- (0,3) -- (2,2) -- (2,0) -- cycle; 
 \draw [very thick] (1,1.38) circle (0.05);
  \draw [very thick] (4.9,0.95) circle (0.05);
 \draw [very thick] (2.5,1) arc (60 :90 :3) ;
  \draw [very thick] (2.3,1.1) arc (240 :260 :3) ;
  \draw[very thick] (3.25, 0.75 ) arc(260: 292: 3);
\draw[pattern=north west lines, pattern color=ovgreen] (4,0) -- (4,2) -- (6,3) -- (6,-1) -- cycle;
 \draw[->, very thick] (-4,0) -- (-2.,0) ; 
  \draw[->, thick] (-4,0) -- (-4, 1.5) ; 
   \draw[->, thick] (-4,0) -- (-2.7, 0.5) ;
   \draw (-4, 1.9 ) node{$x^1$}; 
   \draw (-2.7, 0.9 ) node{$x^{2, \cdots ,p}$}; 
    \draw (-2., -0.4 ) node{$x^{p+1, \cdots ,d-1}$}; 
\end{tikzpicture}
\end{center}
\vskip 5mm
\centerline{\small {\bf Fig. 1}\ \ An open string stretching between two parallel D$p$ branes.}

 \vskip 6mm
 
The mass-shell condition for open strings is obtained as before by integrating 
 eqs.(\ref{LCgauge}). 
For  a string stretching between  static parallel  D$p$ branes it  reads
\be 
 {\textcolor{darkgray}{\rm\bf  open\  (D{\it   p\,}  D{\it  p}) }} :\quad 
  M^2 =   -P^\mu P_\mu = 
   {1\over  2 \alpha^\prime} \sum_{j=2}^{d-1} \, \sum_{n\not=0} a_{-n}^j  a^j_{n} \    { +\,  \vert T_F \Delta \vec x \vert^2} \ ,  
   \label{Viropen}  
 \ee
where $\mu$ runs over the $p+1$  Neumann directions only. The last term on the right 
can be recognized as  the  mass squared of an open string stretching linearly between the two D-branes. 
It enters in the mass formula like  momentum in some  hidden dimension. 
As we will later see, this is not a coincidence but
 a consequence of a deep symmetry of string theory called T-duality. 
 Comparing with eqs.\,(\ref{Virclosed}) suggests that  a closed string is the same as  two open strings tied together. There is actually  more to  this remark  than meets the eye.

   
\setcounter{figure}{1}

  More generally, the two D-branes at the string endpoints can be different.
  They may have,  in particular,  a different orientation  or dimension $p$,  and uniform relative  
  
       \begin{wrapfigure}{r}{0.42\linewidth}
\centering
\vskip -0.1cm
\includegraphics[width=0.54\textwidth]{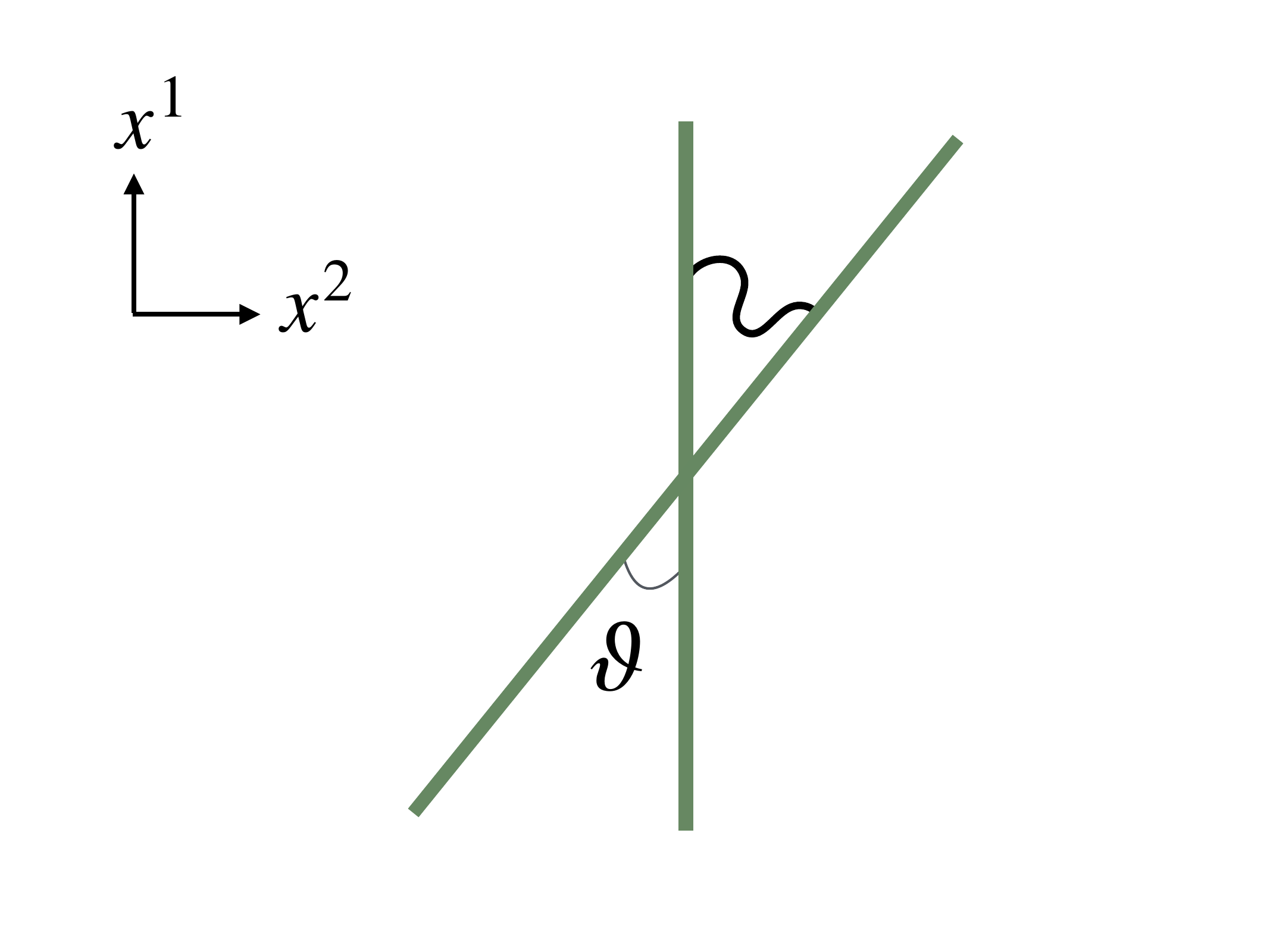}
\caption{An open F-string   between two straight D-strings making  an angle $\vartheta$.}
\label{fig:myfig}
\end{wrapfigure}

  \noindent  motion. In  all  these situations  the boundary conditions remain  linear
  and the problem 
 can be  readily  solved \cite{Bachas:1995kx,Berkooz:1996km}. 
  As an example, consider  two  straight static D1 branes, alias D-strings, that make  an angle $\vartheta$ in the
$(x^1,x^2)$ plane as illustrated   in figure \ref{fig:myfig}. 
Let the first 
 D-string  be oriented along the $x^1$ direction  so that
 $X^1$ obeys a Neumann condition and $X^2$ a Dirichlet condition at $\sigma^1=0$. 
 At the other  endpoint,   $\sigma^1=\pi$,  the coordinate
 $(X^1 \cos\vartheta   + X^2\sin\vartheta  )$ obeys  a  Neumann condition while the coordinate
 $(X^1\sin\vartheta   - X^2\cos\vartheta  )$ obeys a  Dirichlet condition. To avoid confusion the stretched open string is often referred to as an F-string
 (`F'  for fundamental). The general solution with these boundary condition is given by eq.\,(\ref{modesDDa}) for all   $j=3, \cdots, d-1$ since these
 are Dirichlet at both endpoints,  and by
   \be\label{openrotated}
X^1 + i X^2 =    \sqrt{\alpha^\prime\over 2}\left[\,  \sum_{r}  
{1\over r}  (a_r \,  e^{- ir\sigma^-}  +  a_r^*    e^{ ir\sigma^+}) + 
  \sum_{s}  
{1\over s}  (b_s^* \,  e^{ is\sigma^-} +  b_s   e^{- is\sigma^+})\, \right] \  , 
\ee
where  the oscillation frequencies are now $r \in \mathbb{Z} + {\vartheta\over \pi}$ and $s \in \mathbb{Z} - { \vartheta\over \pi}$. 
One can check  indeed that
at $\sigma^1=0$ the complex coordinate
$X^1 + i X^2$ is real so that $X^2$ obeys a  Dirichlet condition, while at $\sigma^1=\pi$ it is the combination
 $e^{-i\vartheta} (X^1+iX^2)$  that becomes  real  as required by  the Dirichlet condition 
$X^1\sin\vartheta   - X^2\cos\vartheta  =0$. Note that for $\vartheta= \pi/2$, i.e. for mixed Dirichlet-Neumann boundary conditions, 
all oscillation frequencies are half-integer. 

 The  expansion (\ref{openrotated}) has no zero mode, so the centre of mass of the F-string is localized at the intersection of the D-strings,  at $X^1=X^2=0$. 
Escaping  to infinity requires infinite energy and is   forbidden.

 
\subsection{\textit{Quantization}}

In the classical theory the amplitudes $a_n^i $ are continuous complex variables, and  the  spectrum,   
 eq.\,(\ref{Virclosed}) or  eq.\,(\ref{Viropen}),  is continuous and positive. 
 In the quantum theory the amplitudes become operators that obey the commutation relations
\be\label{bosoniccom}
[ a_n^i , (a_m^j)^\dagger ] = [ \tilde a_n^i , (\tilde a_m^j)^\dagger] = 
   n \delta_{n-m,0}\,  \delta^{ij} \ . 
\ee
The ground state is annihilated by all $n>0$ operators, 
while the conjugate operators $a_{-n}^j = (a_n^j)^\dag$ and $\tilde a_{-n}^j = (\tilde a_n^j)^\dag$  create excited string states. 
Furthermore,  normal ordering  the infinite sums in eq.\,(\ref{Virclosed}) or  eq.\,(\ref{Viropen}) shifts the spectrum by an infinite constant 
which  must be renormalised. 

 For concreteness we consider the open string  attached to two parallel D$p$ branes. Upon  quantization eq.\,(\ref{Viropen}) becomes
 \be\label{massopenQ}
 {\textcolor{darkgray}{\rm\bf  open\  (D{\it   p\,}  D{\it  p}) }} :\qquad 
  \alpha^\prime M^2 =     \,  \hat N \, + \  {\vert \Delta {\rm\bf x} \vert^2 \over 4\pi^2\alpha^\prime}   - {d-2\over 24 } \  ,  
\ee  
where $\hat N$\,  is the normal-ordered `level operator'
  \be\label{level}
  \hat N\ = \  \sum_{j=2}^{d-1}\,\, \sum_{n>0} a_{-n}^j  a^j_{n} \   \ 
 \Longrightarrow \ \ [\hat N, a_n^\dagger] =  n a_n^\dagger\ \ {\rm and}  \ \ [\hat N, a_n] =  -n a_n\ ,  
 \ee  
and the last term in (\ref{massopenQ}) is the zero-point contribution. 
This latter comes from  a collection of harmonic oscillators, one for each transverse coordinate and for each  frequency $\hbar\omega= n$.  
Using  the zeta-function regularization  we find
 \be
 (d-2) \sum_{n=1}^\infty {n\over 2}  = {(d-2)\over 2} \zeta(-1) = -{(d-2)\over 24}\   
 \ee  
which is the result announced in eq.\,(\ref{massopenQ}). 

 It is important  that the  zeta-function   respects locality on the string  worldsheet. To see why  
 let  $\sigma^1\in [0, \pi L]$,  where $L$ is some irrelevat worldsheet-distance scale which  will drop   out in the end from the expression for $M^2$. 
 The zero-point contribution to $ \alpha^\prime M^2/L$,   regularized by  a  short-distance cutoff $\epsilon$,   reads 
  \be \label{trick}
  \sum_{n=1}^\infty 
 {n\over 2L}  e^{- n\epsilon/L}  = {L\over 2\epsilon^2} - {1\over 24\, L} + O({\epsilon^2\over L^2})\ . 
 \ee 
A local subtraction must be proportional to $L$, so it  removes precisely the leading $\epsilon\to 0$  term {\small Q.E.D.}

 It is straightforward to extend  the analysis  to open strings  stretching between non-identical D-branes. Each (NN) or (DD) coordinate contributes as above,
while each  complex coordinate like  (\ref{openrotated}) has modes with fractional frequencies $n\pm  {\vartheta\over \pi}$  and gives  a zero-point contribution
 \bea
  {1\over 2} \sum_{n=1}^\infty (n- {\vert \vartheta\vert \over \pi})  +   {1\over 2} \sum_{n=0}^\infty (n + {\vert \vartheta\vert \over \pi})  = -{1\over 12} + {1\over 2} ({\vartheta\over\pi})^2\ . 
 \eea
 Here  $\vartheta \in (-\pi, \pi)$ and to  obtain this  result we followed the same logic as in (\ref{trick}). 
 Note that for $\vartheta = {\pi\over 2}$ the result is ${1\over 24}$, so the zero-point contribution  of each  (ND) or (DN) coordinate is ${1\over 48}$.  
\smallskip 
 
Closed strings have periodic coordinates which are integer-modded. 
Level operators can be defined separately in  the left-moving and right-moving sectors and they are matched by 
eqs.\,(\ref{Virclosed}). After quantization these constraints read
   \be\label{massclosedQ}
 {\textcolor{darkgray}{\rm\bf  closed }} :\qquad 
  {1 \over 4}\alpha^\prime M^2 \ = 
      \  \hat N_L \,   - {d-2\over 24 } =     \  \hat N_R \,   - {d-2\over 24 }\  .  
\ee   
Winding, which is analogous to $\Delta {\bf x}$,  and fractionally-modded coordinates arise when space is compactified on tori or on orbifolds.
In these cases eqs.\,(\ref{massclosedQ}) must be appropriately modified.

 
\subsection{\textit{Mass spectra and critical dimension}}

The  mass eigenstates of the string are  constructed by acting with creation operators,  $(a_{n}^j)^\dagger$ with $n>0$,  on the ground state $\vert 0\rangle$. 
From the commutation relations (\ref{level}) we see that the level of a state    is   the sum of the frequencies of {all} the  creation operators applied to the ground state. 
We have also seen that  $\alpha^\prime M^2 =  \hat N + \alpha^\prime M^2_0$ for open strings and 
${1\over 4} \alpha^\prime M^2 =  \hat N_L  + \alpha^\prime M^2_0 = \hat N_R  + \alpha^\prime M^2_0$ 
for closed strings,  with $M_0^2$   (or $4M_0^2$) the mass squared  of the ground state.

Let us first discuss an open string with freely-moving endpoints,  i.e. `attached'  to space-filling D$(d-1)$ branes. 
Mass eigenstates should  transform in representations of the unbroken Lorentz symmetry $SO(1,d-1)$. 
The lowest-lying  state is a scalar with mass  $\alpha^\prime M_0^2 = -(d-2)/24$. This  is negative  
  for $d>2$, so the state  is a {\it tachyon}. 
  A  theory with spontaneously-broken gauge symmetry  has a tachyon if expanded around the symmetric vacuum. Tachyons are 
  bad  but not   a priori fatal -- they may just  signal a wrong choice of the vacuum.

  The first excited states  $a^j_{-1}\vert 0\rangle$ of the open string,  in light-cone quantization,   form a vector of $SO(d-2)\subset SO(1,d-1)$. 
This  is the little group of a massless spinning particle --  a massive one  has one more polarization state.  Changing  the vacuum cannot save the day since  the 
 number of  degrees of freedom is preserved. To avoid a contradiction  these states must thus  be  massless, 
   \bea
  0 = 1 - {(d-2)\over 24} \ \Longrightarrow\ d=26\ . 
  \eea
 This is the critical dimension of the bosonic  string.  
When $d\not= 26$ the algebra  of the Lorentz-group  generators
$
   J^{\mu\nu} = \int d\sigma^1 (X^\mu \partial_0 X^\nu - X^\nu \partial_0 X^\mu) \   
$
  has a quantum anomaly and the symmetry is broken.\footnote{The anomaly  arises in the commutator of two  $J^{\, k\, -}$.
  In the alternative covariant quantization the missing states are provided by the Liouville mode which does not decouple when $d\not= 26$ \cite{Polyakov:1981rd}.
  Theories with a dynamical Liouville mode are called
{non-critical} string theories.}
 
    There are no further problems at the higher levels.  At   level two,  for example,  one  finds a vector $a^j_{-2} \vert 0\rangle $
   and a 2-index symmetric tensor  
      $a^j_{-1} a^k_{-1} \vert 0\rangle$ of $SO(24)$. They  combine nicely  into a symmetric traceless  2-index tensor of $SO(25)$ --  the little group
      for a massive particle.   These facts are summarized in  table \ref{table11}. 
    \smallskip
      
        The interested reader can consult ref.\,\cite{Hanany:2010da} for 
    a  discussion   of how massive string  states organize into $SO(25)$ representations. Here we will only count their  number, 
   ${\cal N}(\hat N)$, at each level $\hat N$.  For  one transverse coordinate  this  would have been  the number of partitions of $\hat N$ into positive integers, 
    with   generating function  $ \sum  {\cal N}(\hat N) q^{\hat N} = 
    \prod_{n=1}^\infty (1-q^n)^{-1}$.  For several coordinates the generating functions multiply. Using also the relation
    $\alpha^\prime M^2= \hat N-1$  we finally get 
     \be
  Z(q) =  \sum_{\rm \hat N=0}^\infty  {\cal N}(\hat N)\,  q^{\alpha^\prime M^2} =  q^{-1} \prod_{n=1}^\infty  (1-q^n)^{-24} = q^{-1} + 24 +  324 q + \cdots\ 
  \ee
 
\begin{table}[t]
\begin{center}
\vskip 3mm
 \begin{tabular}{|c|c|c|}
  \hline
      \,  & \, & \\ 
 \  {\normalsize $\alpha^\prime M^2$} \  &{\normalsize \bf States} & {\normalsize\bf   Representation} \\
      \,  & \, & \\
  \hline \hline
    \,  & \, & \\
 {\normalsize -1} & {\normalsize $\vert 0\rangle$ }   & \ {\normalsize scalar} \\
  \,  & \, & \\
  {\normalsize 0 } &  {\normalsize $a^j_{-1} \vert 0\rangle$}  & {\normalsize  transverse vector} \\
    \,  & \, & \\
  {\normalsize 1}  &  {\normalsize $\ a^j_{-1} a^k_{-1} \vert 0\rangle , \, a^j_{-2} \vert 0\rangle $}  \  &\ \  {\normalsize  symmetric traceless  tensor}\ \  \\
     \,  & \, & \\
   \hline   
   \end{tabular}
\end{center} 
  \caption{\footnotesize The first three levels of the bosonic open  string with freely-moving endpoints in $d=26$, and the corresponding representation of the Poincar\'e group.}
   \label{table11}
\end{table} 
 

 Note that $Z(q)$ should  not be confused with the canonical partition function of the  string 
 in which states are  weighted with the Boltzmann factor   $\exp(-\beta M)$. 
 The number of states   of a highly excited string can be extracted 
   from the saddle point approximation with the result
\bea\label{saddle}
  {\cal N}(\hat N) \, = \, \oint {dq \over 2\pi i }\, \,  q^{- \hat N} Z(q)\  \sim   \   \   e^{4\pi\sqrt{\hat N} }\ \ . 
\eea  
To perform the calculation one writes $Z(q)$ as a power of  Dedekind's  eta function, \vskip -1mm  
   \be\label{Dede}
 Z(q) = \eta(q)^{-24} \qquad {\rm where}\quad 
  \eta(q) \equiv  q^{1/24} \prod_{n=1}^\infty  (1-q^n) \ .  
  \ee
This function, which we will encounter  later in the cylinder amplitude,   
 transforms as a  
 {modular} form of weight $1/2$ under  fractional linear transformations of $\tau$,
 where $q= \exp(2\pi i \tau)$. Under inversion of $\tau$ in particular
   \be
  \eta\left(-{1\over \tau}\right) = \sqrt{-i\tau}\,\,  \eta(\tau)\ \ . 
  \ee
To  obtain  the estimate (\ref{saddle}) one uses that fact that 
for large $\hat N$ the contour  integral   is dominated by a saddle point
at  small  Im$\tau$ where 
$$\eta(\tau) = \sqrt{-i\tau}\,\, \eta(-{1\over \tau}) \sim\   e^{-{ i \pi  /12\tau}}\ . $$
The linear  growth of the entropy with mass, 
  $ {\cal N}(\hat N) \sim \exp (4\pi \sqrt{\alpha^\prime} M)$,  means that the canonical partition function of a free string 
  cannot be defined beyond the  limiting {\it Hagedorn} temperature   $\beta_H =  4\pi\sqrt{\alpha^\prime}$.

\smallskip

 It is straightforward  to extend the above analysis to  closed strings. From eq.\,(\ref{massclosedQ}) we see 
 that  in the critical dimension $d=26$ the  ground state is a tachyon  with $\alpha^\prime M^2 =  -4$.
  The first excited  states $a^j_{-1} \tilde a^{\,k}_{-1} \vert 0\rangle$ are massless. 
  They  transform as a general   2-index tensor of $SO(24)$, whose
  traceless symmetric, anti-symmetric and trace parts are  identified in the second-quantized string theory with 
     fluctuations of the space-time metric $G_{\mu\nu}$, an  antisymmetric
  field $B_{\mu\nu}$  and a scalar field   $\Phi$. We will discuss these in more detail after introducing the superstring. 
More generally the states of a closed string  transform
   as tensor products of two  copies of the open string  at level $\hat N= \hat N_L= \hat N_R$. Since $ {\cal N}_{\rm closed}(\hat N) =  {\cal N}_{\rm open}(\hat N)^2$
   and  $M_{\rm closed} = 2 M_{\rm open}$ at any given level, the Hagedorn temperature stays  the same. This is to be expected  because  the high-energy
   fluctuations of the string should depend only mildly  on boundary conditions.


\section{Superstrings}\label{sustrings}

Supersymmetry cures the problem of  the tachyon and introduces  string states that  transform in spinor representations of the Lorentz group, i.e.  they are  spacetime fermions. 
We  will employ  the covariant Neveu-Schwarz-Ramond (NSR)  formalism following at first closely 
the classic textbook  \cite{Green:1987sp}
which predates  D-branes.  
 Readers interested in the alternative  Green-Schwarz formulation of the superstring  can also consult  this textbook.   
   
\subsection{Worldsheet supergravity}

   The Polyakov action  eq.\,(\ref{Polyakov})  describes a  conformal field theory (CFT) coupled to  gravity in  two (worldsheet)  dimensions. 
Gravity is  non-dynamical  and serves to impose the vanishing of the energy-momentum tensor, $T_{\alpha\beta}=0$. 
 Since  $T_{\alpha\beta}$ is traceless for  a CFT\,\footnote{On curved worldsheets the Weyl anomaly implies $T_{\ \alpha}^\alpha = {c\over 12} R$ 
  where $R$ is the Ricci scalar and $c$ the central charge of the 
  `matter' CFT  ($c=1$ for a free scalar field and $1/2$ for a free Majorana fermion).
   In critical string theories this anomaly 
   cancels  between `matter' and ghost fields. }
we are left with two equations,  the Virasoro conditions.

  Superstring theory can be likewise  described as  two-dimensional  {supergravity} coupled to a  {superconformal} field theory (SCFT). 
The simplest  SCFT has  a free massless boson $X$ and a free massless Majorana fermion $\psi$ with action
\be\label{fermaction}
S = \int d^2\sigma  \left( -{1\over 2} \partial_\alpha X \partial^\alpha X + {i\over 2} \bar\psi \rho^\alpha\partial_\alpha \psi \right)\  . 
\ee
 We set for now $2\pi\alpha^\prime = 1$.  The    Dirac algebra  $\{ \rho^\alpha , \rho^\beta\} = -2\eta^{\alpha\beta}$  can be  represented
 by purely  imaginary matrices in two dimensions, 
 \be
 \rho^0 = \left( \begin{array}{cc} 0 & -i  \\  i & 0  \end{array} \right)\ , \qquad 
 \rho^1 = \left( \begin{array}{cc} 0 & i  \\  i & 0  \end{array} \right)\ . 
 \ee
 The Lorentz-boost generator $J^{01} = {i\over 2} \rho^0\rho^1$ is thus also  imaginary and  the  spinor  $\psi = (\psi_R, \psi_L)$
 is  real. 
 From  Dirac's  equation  
 \be\label{Dirac}
 0 =\rho^\alpha \partial_\alpha \psi =  \left( \begin{array}{cc} 0 & -2i \partial_-  \\ 2 i \partial_+ & 0  \end{array} \right) 
  \left( \begin{array}{c} \psi_R \\ \psi_L\end{array} \right) 
 \ee
we see  that $\psi_R$ is  a function of $\sigma^-$ and $\psi_L$ is a function of $\sigma^+$. 
The action  (\ref{fermaction})  is invariant under the global supersymmetry transformations
\be
\delta X = \epsilon^T \rho^0  \psi\ , \qquad \delta\psi = -i \rho^\alpha\partial_\alpha X  \epsilon\ 
\ee
with $\epsilon$ a  Majorana spinor.\footnote{This is called  $N=(1,1)$ supersymmetry because there are two  transformations acting  separately on the left-moving and
right-moving sectors.}
Noether's theorem gives  the conserved  {supercurrent}
\be
J_\alpha = - {1\over 2}   \rho^\beta \rho_\alpha \psi \partial_\beta X\   
\ee
which carries  (in addition to the vector)  a  spinor index that was here suppressed. 
Thus  $J$  
 has four  components, but 
two  of them  vanish  automatically by virtue of  the   identity  $\rho^\alpha \rho^\beta \rho_\alpha = 0$ which
 implies   $\rho^\alpha J_\alpha = 0$. This  is the supersymmetric partner  of the zero-trace condition $T_\alpha^{\ \alpha}=0$ for  a SCFT. 
The  remaining components can be recast into the combinations 
 \bea
 J_+ = {1\over 2} (J_0 + J_1) = \,   \left( \begin{array}{c} 0  \\ \psi_L\partial_+X \end{array} \right) 
   \quad  {\rm and}\quad  
 J_- = {1\over 2} (J_0 - J_1) =    \left( \begin{array}{c}   \psi_R\partial_-X \\ 0  \end{array} \right) \  .
 \eea
If we endow  $X$ and $\psi$ with a spacetime-vector index $\mu$, as in the   bosonic string,  the non-vanishing supercurrents read
 $ \psi_L^\mu \partial_+X_\mu$ and $\psi_R^\mu \partial_-X_\mu$. 
The idea is to couple  the SCFT   to non-dynamical supergravity in order  to impose the constraints  $J_\pm =0$
from an action principle \cite{Brink:1976sc}.


 Let us see how this works.
  Coupling a spinor  to gravity requires  an orthonormal frame (or zweibein)   $e^\alpha_a$ where $a$ is a flat-space  index.
   We have  $e_a^\alpha e_b^\beta g_{\alpha\beta} = \eta_{ab}$ and  $e^a_\alpha e^b_\beta\eta_{ab} =  g_{\alpha\beta}$ for the inverse zweibein. 
   Spinors transform under Lorentz
   rotations of the local frame, but they are scalars under diffeomorphisms. 
 Their coupling to gravity is through  the {spin
   connection}\, $\omega_{\alpha\  b}^a$ which can be expressed in terms of the zweibein  if one  insists 
    that the frame be covariantly constant,
   \be
   0 = D_\alpha e^a_\beta\  \Longrightarrow  \omega_{\alpha\  b}^a = e_b^\beta (\partial_\alpha e^a_\beta - \Gamma_{\alpha\beta}^\gamma e_\gamma^a)\ . 
   \ee
 Here $\Gamma_{\alpha\beta}^\gamma$ is the affine connection, and the
 spin connection  $\omega_{\alpha}^{ab}$ is antisymmetric in the flat indices  $(ab)$. 
  The coupling of the free massless  multiplet to  gravity reads 
  \be
  S_2 = -{1\over 2} \int d^2\sigma \sqrt{-g} ( g^{\alpha\beta} \partial_\alpha X \partial_\beta X  -i \bar\psi \rho^\alpha D_\alpha \psi )\ , 
  \ee
 where $D_\alpha\psi \equiv  (\partial_\alpha + {1\over 4}\omega_\alpha^{ab} \rho_{ab} ) \psi$ , $\rho_{ab} \equiv {1\over 2} [\rho_a , \rho_b ]$, and
 $\rho^a = e^a_\alpha \rho^\alpha$ are the flat-space Dirac matrices.  Under local supersymmetry transformations
 \be
 \delta S_2 = -2 \int d^2\sigma \sqrt{-g} (D_\alpha \bar\epsilon ) J^\alpha\ . 
 \ee
This  can be cancelled by the variation of a gravitino term, 
  \be
 S_3 =   2 \int d^2\sigma \sqrt{-g}\,   \bar \chi_\alpha J^\alpha\   
 \ee
with   $\delta\chi_\alpha = D_\alpha\epsilon$. 
 But this is not  the end of the story because $J^\alpha$  
transforms under   local supersymmetry transformations. The extra variation is however cancelled by the additional term
\be
S_4 = -{1\over 4} \int d^2\sigma \sqrt{-g} \, (\bar\psi \psi ) (\bar\chi_\alpha \rho^\beta\rho^\alpha\chi_\beta)\ , 
\ee
and by modifying the supersymmetry transformations   as follows:
\bea\label{localsusy}
\,&\delta X = \bar\epsilon\psi\ , \qquad \delta\psi = -i\rho^\alpha\epsilon (\partial_\alpha X - \bar\psi \chi_\alpha)\,  \nonumber  \\ 
\,&\delta e^a_\alpha = -2i \bar\epsilon\rho^a \chi_\alpha\ , \qquad \delta\chi_\alpha = D_\alpha\epsilon\ . 
\eea

The action $S_2+S_3+S_4$ is invariant under diffeomorphisms, plus  local Lorentz and  supersymmetry transformations. 
It  has in addition three  `accidental'  local symmetries. 
  The Weyl   symmetry already encountered in the Polyakov string, 
\be
X \to X\ , \quad \psi \to \Omega^{-1/2}\psi\ , \quad e^a_\alpha \to \Omega e^a_\alpha\ , 
\quad \chi_\alpha \to \Omega^{1/2} \chi_\alpha\ ,  
\ee
and its super-extension which only transforms  the  gravitino field, 
\be\label{superWeyl}
\delta\chi_\alpha = i\rho_\alpha \eta\ , \qquad \delta ({\rm rest}) = 0\ . 
\ee
Using  local supersymmetry  
one can  set $\chi_\alpha = i \rho_\alpha \chi$.  This is the extension of the conformal gauge for the bosonic string, with
  $\chi$  the superpartner  of the Liouville field $\phi$ . 
Thanks to superconformal  invariance  $\phi$ and $\chi$ drop out   from the classical supergravity action since
the  corresponding components of the energy-momentum tensor and supercurrent,
   $T_{+-}, J_{+R}, J_{-L}$,   are identically  zero.

  After  the dust has settled,  the  entire supergraviton  multiplet can be   gauge fixed away 
 leaving us with   free fields $(X^\mu,\,\psi^\mu)$. The supergravity equations  impose the vanishing of the remaining four components of $T_{\alpha\beta}$ and $J_\alpha$
on the initial data. More explicitly, these
   `super-Virasoro'  conditions read
    \begin{equation}\label{sVir1}
\eta_{\mu\nu}  (\partial_+ X^\mu \partial_+ X^\nu + {i \over 2}\, \psi_L^\mu \partial_+ \psi_L^\nu)\,=\,
\eta_{\mu\nu}  (\partial_- X^\mu \partial_- X^\nu + {i \over 2}\, \psi_R^\mu \partial_- \psi_R^\nu) \, =\, 0
\end{equation}
\begin{equation}\label{sVir2}
{\rm and}\qquad \eta_{\mu\nu} \,\psi_L^\mu \partial_+X^\nu \, =\, \eta_{\mu\nu} \, \psi_R^\mu \partial_-X_\nu \  =\  0
 \ . 
\end{equation}
They can be solved explicitly by going to the light-cone gauge,  as we did for the bosonic string in section  \ref{sec:2}. 
Indeed,  using the residual freedom that is not fixed by  the superconformal gauge  
one can set\,\footnote{The residual supersymmetry transformations 
correspond to  arbitrary  $\epsilon_R(\sigma^-)$  and $\epsilon_L(\sigma^+)$. From  $\delta \psi^+  =  -i  \rho^+ \epsilon \partial_+ X^+   -i  \rho^-  \epsilon \partial_-  X^+
= \alpha^\prime p^+ (\epsilon_R,  - \epsilon_L)$ we see that these suffice 
 to set the left- and right-moving  components of the on-shell fermion  $\psi^+$ to zero.}
\bea
X^+ = \alpha^\prime P^+ \sigma^0\quad {\rm and}\quad  \psi^+_L=\psi^+_R = 0\ ,   
\eea
whre  $X^\pm = X^0\pm X^1$ and  $\psi^\pm  = \psi^0 \pm  \psi^1$. With this choice
 equations (\ref{sVir1})  and (\ref{sVir2}) become    linear in $X^-$ and $\psi^-$. They  can be used to express  these coordinates  in terms of the independent 
 data $\{X^j, \psi^j\}$ for $j=2, \cdots, d-1$\,.
   
   
\subsection{Neveu-Schwarz and Ramond sectors}\label{NSR2}

 The coordinates $X^\mu$ of the superstring have the same mode expansions as those of the bosonic string, and they obey
   the same canonical  commutation relations (\ref{bosoniccom}). 
   So let us focus on the fermionic coordinates. For closed superstrings the left and right modes are independent,  
  \be\label{modesclosedf}
 {\textcolor{darkgray}{\bf closed}}: \qquad 
  (\psi_R^\mu ,   \psi_L^\mu ) =    (\, \sum_{r}    \psi_r^\mu\,  e^{-ir \sigma^- } \, , \, 
 \sum_{\tilde r}  \tilde \psi_{\tilde r}^\mu\,  e^{-i{\tilde r }\sigma^+ } ) \ . 
  \ee
Reality of the Majorana spinors implies
 $(\psi_r^\mu)^\dagger  = \psi_{-r}^\mu$ and  $(\tilde \psi_{\tilde r}^\mu)^\dagger  = \tilde \psi_{-\tilde r}^\mu$\,  
and the   canonical anti-commutation relations read
\bea
\{ \psi_r^\mu , \psi_s^\nu \} =\{ \tilde\psi_r^\mu , \tilde\psi_s^\nu \} =  \delta_{r+s, 0}\eta^{\mu\nu}\  . 
\eea
  Here comes now  an important new feature. Since all observables contain an even number of fermions, 
 the $\psi^\mu$ can have  either periodic 
 or antiperiodic  conditions. 
Thus  the mode frequencies $r$ and $\tilde r$ in  the expansions  (\ref{modesclosedf}) can a priori  be  either all integer or  all half-integer.
 We must  however make sure  that   the two supercurrents   $  J_{-R} =   \psi_R^\mu  \partial_- X_\mu$ and  $ J_{+L}= \psi_L^\mu  \partial_+ X_\mu$,  
 which  generate 
 residual gauge symmetries,  are well defined modulo a sign. 
  This means  that 
 \bea
   J_{-R}(\sigma^1 + 2\pi) = \eta_R \, \,  J_{-R}(\sigma^1)
 \quad {\rm and} \quad 
   J_{+L}(\sigma^1 + 2\pi) = \eta_L \, \,  J_{+L}(\sigma^1)
\eea
    where $\eta_L,    \eta_R $ are signs. Put differently,  all  the $ \psi^\mu_R $ must be simultaneously  periodic or antiperiodic,   and likewise for the
 $\psi^\mu_L$.  The choices $\eta =+$ and $\eta = -$   are called  the Ramond ({\bf R})  and the
Neveu-Schwarz ({\bf NS}) boundary conditions. 
Since for the  closed superstring $\eta_L$ and $\eta_R$ are independent  there exist  four possibilities:  {\bf NS}-{\bf NS}, {\bf NS}-{\bf R}, {\bf R}-{\bf NS} and {\bf R}-{\bf R}. 
We will  see that all four are   needed.
\smallskip

What about the open superstrings\,? 
 Varying  the  action of a Majorana fermion on an open   worldsheet  produces  a boundary term
\be
\delta S  = 
 {i\over 2} \int_{\partial\Sigma}  d\sigma^\alpha     \epsilon_{\alpha\beta} \,  \psi^T\rho^0   \rho^\beta  \delta \psi \ . 
\ee
Taking the  boundary along  the $\sigma^0$ direction, this variation vanishes provided that
\be
  \psi^T \rho^0 \rho^1 \delta\psi =  \psi_R\delta\psi_R  - \psi_L\delta\psi_L\bigl\vert_{\partial \Sigma} \, =0
   \quad 
 \Longleftrightarrow
    \quad 
\psi_R  =  \pm    \psi_L \bigl\vert_{\partial \Sigma}\  .  
\ee
The left- and right-moving  components of the Majorana field are  therefore identified  on the boundary up to  a sign. 
As in the case of the closed string we must however make sure that the boundary conditions  preserve superconformal invariance.\footnote{\,This 
is necessary because superconformal invariance is a residual  {\it gauge} symmetry. There is no such requirement for global symmetries which may be
broken by the boundary conditions. An example is  Poincar\'e symmetry which is (partially)  broken by   localized D-branes.} 
 The existence of a conserved supercurrent requires that   $J_{+L} =  \eta  J_{-R} $\  on $\partial \Sigma$, or explicitly
  \bea
   \psi_L^\mu  \partial_+ X_\mu    
  = \  \eta\,  \psi_R^\mu  \partial_- X_\mu   \ \bigl\vert_{\partial \Sigma}\  .  
  \eea
It follows that  $\psi_L^\mu = \eta \psi_R^\mu$  if  $X^\mu$ obeys a Neumann condition, and  
 $\psi_L^\mu = -\eta \psi_R^\mu$  when  $X^\mu$ obeys a Dirichlet condition. 

 An open string has two boudaries, at $\sigma^1= 0, \pi$. Since one sign  can be absorbed by redefining   the supercurrent,
only the relative sign $\eta_0\eta_\pi \equiv \eta$
is relevant. 
There are thus only two sectors of the  open superstring, the Neveu-Schwarz sector ($\eta= - $) and the Ramond  sector ($\eta= + $). 
Choosing    $\eta_0=+$   leads to  the following 
 mode expansion of   fermionic coordinates with   (NN) or (DD)
boundary conditions
 \be\label{modesopenf}
 {\textcolor{darkgray}{\bf open}}: \qquad 
  (\psi_R^\mu ,   \psi_L^\mu ) =  \begin{cases}    \sum_{r}    \psi_r^\mu\,  ( e^{-ir \sigma^- } \, , \, e^{-i{ r }\sigma^+ }) \ & {\rm (NN)}
\\ 
   \sum_{r}    \psi_r^\mu\,  ( e^{-ir \sigma^- } \, , \, -  e^{-i{ r }\sigma^+ } ) \  & {\rm (DD)}
   \end{cases}
  \ee
with $r\in \mathbb{Z}$ in  the Ramond sector and $r\in \mathbb{Z}+ {1\over 2}$ in  the Neveu-Schwarz sector. 
\smallskip
 
 These expansions are all one needs for  parallel identical 
 D$p$ branes. The
  analysis can be extended readily to the case of open superstrings stretching between different D-branes.
  In a nutshell, the fermionic coordinate
   goes  along with its  bosonic partner  for the ride. If this latter obeys 
 the  generic  boundary condition  eq.\,(\ref{openrotated}), the boundary condition of 
  $\psi^1+  i\psi^2$ is the same at $\sigma^1=0$, and the same up to the  sign $\eta$ at $\sigma^1=\pi$. Thus in 
 the Ramond sector   fermionic and bosonic 
   modes have the same mode frequencies, while in the Neveu-Schwarz sector the fermionic ones are shifted compared to the bosonic ones by a factor of ${1\over 2}$. 

 Without further ado we  turn now  to   the spectrum of  superstrings, beginning  with the  closed ones. 

 
 \subsection{GSO and type-II  superstrings}

 The  mass-shell  and level-matching conditions for a  closed superstring are obtained 
   by integrating  the Virasoro constraints (\ref{sVir1}), 
  \be\label{masclosed}
 {\textcolor{darkgray}{\rm\bf  closed\    }} :\qquad 
   \!\begin{aligned}
&{\rm ({\bf NS})} & \hat N_R \, \,  - \scalebox{1.1}{${d-2\over 16}$} \\  
& \, \, {\rm ({\bf R})} & \hat N_R\ \ \ \ \  \ \
\end{aligned}
\Biggr\}  = \
  {\alpha^\prime\over 4} M^2 = \
  \Biggr\{\begin{aligned}
&
 \hat N_L\, \,  - \scalebox{1.1}{${d-2\over 16}$}
 &{\rm ({\bf NS})} 
  \\  
&  \ \ \ \ \  \ \ \hat N_L &  \, {\rm ({\bf R})}
\end{aligned}
 \! 
  \ee    
 where the level operators that measure the total oscillator frequency read
 \be\label{levels}
  \hat N_R\ = \  \sum_{j=2}^{d-1}\,\left(  \sum_{n>0} a_{-n}^j  a^j_{n} \  + \sum_{r>0}  r \psi_{-r}^j  \psi^j_{r} 
  \right)\ ,
 \ee
 and likewise  for $\hat N_L$  with $\{a^j_{n}, \,\psi^j_{r} \}$ replaced by the left-moving amplitudes $\{\tilde a^j_{n}, \, \tilde \psi^j_{r} \}$. 
The sum over $r$ runs over  positive integers in the Ramond sector and over positive half-integers in the Neveu-Schwarz sector.
 Zero-point oscillations cancel  between bosons and fermions for  Ramond, while for Neveu-Schwarz one finds
 \bea\label{DN=8}
  (d-2)(\sum {n\over 2}  - \sum {r\over 2}) =  (d-2)(-{1\over 24} - {1\over 48}) = - {d-2\over 16}\  \ . 
\eea
 

\noindent From eqs.\,(\ref{masclosed}) we can now obtain the spectrum of the superstring in each of the four 
sectors of  the worldsheet fermions. 
\smallskip

The {\bf NS}-{\bf NS} ground state \ $ \vert 0\rangle_{\rm NS}\otimes  \vert   0\rangle_{{\rm NS}}$\   is a  tachyon for $d>2$, 
and  the first excited states 
$\psi^j_{-1/2}\vert 0\rangle_{\rm NS}\otimes \tilde \psi^k_{-1/2} \vert   0\rangle_{{\rm NS}}$\ 
transform as a
general   2-index tensor of $SO(d-2)$. 
For consistency this latter must be the little group of  massless particles,  
  \bea
  0 = {1\over 2}  - {(d-2)\over 16} \ \Longrightarrow\ d=10\ . 
  \eea
 Critical superstrings live therefore  in {\it ten} spacetime dimensions. In the second-quantized  theory each string state corresponds to a spacetime field.
 The covariant fields that create the above states are the tachyon $T$, the string-frame metric  $G_{\mu\nu}$, 
 the antisymmetric Kalb-Ramond  field $B_{\mu\nu}$,  and the dilaton $\Phi$. 
  All other  {\bf NS}-{\bf NS} states are massive  and decouple from the effective field theory at 
  energies much below the string scale $\alpha^{\prime\  -1/2}$. 
  
 The lowest   states in the  {\bf R}-{\bf NS} sector   are 
$ \vert 0 \rangle_{\rm R} \otimes \tilde \psi^j_{-1/2} \vert   0\rangle_{\rm NS}$ and they are massless. 
They obey level matching in the critical dimension  $d=10$  as  seen from  eqs.\,(\ref{masclosed}). 
 In  contrast however  with the {\bf NS}  ground state which is a  scalar, the {\bf R}   ground state is not unique but it   transforms  non-trivially under  $SO(8)$. 
   The reason is that it  must represent the algebra of fermionic zero modes 
  which commute with $\hat N_L$  and hence with $M^2$. 
   The  canonical anticommutation relations  $\{ \psi_0^j , \psi_0^k\} = 
  \delta^{jk}$ are the same as  the algebra   of  Dirac matrices $\Gamma^j$, 
 so  the  {\bf R} ground states    
 form   an  $SO(8)$  spinor. We conclude that the massless {\bf R}-{\bf NS} carry 
 one  vector index  and one  spinor index, so  they are  {spacetime fermions}. 
The corresponding   fields are a gravitino $\Psi^\mu$ that obeys the trace  condition $\Gamma_\mu \Psi^\mu=0$, 
and a dilatino  ${\Xi}$ (spinor indices are here suppressed). All other  {\bf R}-{\bf NS} states  are massive 
and, since they carry  one  spinor and any number of vector indices, they are all
 spacetime  fermions.

   The {\bf NS}-{\bf R} sector is identical and  contributes
 a second gravitino  field $\widetilde \Psi^\mu$ and a second dilatino ${\widetilde\Xi}$\,,  plus  massive states. Finally the 
 {\bf R}-{\bf R}  ground states transform  as a bispinor of $SO(8)$, i.e. as a matrix with two spinor indices, so they are spacetime bosons.  
 We   denote the corresponding  
 bispinor  field by \textbf{\textit{C}}.  It  can be  decomposed in terms of  $n$-form gauge
  fields as we will  discuss  in a moment.

\smallskip

   The  theory  constructed so far still includes a tachyon. But contrary to the bosonic string,    this instability can here be cured
   by imposing   the  Gliozzi-Scherk-Olive  {(GSO)}\,projections \cite{Gliozzi:1976qd}. 
   These only keep states of even  worldsheet-fermion parity,  separately in  the left- and in the  right-moving sectors. The fermion-parity operators are denoted $(-)^F$ and 
   $(-)^{\tilde F}$ and they obey
   \bea\label{parit}
    \{ (-)^F , \psi^j_r \} = \{ (-)^{\tilde F} , \tilde \psi^j_r \} = 0 \ . 
   \eea
  When strings join or split  parity
 is multiplicative, so  only the even projections are  consistent in the interacting theory. 
   What is less obvious at first   is   that the parity of
 the {\bf NS}  ground state  is odd, so that GSO projects indeed out  the tachyon. A
  proper explanation  involves the construction of vertex operators for the emission of  string 
 states (see e.g.\,\cite{Polchinski:1998rq,Kiritsis:2019npv}). 
 The tachyon vertex operator is odd under both $(-)^F$ and $(-)^{\tilde F}$  
 and is hence projected out as advertized.\footnote{\,More generally, adhoc truncations of the free-string spectrum violate  the symmetry under global reparametrizations of the  worldsheet,
 also called
 {\it modular invariance}, 
which guarantees ultraviolet  finiteness. 
 This  excludes for example  theories  with only  {\bf NS}-{\bf NS}
  sectors,  or  with no  fermion-parity projections at all.
     One  choice that respects  modular invariance  is to keep only  the {\bf NS}-{\bf NS} and {\bf R}-{\bf R} sectors and to perform
    an {overall}   projection $(-)^F(-)^{\tilde F} =+$.
      The ensuing  theories called type-0A or type-0B
       have no spacetime fermions but they are still  tachyonic.  }


     On the Ramond ground states, on the other hand, the parity  operator acts like the product   $\Gamma = \prod_{j=2}^{9} \Gamma^j$. 
 This follows from the anticommutation relations 
  (\ref{parit}) for  $r=0$. The GSO-projected Ramond ground state is therefore a Weyl spinor of $SO(8)$ and it is also Majorana because
  the fermionic zero modes are real. 
 There exist two inequivalent Weyl-Majorana
     representations of $SO(8)$   denoted  {\bf 8}$_s$ and {\bf 8}$_c$.
  Which one we 
     declare  to have even  parity  is a matter of convention, it depends on  whether we identify $(-)^F$ 
     and $(-)^{\tilde F}$
     with $\Gamma$ or with $-\Gamma$. 
     What is  physically  relevant  is the {\it relative}  chirality of  the left- and right-moving 
     Ramond ground states. The theory where  the two  
  have opposite chirality is called { {type IIA}},
     the one with the same chirality is called {{type IIB.}}   

      What do the GSO projections  imply for the  {\bf R}-{\bf R} gauge fields? Acting on the bispinor   \textbf{\textit{C}} they  read
 \bea\label{chirl}  
\textbf{\textit{C}} =  \Gamma \textbf{\textit{C}}   = 
 \  \begin{cases} \, -  \textbf{\textit{C}} \,\Gamma  \quad &{\rm for \ IIA}\ , \\
 \,  +\textbf{\textit{C}} \,\Gamma
 \quad &{\rm for \ IIB}\ . 
 \end{cases}
  \eea
Now a complete basis of matrices in spinor space is given by all  antisymmetric products\  
  $\Gamma^{i_1 i_2 . .  i_n}\equiv \Gamma^{[\,i_1}\Gamma^{i_2} \cdots \Gamma^{i_n]} $. 
    Thus the bispinor
 can be decomposed into  a sum of  antisymmetric $n$-form fields 
  \bea
  C_{i_1 i_2 . .  i_n} \ \equiv \  {\rm tr} (\,\textbf{\textit{C}}\, \Gamma_{i_1i_2. . i_n}) \  . 
 \eea
The projections  (\ref{chirl}) then imply that there are only odd-$n$ forms in the IIA theory, and only even-$n$
forms in the  IIB theory. In addition  the   identities 
\bea\label{gammasign}
 \Gamma \, \Gamma_{i_1i_2. . i_n} =  {(-)^{n(n+1)/2}\over (8-n)!}
\, \epsilon_{i_1 i_2 . . . i_8} \Gamma^{i_{n+1} . . . i_8}
\eea
 relate  $n$-forms and   $(8-n)$ forms. 
 Thus the type-IIA theory has independent 1-form and 
3-form   {\bf R}-{\bf R} fields  $C_\mu$ and $C_{\mu\nu\rho}$, 
 while the type-IIB theory has 
  a 0-form (or scalar) $C$, a 2-form $C_{\mu\nu}$ and a self-dual  4-form  $C^{\,{\rm s.d.}}_{\mu\nu\rho\tau}$\,. 
All this is summarized in   table  \ref{table22} below.

\begin{table}[h]
\begin{center}
\vskip 3mm
 \begin{tabular}{|c|c|c|}
  \hline
      \, &   \, & \\ 
 {\normalsize \bf string\  states} & {\normalsize\bf   IIA \ theory}& {\normalsize\bf   IIB \ theory}  \\
      \, &.  \, & \\
  \hline \hline
     \, &  \, & \\
\  {\normalsize $ \psi^j_{-1/2}\vert 0\rangle_{\rm NS}\otimes \tilde \psi^k_{-1/2} \vert   0\rangle_{\rm NS}  $\ }
   & \  \, 
  {\normalsize $G_{\mu\nu}, B_{\mu\nu}, \Phi $}   & 
  {\normalsize $G_{\mu\nu}, B_{\mu\nu}, \Phi $}  
   \\
   \, & \, &  \\
    {\normalsize $ \vert 0 \rangle_{\rm R} \otimes  \vert 0 \rangle_{\rm R} $}  & \ {\normalsize  $C_\mu , \,C_{\mu\nu\rho}$} 
    &\  {\normalsize  $C ,\, \,C_{\mu\nu}, \,C^{\,{\rm s.d.}}_{\mu\nu\rho\tau} $} \\
     \, &  \, &  \\
     \hline
       \, & \, &  \\
 {
$ \begin{aligned} & \scalebox{1.15}{$ \vert 0 \rangle_{\rm R} \otimes \tilde \psi^j_{-1/2} \vert   0\rangle_{\rm NS}$}
 \\[1ex] 
 &
 \scalebox{1.15}{$ \psi^j_{-1/2} \vert   0\rangle_{\rm NS}  \otimes  \vert 0 \rangle_{\rm R} $}
 \end{aligned}
 $
 }
  \  &\ \ 
  {
$ \begin{aligned} &  \scalebox{1.2}{opposite\  chirality} \\[1ex]    &\scalebox{1.2}{$\Psi^\mu, \Xi\ \ \&\ \ \widetilde \Psi^\mu, \widetilde\Xi$}  
 \end{aligned}
 $
 }
\ \  &\ \  
  {
$ \begin{aligned} &   \  \ \scalebox{1.2}{same\  chirality} \\[1ex]   &\scalebox{1.2}{$\Psi^\mu, \Xi\ \ \&\ \ \widetilde \Psi^\mu, \widetilde\Xi$}  
 \end{aligned}
 $
 }
   \\
     \, &  \, & \\
   \hline   
   \end{tabular}
\end{center} 
  \caption{\footnotesize The massless fields of the type-II supergravities and the corresponding string states.}
   \label{table22}
\end{table} 


 The massless states of the type-II superstrings are in one-to-one correspondence with  the fields of the two maximal supergravities in ten dimensions. 
The non-chiral type-IIA theory can be obtained from the eleven-dimensional supergravity \cite{Cremmer:1978km} by dimensional reduction. 
The three {\bf NS}-{\bf NS} fields of string theory combine into the eleven-dimensional metric $G_{MN}$,  and the two  {\bf R}-{\bf R}  fields into an antisymmetric  3-form $A_{MNR}$. 
 The chiral type-IIB supergravity \cite{Schwarz:1983qr,Howe:1983sra} does not descend from higher dimensions, and its self-dual 4-form\,\footnote{The field strength of a 4-form gauge potential
 is a 5-form and the self-duality condition reads $\,^*F_{\mu_1\cdots \mu_{5}}
  = {1\over 5 !}\epsilon_{\mu_1\cdots  \mu_{10}}\, 
 F^{ \mu_{6} \cdots \mu_{10}}\ . $ 
 With  Lorentzian
 signature $\,^*(\,^*F) = (-)^{(d-2)/2} F$, so  real self-dual form fields only exist in 2 (mod 4)  spacetime dimensions.
 }
  hinders a covariant 
 Lagrangian description.  Both theories  have 128 bosonic and 128  fermionic on-shell  states. This is the content
 of the maximal supergraviton multiplet whose decomposition in terms of  $SO(8)$ representations is
 $({\bf 8}_v \oplus {\bf 8}_s)\otimes ({\bf 8}_v \oplus {\bf 8}_c)$ for type-IIA and $({\bf 8}_v \oplus {\bf 8}_s)\otimes ({\bf 8}_v \oplus {\bf 8}_s)$ for type-IIB,  where
 ${\bf 8}_v$ stands for   the vector representation.

 
 \subsection{D-branes and spacetime supersymmetry} \label{susyDbr}
 
Superconformal transformations of the string worldsheet  are  gauge  symmetries whose generators  
 annihilate physical states. We argued  in section \ref{NSR2}
that boundaries must preserve a diagonal subgroup of these symmetries in order to eliminate  negative-norm states from the open-string Hilbert space. 
 
  There is no such restriction for {\it global} worldsheet symmetries which may be completely broken by the boundary conditions.
  Consider for example the symmetry under translations of a   bosonic coordinate $X$.  
  The corresponding Noether charge,   $\oint d\sigma^\alpha \epsilon_{\alpha\beta} \mathfrak{J}^\beta$
  with  $\mathfrak{J}_\alpha = \partial_\alpha X$,  is conserved on a closed woldsheet.  There is actually a second  current,  $ \mathfrak{W}^\alpha= 
  \epsilon_{\alpha\beta}
    \mathfrak{J}^\beta$, 
  whose continuity equation 
   $\partial_\alpha  \mathfrak{W}^\alpha= 0$ is an identity,  
    so that the  corresponding charge  
    $  \oint d\sigma^\alpha  \mathfrak{J}_\alpha$ is also  conserved. These two charges are the center-of-mass momentum and winding of a closed string,
     both non-trivial when    $x$ is a compact dimension. 
 
     Now consider an open world sheet. For a  contractible contour  $\oint d\sigma^\alpha \epsilon_{\alpha\beta} \mathfrak{J}^\beta=0$, 
 
        \begin{wrapfigure}{r}{0.42\linewidth}
\centering
\vskip -0.1cm
\includegraphics[width=0.57\textwidth]{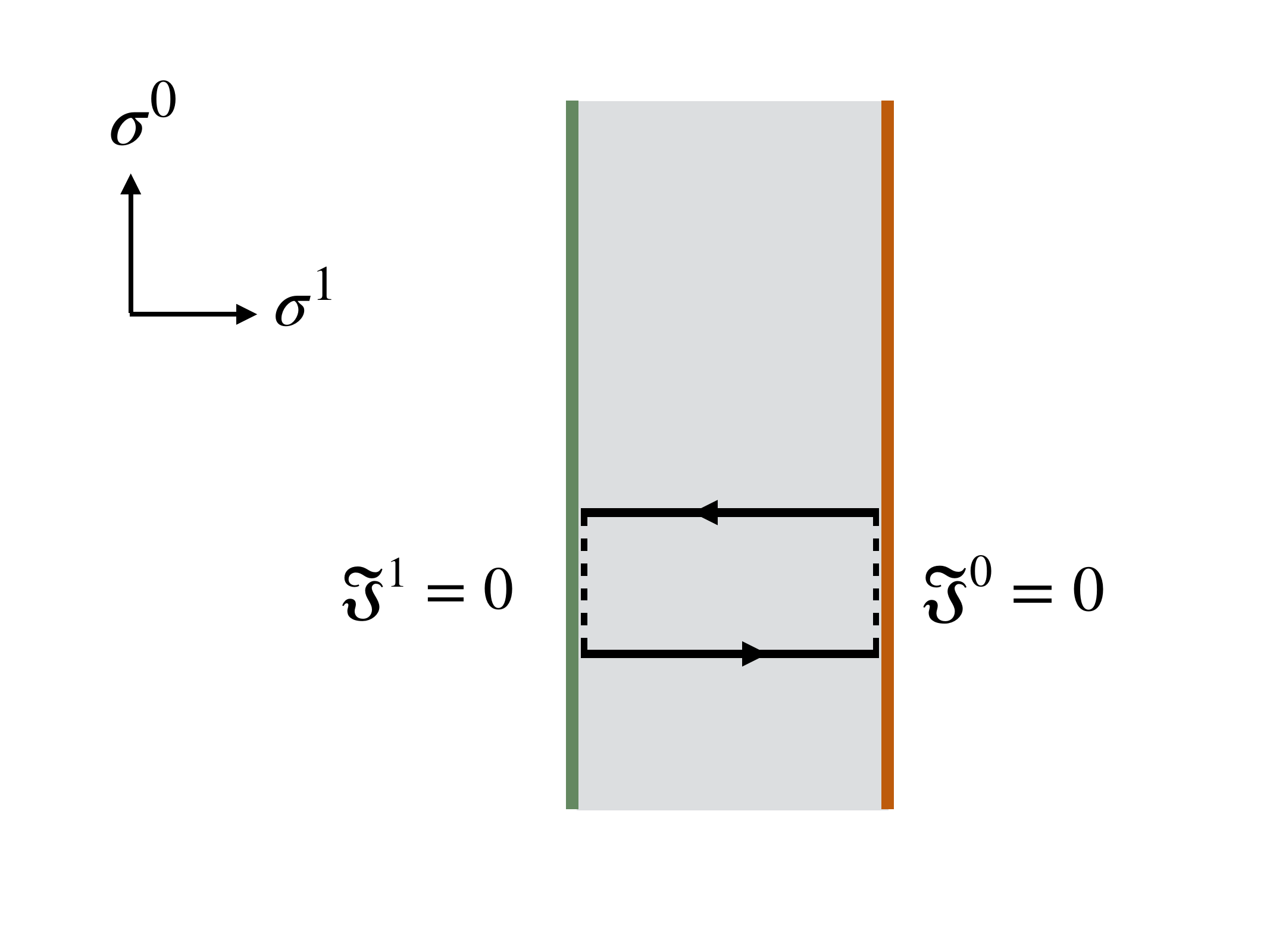}
\vskip -0.5cm
\caption{For a  coordinate   with mixed (ND) boundary conditions neither of the two contour integrals 
vanishes on both boundaries of the open-string worldsheet.}
\label{fig:ws}
\end{wrapfigure}

   \noindent     but to convert this to a conservation law we need $\mathfrak{J}^\perp=0$  on the boundaries, i.e. Neumann boundary conditions.  
    Momentum is indeed conserved when the string endpoints move freely in the $x$ direction. With Dirichlet  conditions at both endpoints
    the conserved charge,  $\int d\sigma^1 \mathfrak{W}^0 = \int d\sigma^1 \mathfrak{J}^1$,  is the minimal  length  of the fundamental 
     string stretching between two localized D-branes.
    For a string with one free and one fixed endpoint, as in   figure \ref{fig:ws}, none of the two charges is conserved. Two D-branes,
    one localized and one extending along the   direction $x$,  force indeed both the  momentum and the stretching of the F-string along $x$  to vanish.

       The discussion can be readily extended  to  spacetime supersymmetry. The corresponding conserved currents    $Q_\alpha$ and 
       $ \epsilon_{\alpha\beta} Q^\beta$,   or the right-   and left-moving combinations $(Q_R, 0)$ and $(0, Q_L)$, 
        carry a 10$d$ Majorana-Weyl
       spinor  index that we suppress. 
    An unpleasant 
       feature of  the NSR formalism is that spacetime supersymmetry is not manifest because
        the explicit expression of  these currents  is somewhat involved. 
All that is needed,  however,  for our purposes here  is   that $Q_R$ is  constructed from the bosonized fermions $\psi^\mu_R$,   and $Q_L$ from  the bosonized  
$\psi^\mu_L$.  Each complex chiral fermion 
$\psi_R^1+ i \psi_R^2$ corresponds to  a free chiral boson $i\psi_R^1\psi_R^2 =  \partial_-\phi_R$\, which contributes a factor $\exp(\pm{i\over 2}\phi_R)$ to $Q_R$,  and likewise
for left movers. We refer the reader to
 \cite{Friedan:1985ge,Polchinski:1998rq} for   details.

On  a closed worldsheet  the  conserved  charges $\oint Q_R$ and $\oint Q_L$ are  independent. They 
    generate  the   ${\cal N}=32$   supersymmetries  of  the  type-II supergravity theories.     
       Consider next an open string with Neumann conditions  for 
  all  $X^\mu$. 
       As explained in section  \ref{NSR2}, their fermionic partners    obey the boundary  conditions  $\psi_R^\mu =   \psi_L^\mu$\,.\footnote{Modulo   an   irrelevant 
       overall sign.}
   The bosonization  formula then shows that,  before the GSO projections, 
   $Q_R=Q_L$ on the boundary,  and so $\int d\sigma^1  (Q_R-Q_L)$ is a candidate for a conserved charge. 
  To  survive,  however,    the  GSO projections $Q_L$ and $Q_R$ must have the same chirality which is the case for the  type-IIB superstring. 
  We  conclude that  supersymmetric space-filling D9 branes  can exist  only  in the type-IIB string theory. 
      
        What about   D$p$ branes with other values of   $p$\,?  If $X^j$ is a  transverse coordinate  obeying  the Dirichlet   condition  $\partial_-X^j = - \partial_+X^j$, 
      then $\psi^j_R = - \psi^j_L$ on the boundary. Thus $Q_R$ is now  identified with  $(-)^{F_j} Q_L$ where  $(-)^{F_j}$  anticommutes with   $ \psi^j_L$ and commutes with
      all the other fermions. Acting on spinors this is the
      operator   $\Gamma \Gamma^j$.
       The  supersymmetries left unbroken by  a D$p$ brane  are therefore in   correspondence  with those conserved  charges 
       \bea\label{Gperp}
        \int_0^\pi d\sigma^1 \left( Q_R - \Gamma_\perp Q_L \right)\ , \quad {\rm with}\quad \Gamma_\perp = \prod_{j\in \perp} \Gamma \Gamma^j\  ,  
       \eea
   that survive the GSO projections.  Since $\Gamma_\perp$ flips (does not flip)  the spinor chirality when $p$ is even (odd),  
 supersymmetric D$p$ branes only exist  in  type-IIB string theory for  $p$ odd  and in  type-IIA string theory for  $p$ even.\footnote{\,If we relax the requirement  of
 supersymmetry   odd-$p$ branes also exist in type IIA and even-$p$ branes in type IIB. These branes are unstable  unless one applies an extra 
 orbifold or orientifold projectionion  \cite{Sen:1999mg} . 
 }The number of  supersymmetries left unbroken by such D-branes is ${\cal N}=16$, i.e. half of the ${\cal N}=32$ supersymmetries of the closed-string theory in the bulk.

    Composite D-branes  break more supersymmetries. Consider  for example a pair of orthogonal D-branes, i.e. such  that all coordinates
    have (NN), (DD),  (ND) or (DN) boundary conditions. The boundary conditions for  supersymmetry currents are $Q_R = \Gamma_\perp Q_L\vert_{\sigma^1=0}$ 
    and $Q_R = \Gamma_\perp^\prime  Q_L\vert_{\sigma^1=\pi}$\,,  where $\Gamma_\perp$ and $\Gamma_\perp^\prime$ are the operators defined in (\ref{Gperp})
    for   the two D-branes at the string endpoints. The unbroken supersymmetries correspond therefore to   solutions of the spinor equation
\bea\label{proj}
  ( \Gamma_\perp )^{-1} \Gamma_\perp^\prime\  {\rm S} = {\rm S}\ . 
\eea
By doing the Dirac-matrix algebra one finds  that ${\cal N}$  depends on the number of mixed (DN) and (ND) dimensions. 
If this number is zero then ${\cal N}=16$, if it is four or eight  ${\cal N}=8$, and in all other cases ${\cal N}=0$, i.e. spacetime supersymmetry is completely broken
unless the number of (DN) or (ND) coordinate is $0$\,(mod$(4)$.

   The generalization to rotated branes is straightforward: One must conjugate $\Gamma_\perp$ and $\Gamma_\perp^\prime$ in eq.\,(\ref{proj}) with the corresponding rotation matrices
   in spinor space. The conditions for unbroken  supersymmetry can be easily worked out, see \cite{Berkooz:1996km}.


\section{{D-branes as solitons}}\label{sec:4}

   Many relativistic field theories  have classical localized solutions that describe 
       non-dissipative   excitations, alias solitons.  Familiar examples are the   monopoles, 
   cosmic strings or domain walls of  Grand Unified field theories.  D-branes can be considered as solitonic excitations of closed-string theory.
  To introduce some basic facts about solitons we will use  a  theory that plays an important role in the study of  D-branes also for other  reasons. This   is the 
   $N=4$
  supersymmetric Yang-Mills (SYM) theory  in four dimensions \cite{Brink:1976bc} (for a review see \cite{Harvey:1996ur}).

 \subsection{$N=4$ super Yang-Mills}\label{sec:4.1} 

A convenient  starting point is the ten-dimensional  SYM  
\bea\label{10dsYM}
 S_{10}^{\rm YM} =   -{1\over 2 g^2} \int  d^{10}x\,  \,{\rm tr} \bigl( F_{\mu\nu} F^{\mu\nu}  +  2i
 \bar\lambda \slashed{D} \lambda
 \bigr)\ , 
\eea
where $\lambda$ is a Weyl-Majorana fermion in the adjoint representation of the gauge group $SU(n)$.
The action  (\ref{10dsYM})  is invariant under the supersymmetry transformations $\delta A_\mu = - i\bar\epsilon \Gamma_\mu \lambda$
  and $\delta\lambda = {1\over 2} \Gamma_{\mu\nu}F^{\mu\nu} \epsilon$,   with $\epsilon$  a 10$d$  Weyl-Majorana spinor. 

\smallskip
Upon  reduction to $p+1$  dimensions the $SO(1,9)$  Lorentz symmetry   breaks to $SO(1,p)\times SO(9-p)$,  where
$SO(9-p)$ is  called the  R-symmetry. 
The gauge  field decomposes into  a  $(p+1)$-dimensional  gauge field $A^{\mu=0, \cdots , p}$ and 
$9-p$ scalars $A^{p+1, \cdots ,  9} \equiv Y^{1, \cdots, 9-p}$. The  gaugino   $\lambda$ 
  reduces to  a number of  $(p+1)$-dimensional gaugini transforming as a spinor  of  the R-symmetry group.\footnote{\,In  $p+1=6$ dimensions  the gaugini transform as  a (pseudoreal) 
doublet of $SU(2)\subset SO(4)_R$, whereas  in  $p+1=4$ dimensions
they  transform as a vector $\lambda^a$ of  $SU(4) \simeq SO(6)_R$.} 
The scalars $Y^j$ have a  potential 
$V=  - \sum_{i,j}{\rm tr}([Y^i , Y^j]^2)/2g^2
 $ 
 descending from the commutator term in $F_{\mu\nu}$ 
 which
 vanishes when the vacuum expectation values $\langle Y^j \rangle$ are  mutually commuting. At a generic point of  this so-called  {{Coulomb branch}}
the unbroken gauge symmetry is  abelian,   $SU(n) \to U(1)^{n-1}$.

 This maximal SYM  theory has some remarkable properties in four dimensions.  First it  is  conformal  because the  coupling $g$ has vanishing beta
 function.  
Second, the coupling  can be  complexified by adding to the action
a {topological term}, 
\bea
\tau = {\theta\over 2\pi} + {4\pi i\over g^2}\  \quad {\rm with}\quad  \Delta S  =  -{\theta\over 32\pi^2} \int d^4x\, 
\, {\rm tr}  (  F_{\mu\nu} \,^* F^{\mu\nu})\ . 
\eea
Here $\,^* F$ is the dual field strength,  and  $\Delta S$ is proportional to the Pontryagin index or  instanton number 
of the gauge field.  Last but not least,  $N=4$ SYM  is conjectured  to be invariant \cite{Montonen:1977sn} under the 
SL(2, $\mathbb{Z}$) duality transformations   
\bea
\tau \to {a\tau + b\over c\tau + d} \ , \qquad 
\eea
where  $a,b,c,d$ are integers with   $ad-bc=1$. 

 Take the gauge group  $SU(2)$. 
  By an $SO(6)_R$ rotation we can bring any point  on the Coulomb branch to
   $\langle Y^1_0 \rangle = gv$, where $0, \pm$ label the neutral and charged components 
 of the adjoint   triplet under the unbroken $U(1)$. Charged fields  
get a mass $M_e= gv$. 
 There exist also classical solutions, the 't Hooft-Polyakov monopoles,   which  far from a smooth core 
 look like   a Dirac  monopole of   $U(1)$.
 Their  mass  $ M_m =  4\pi v/g$ is mapped to that of the   electric charges under the strong-weak   duality $g\to 4\pi/g$. 
 More generally,  SL(2, $\mathbb{Z}$) duality predicts that the $N=4$
  SYM has  dyonic excitations whose  electric and magnetic charges $Q_e, Q_m$ and   mass $M$ are 
   \bea\label{dyoncharges}
  \left(\begin{matrix}
  Q_e \cr -Q_m \end{matrix}\right) = \ \sqrt{4\pi \over {\rm Im}\tau}
   \left(\begin{matrix}  1 & -{\rm Re}\tau  \cr 0 & {\rm Im}\tau  \end{matrix}\right)
    \left(\begin{matrix}  n_e \cr - n_m \end{matrix}\right)  
    \  ,\quad
    M =   v\,\sqrt{Q_e^2 + Q_m^2}\ ,  
  \eea
  where $n_e, n_m\in \mathbb{Z}$ are relatively-prime  integers. 
   Note that for $\theta= 2\pi \,{\rm Re}\tau \not=0$ the magnetic monopole acquires an electric charge through  Witten's  effect \cite{Witten:1979ey}.
 The   dyon mass   is invariant if  the integer charges transform as an   SL(2, $\mathbb{Z}$) doublet, 
 \bea
  \left(\begin{matrix}  n_e \cr -  n_m \end{matrix}\right)  \to   \left(\begin{matrix}  a & b  \cr c & d  \end{matrix}\right) \left(\begin{matrix}  n_e \cr - n_m \end{matrix}\right) \ . 
 \eea

Consider next the low-energy  excitations in  the monopole background. 
They include the  massless neutral fields $(A^\mu, Y^i, \lambda^a)$ living  in the bulk, and  the collective coordinates  $\vec X(t)$ that 
describe  the slow  motion  of the heavy monopole in  space. The  latter are the Goldstone bosons of   broken translation symmetry. 
Since the monopole breaks also half of the ${\cal N}=16$ supersymmetries, there are in addition localized  fermionic zero modes which 
provide  the full spin content of a vector multiplet 
as expected from  duality.

 The effective low-energy action  has  two terms,  $S= S_{\rm bulk} + S_{\rm monopole}$ 
with  $S_{\rm bulk}$   the  supersymmetric Maxwell action  and $S_{\rm monopole}$  the  point-particle action of  the   monopole. 
Keeping  only the  bosonic fields these read
\bea\label{Eftmonopole}
S  =     -{1\over 2} \int d^4x\, \bigl[{1\over 2}F_{\mu\nu} F^{\mu\nu}  + (\partial_\mu \bm{Y})^2\bigr]  
   -{ 
   \scalebox{0.9}{$4\pi$}
   \over  g} \int dt   \left [\, \vert  {\bm Y} \vert  \sqrt{-\dot X^\mu \dot X_\mu}  \,+ \,  
  \tilde A_\mu \dot X^\mu\,  \right] \  
  \eea 
 where $\bm Y$ stands for $(Y^1_0, \cdots , Y^6_0)$, \,$F_{\mu\nu}$ is the  field strength of the unbroken $U(1)$   
 and $\tilde A_\mu$   its  magnetic potential. The bulk fields $\bm Y$ and $\tilde A_\mu$   in  the second term are evaluated on the particle  worldline   $X^\mu(t) = (t, \vec X)$. 
 They were rescaled  in order to have a  canonically normalized  bulk action. Dots stand for  $t$-derivatives. 
 \
       
     By linearizing  this  action around the vacuum 
     $ \langle \bm Y\rangle  = (v, \, 0 \cdots, 0)$ 
       one can show  that there is no  force between  two heavy   monopoles at rest. 
       Put differently  the  Coulomb repulsion cancels precisely  the  Yukawa  attraction  mediated by the Higgs scalar $Y^1$.
       The absence   of a  static force  persists in the complete   theory.
  It  is a consequence of the fact that the 't Hooft-Polyakov  monopoles   leave 
   ${\cal N} = 8$ unbroken supersymmetries 
   and   saturate the 
   Bogomolny-Prasad-Sommerfeld (BPS)   bound,  $M \geq Q_m v$.\footnote{\,The terms `supersymmetric'  and `BPS'   are used interchangeably in the literature though they are not,  strictly speaking, equivalent.}
    There actually exists   a  moduli space
  of  classical multi-centre monopole  solutions with  a   metric  that   gives the leading velocity-dependent force between
    slowly-moving monopoles \cite{Manton:1981mp}.


 \subsection{Effective actions }\label{sec:4.2} 

     The idea is now  to consider the  D-branes as semiclassical solitons of  closed-string theory.
      Whereas the closed strings move freely in the bulk, the open strings describe 
 \begin{wrapfigure}{r}{0.45\linewidth}
\centering
\vskip -0.3cm
\includegraphics[width=0.62\textwidth]{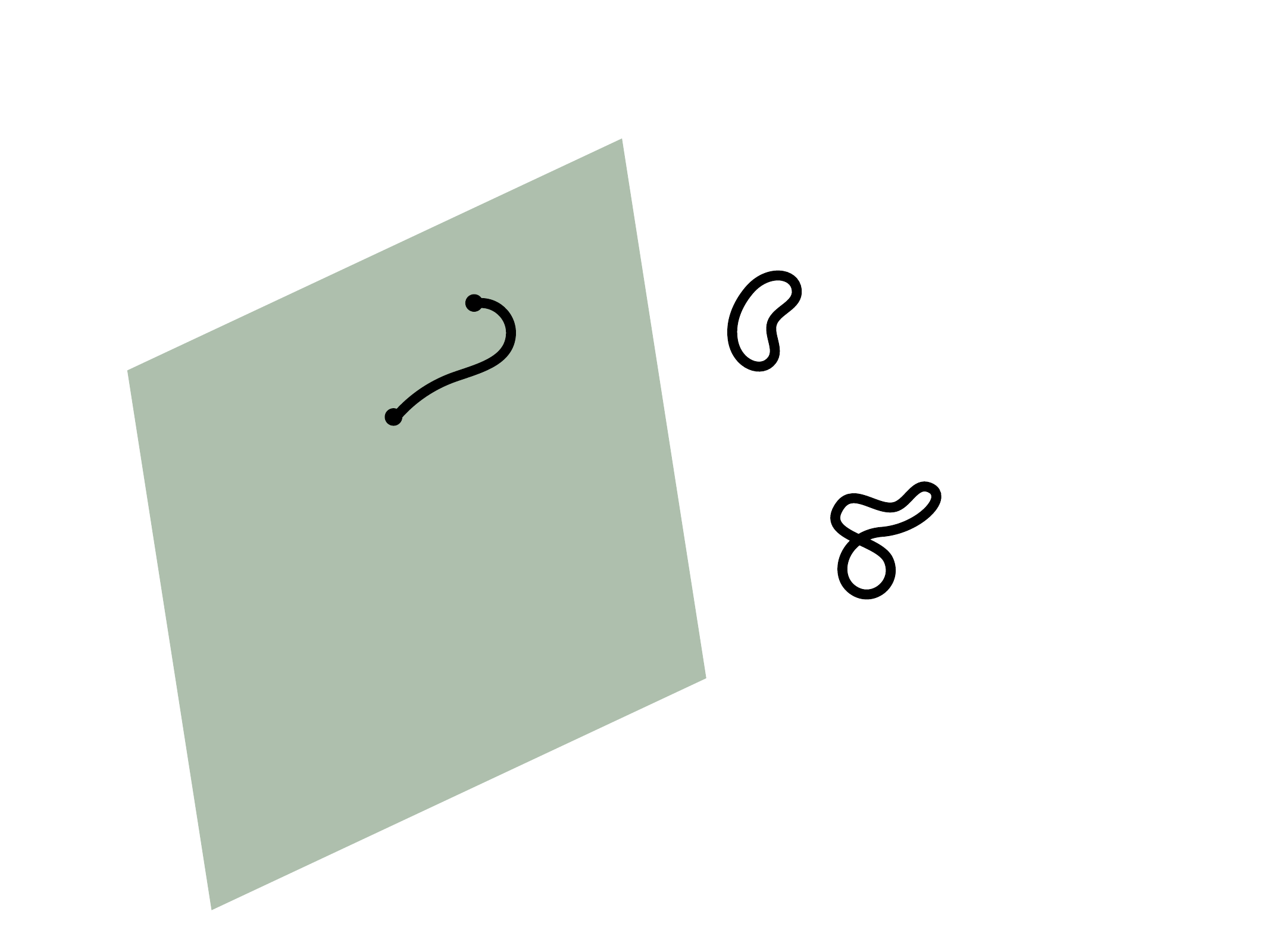}
\vskip 0.3cm
\caption{The D-brane degrees of freedom are open strings that can split or join with closed strings in the bulk.   }
\label{fig:co}
\end{wrapfigure}

 \vskip -4mm \noindent  
  collective coordinates  of the D-brane on which they  attach,  see figure \ref{fig:co}.  The low-energy excitations are  the 
  massless states of both open and closed 
 strings. We will see that  things   fall nicely into place from  this perspective.
 
 In  the   example  
 of  the previous section the effective bulk theory  at energies $E \ll gv$   
 is the   $4d$, $N=4$  supersymmetric Maxwell theory. The  't Hooft-Polyakov monopole on the other hand reduces to a   point particle with magnetic charge. 
In   the case at hand the effective 
  theory in the bulk at $E \ll \sqrt{T_F}$   is one of the type-II   supergravities in ten dimensions.  
The D-brane   degrees of freedom on the other hand are described by a   theory in $(p+1)$ dimensions: 
quantum mechanics for the D-particle, a two-dimensional sigma model for the D-string, or a four-dimensional field theory for the D3-brane. 
At  the  linear  level all these theories turn out to be   our good old friend, the  supersymmetric  Maxwell theory
 reduced from 10    to   $(p+1)$ dimensions.  
 
 This can be seen  with  the methods developed in  section \ref{sustrings}.  The spectrum of an open superstring with only (NN) or (DD) coordinates
 is  basically   isomorphic to the left- or right-moving sector of the closed superstring.  In particular, 
      the massless states of an open string living on a  D$p$ brane are 
  \bea
  \psi^{\,\mu=0, \cdots, 9}_{-1/2}\vert 0\rangle_{\rm NS}\,   \quad {\rm and} \quad
  \vert 0\rangle_{\rm R}\ ,  
  \eea
 corresponding to a  ten-dimensional vector field  $A^\mu$ and a  Weyl-Majorana fermion  $\lambda$. 
 These fields are reduced to $(p+1)$ dimensions, as described below eq.\,(\ref{10dsYM}), 
    because the open string has no centre-of-mass momentum   in the  (DD) directions transverse to the brane. 
    
The scalars $Y^i$ are
 precisely the Goldstone bosons of  broken Poincar\'e symmetry,  and 
  the   gaugini  $\lambda$ the  Goldstini of   broken supersymmetries.  
As for  the gauge field $A^\mu$, it is  the  Goldstone  boson of the broken topological symmetry whose conserved charge is the closed-string  winding number. 
Indeed  a closed string can break up on a D-brane wrapped around  a compact dimension and undo  its winding. 
 Thus the  entire supermultiplet  $(A^\mu, Y^i, \lambda^a)$ is 
 a localized Goldstone supermultiplet   whose existence could be predicted  from symmetries alone. 
 \smallskip

 Let us consider now the effective theory in more detail. We begin with 
   type-IIA supergravity  that can be more succinctly  described  by lifting  to  eleven  dimensions. The  bosonic part of the 11$d$ action reads \cite{Cremmer:1978km}
 \bea\label{Mthry}
 S_{11} =  {1\over 2\kappa_{11}^2} \int \hskip -0.8mm
 d^{11}x   \left[ \sqrt{-G}   ( R   - {1\over 2} \vert F_4\vert^2 ) - {1\over 6} A_3 \wedge F_4\wedge F_4
 \right] \ 
\eea
 with  $A_3$  a 3-form gauge potential,   $F_4=dA_3$ its 4-form  field strength, $G_{MN}$ the metric and 
  $\kappa_{11}$ the gravitational coupling.  
  The normalization of $\vert F\vert^2$ is such that each component squared appears with unit coefficient. 
  When no confusion is possible subscripts will henceforth
  indicate  the rank of a form. 
    The Chern-Simons  term at the  end  is the integral of an 11-form,  and  $\wedge$ stands for wedge product. 

 Reduction  on the  circle $x^{10}= x^{10} + 2\pi$ gives the type-IIA action. The natural fields  on the string worldsheet (the  `string-frame'  fields) 
 are given by the following combinations 
$$
    G_{MN} \,dx^M dx^N =  e^{-2\Phi/3} G_{\mu\nu}\,  dx^\mu dx^\nu + e^{4\Phi/3} [dx^{10} + e^{-\Phi} C^{\prime}_{\,\mu}dx^\mu\,]^2
$$ \vskip -8mm
\bea\label{redef} 
A_3 = B_{2} \wedge dx^{10} \, + \,  e^{-\Phi} C^\prime_{3}
\eea
where $M= 0, \cdots , 10$ and  $\mu=0, \cdots , 9$\,. 
We can  recognize in this decomposition  the massless fields of type-IIA string theory, see table \ref{table22}: the 
NS-NS 2-form $B_{2}$,   the R-R  forms  $C^\prime_{1}$ and $C^\prime_{3}$,  and the dilaton $\Phi$. The reason for the prime will be clear in a minute. 
Inserting   (\ref{redef}) in  $S_{11}$  and dropping derivatives with respect to  $x^{10}$ gives the type-IIA supergravity action,  
\ $
S_{\rm IIA} = S_{\rm NS-NS} + S_{\rm RR} + S_{\rm CS} 
$
with \smallskip
$$
S_{\rm NS\,NS} = {1\over 2\kappa_{10}^2} \int \hskip -0.3mm
 d^{10}x  \,  \sqrt{-G} \,  e^{-2\Phi}  \bigl( R    + 4 (d\Phi)^2 - {1\over 2} \vert dB_2\vert^2 \,   \bigr) \ , 
$$\vskip -5mm
$$
S_{\rm RR} =  - {1\over 4\kappa_{10}^2} \int \hskip -0.3mm
 d^{10}x  \,  \sqrt{-G} \,    \Bigl( \,  \bigl\vert\,  d(e^{-\Phi} C^\prime_1) \, \bigr\vert^2  +  \bigl\vert d(e^{-\Phi} C^\prime_3) -  e^{-\Phi}  C^\prime_{1} \wedge dB _{2} \bigr\vert^2  \bigr)\,  \ , 
$$\vskip -6mm
\bea\label{IIAS}
S_{\rm CS} =  - {1\over 4\kappa_{10}^2} \int B_2 \wedge d(e^{-\Phi} C^\prime_3)\wedge d(e^{-\Phi} C^\prime_3)\ \  . 
\eea
The merit of this   clumsy rewriting is to exhibit  an overall factor  $e^{-2\Phi}$. The coupling of a dilaton background to the string is 
via a term\, 
$-\int {d^2\sigma\over 4\pi} \Phi \sqrt{-g} R$ in the worldsheet action. 
As explained  in section \ref{subsec:21},  the factor $e^{-2\Phi}$ then shows    that all terms in (\ref{IIAS})  are classical, i.e. 
come  from closed-string sphere  diagrams. 
Unless otherwise stated the vacuum expectation value of $\Phi$ will  be  absorbed  in   $\kappa_{10}$, so that $\langle \Phi\rangle=0$.

    Now that we have   exhibited the string-loop counting parameter, it  makes more sense to eliminate the mixing of
    $\Phi$ with  other fields by going to the `Einstein frame'
    \bea\label{Einsteinf}
     G_{\mu\nu} =   e^{\Phi/2} g_{\mu\nu}  \   \quad {\rm and} \quad  C^\prime_n = e^{\Phi} C_n \ . 
    \eea
   After this last  field redefinition  the action (\ref{IIAS}) simplifies to
$$
S_{\rm NS\,NS} = {1\over 2\kappa_{10}^2} \int \hskip -0.3mm
 d^{10}x  \,  \sqrt{-g} \,   \biggl( R   -  {1\over 2}   (d\Phi)^2 -  {1\over 2}  e^{-\Phi} \vert dB_2\vert^2 \,   \biggr) \ , 
$$\vskip -6mm
$$
S_{\rm RR} =  - {1\over 4\kappa_{10}^2} \int \hskip -0.3mm
 d^{10}x  \,  \sqrt{-G} \,    \Bigl( \,  e^{-\Phi/2}\bigl\vert\,  dC_1 \, \bigr\vert^2  + e^{-3\Phi/2} \bigl\vert d C_3 -    C_{1} \wedge dB _{2} \bigr\vert^2  \bigr)\,  \ , 
$$\vskip -6mm
\bea\label{IIASe}
S_{\rm CS} =  - {1\over 4\kappa_{10}^2} \int B_2 \wedge d C_3\wedge d C_3\ . 
\eea
This action is invariant under 
 10$d$  diffeomorphisms and  gauge transformations of the form fields. The   transformation  of the 1-form,   
 which is a metric component in eleven dimensions, corresponds to  reparametrizations 
  $x^{10} \to x^{10} + \xi(x^\mu)$.  But as seen  from eq.\,(\ref{redef}) this also transforms the 3-form, so the gauge transformations of the RR fields are
  \bea
\delta C_1 = d\xi\quad {\rm and} \quad  \delta C_3 =  d\Lambda_2 + B_2\wedge d\xi\ . 
\eea
The    field strength  $F_4 = dC_3$  is therefore an exact but not gauge-invariant form, whereas  the modified 
 field strength
 $\tilde F_4 = dC_3 - C_1\wedge dB_2$    is gauge invariant but not exact since 
$d \tilde F_4 = - d C_1\wedge dB_2\not= 0$.  We take  note of this subtlety,  but   
 bypass  it   for the moment  by setting  $B_2=0$.

   Consider  next the effective action of   D-branes, starting  with the more intuitive case of a D-particle. 
    In the spirit of effective theories one  writes   all possible terms   consistent
with symmetries plus any additional  knowledge  of  the   system. For a point particle 
the two most relevant parameters  are its mass  $T_{\rm D0} $ and its charge $\rho_{\rm D0} $, 
and  the leading terms in the effective D-particle action are
 \bea\label{SD0}
 S_{\rm D0} =  - T_{\rm D0}  \int ds \,   \, e^{-\widehat \Phi} \sqrt{-\widehat G_{ss}}  \,  \,+\, \, 
  {  { \rho_{\rm D0}}} \int   \widehat C_{1} \ \  . 
\eea
The   hats  in this expression denote  the pullback of the bulk supergravity fields to  the worldline 
 of the D-particle $Y^\mu(s)$,    explicitly  
   \bea
 \widehat \Phi =  \Phi(Y(s))\ , \quad
\widehat G_{ss} =   G_{\mu\nu}(Y(s)) {dY^\mu\over ds}{dY^\nu\over ds}\ , \quad  \widehat C_{1} = C_\mu(Y(s)) {dY^\mu\over ds} ds\ .
\eea
A convenient  parametrization is $Y^0(s)=s$, this is referred to in the jargon as the static gauge.  
The remaining D-brane coordinates  $Y^1, \cdots , Y^9$ are  the scalars of the supersymmetric Maxwell multiplet reduced from
10  to 0+1  dimension,  which  is   why we use the same symbol for  them.

  The only   thing  not determined   by symmetries in the action (\ref{SD0}), besides $T_{\rm D0}$ and $ \rho_{\rm D0}$, 
   is the coupling of the dilaton. 
 This is  fixed by the  loop-counting argument used   for the supergravity action. The tree-level interactions 
 of closed and open strings come from  worldsheets with
 the disk topology and since $\chi_{\rm disk} = 1$  we expect a factor of  $e^{-\Phi}$  in the  string-frame action (note that the string-frame RR  field is $e^\Phi C_1$). 
  The  coupling of the dilaton in (\ref{SD0}) is  thus a string-theory input.
 \smallskip

 The extension to  D$p$ branes with $p>0$   is  straightforward. 
In terms of the tension $T_{{\rm D}p}$ and charge density $\rho_{{\rm D}p}$ the low-energy action reads
 \bea\label{SDp}
 S_{{\rm D}p} =  - T_{{\rm D}p}  \int [d^{p+1}s]  \, e^{-\widehat \Phi}\sqrt{-\rm det (\, \widehat G_{\alpha\beta}) } \,+\,
  {  { \rho_{{\rm D}p}}} \int   \widehat C_{p+1} \ \  ,  
\eea 
where $(s^0, \cdots , \,s^p)$  parametrize the brane worldvolume  
$Y^\mu(s^\alpha)$,  the pullback of the metric 
is  $\widehat G_{\alpha\beta} = G_{\mu\nu}(Y) \partial_\alpha Y^\mu \partial_\beta Y^\nu$ and there is a similar expression for the pullback of the antisymmetric tensor  $C_{p+1}$.
Similar to  point particles which couple   to 1-form gauge potentials, $p$-branes may  couple minimally to $(p+1)$-form gauge fields. 
Thus  D2 branes can couple to  $C_3$ while D4 branes and D6 branes  can  couple to the dual forms $C_5$ and $C_7$ defined by
\bea
 \epsilon_p \, dC_{p+1} = \,*(dC_{7-p})\   
\eea
with $\epsilon_p$ a sign, see eq.(\ref{gammasign}).  
What about the  D8 brane  of the  type-IIA superstring\,? It   may  couple to a 9-form
 gauge field whose  equation implies that   the  field strength $F_{10}= dC_9$ is everywhere constant except at the position of  the D8 brane. This explains
    why we did not find $C_9$  among the on-shell string  states of table \ref{table22}.
The piecewise constant $F_{10}$ is called  
 Romans mass and parametrizes a  deformation of type-IIA supergravity in ten dimensions \cite{Romans:1985tz}. 
 \smallskip

 The story  for  type-IIB  is basically  the same   with few notable  differences. 
 First,  the theory has  D(-1) branes, i.e. Dirichlet conditions for all   coordinates including $X^0$. 
They are interpreted as spacetime   instantons,  and   eq.\,(\ref{SDp}) as  their  Euclidean action. 
Second,  there are two kinds of string, the D-string and the  fundamental or F-string. They are  charged, respectively, under the RR and NS-NS 2-forms $C_2$ and $B_2$. 
Third, because $dC_4 = \,*(dC_4)$ the D3-branes carry both electric and magnetic charge -- they are dyons. Finally, type-IIB theory admits space-filling D9 branes
which cannot exist  however on their own but require  an exotic object called `orientifold.'  
I will not discuss the ensuing theory of non-oriented strings (called type-I theory). here. A
 comprehensive  review is  ref.\,\cite{Angelantonj:2002ct}.

It  is   useful for the sequel to rewrite  the D-brane action   in the Einstein frame. 
From eqs.\,(\ref{Einsteinf}) and  (\ref{SDp}) we find
 \bea\label{SDpE}
 S_{{\rm D}p} =  - T_{{\rm D}p}  \int [d^{p+1}s]  \, e^{- (p-3)\widehat \Phi/4}\sqrt{-\rm det (\, \widehat g_{\alpha\beta}) } \,+\,
  {  { \rho_{{\rm D}p}}} \int   \widehat C_{p+1} \ \  .   
\eea 
Note in particular that   the D3 brane does not couple to the dilaton, a fact that plays a key role in  the AdS/CFT correspondence.


 \subsection{D-brane tension  and charge}\label{sec:4.3}

 We will now extract  the D-brane tension and charge  by calculating  the force between two identical D-branes. 
 This is  Polchinski's seminal result \cite{Polchinski:1995mt}. Later we will see  a quicker  derivation  using    T-duality.

The static force between two  D-branes  can be computed in the effective  low-energy theory by
    treating  the branes  as  external sources for  the supergravity fields 
    $C_{p+1}$,   $\Phi$ and $ h_{\mu\nu} = g_{\mu\nu} - \eta_{\mu\nu}  $\,.
  At leading order
    in $\kappa_{10}$ we keep only terms linear in these perturbations, 
 \bea
   S_{{\rm D}p} \ =\     \int d^{10}x\,   (  T^{\mu\nu} h_{\mu\nu} + j_\Phi \Phi + j_C C_{01 \cdots p} ) \, + \, {\rm non\ linear}\ . 
   \   
 \eea  
We use as before  the static gauge  which is convenient for a slowly-moving nearly planar D$p$ brane
\bea\label{staticgauge}
Y^\mu = (s^0, s^1, \cdots , s^p, Y^{p+1}(s), \cdots Y^9(s))\ . 
\eea
 Inserting in  eq.\,(\ref{SDpE}) and expanding around $\Phi=0$ gives
\bea
   T^{\mu\nu} = \,  {1\over 2} T_{{\rm D}p} \,\delta ({\bm x}^\perp) \times
 \begin{cases}
 &\eta^{\mu\nu}\quad {\rm for} \quad 
 \mu, \nu = 0, \cdots , p \\
 &0 \qquad {\rm otherwise}  
   \end{cases}
    \eea  \vskip -4mm
   \bea
    j_\Phi =   -{p-3\over 4} \,T_{{\rm D}p} \, \delta ({\bm x}^\perp)   \quad  {\rm and} \quad 
   j_C = \rho_{{\rm D}p} \, \delta ({\bm x}^\perp) \   , 
  \eea
where ${\bm x}_\perp = (x^{p+1}, \cdots, x^9)$ are the coordinates in the normal directions,  the D$p$ brane is located  at ${\bm x}_\perp = 0$,  and we have dropped  terms involving
${\bf Y}_\perp$ which correspond to open-string excitations.

Consider now  two  parallel D$p$ branes  at  $\vec x_\perp = \vec r$ and $\vec x_\perp =
 \vec r^{\,\prime}$.  
   Their  interaction energy is  given at  leading order  by the exchange of a virtual  graviton, dilaton or
  RR gauge field with the result 
  \bea\label{sourcesource}
  {E}_{\rm int}  \,  {\rm T}_{\rm int}  =  -2\kappa_{10}^2 \int d^{10}x \int d^{10}x^\prime \left[ j_\Phi \Delta  j_\Phi^\prime 
  - j_C \Delta j_C^\prime  + T_{\mu\nu} \Delta^{\mu\nu , \rho \sigma} T_{\rho\sigma}^\prime
  \right]\ ,  
  \eea
where  ${\rm T}_{\rm int}$ is the total   time of interaction,   $\Delta(x, x^\prime)$ is the scalar propagator and  
 $\Delta^{\mu\nu , \rho \sigma}(x, x^\prime)$   the
   propagator of the graviton. 
 The negative sign in   
 the contribution of the antisymmetric RR field  is due to the fact that the   `exchanged component'   $C_{01 \cdots p}$\, 
 is timelike\,.\footnote{This is why the usual Yukawa and Coulomb forces have opposite signs.}

  The
  massless propagators in  10$d$ Minkowski spacetime  read 
$$\hskip -5mm
  \Delta(x, x^\prime) =  \int {d^{10}k \over (2\pi)^{10}}\,{e^{\, ik_\mu (x-x^\prime)^\mu}\over k^2}\  \qquad {\rm and}
$$\vskip -5mm
\bea\label{sprop}
   \Delta^{\mu\nu , \rho \sigma}(x, x^\prime) = \left( \eta^{\mu\rho}\eta^{\nu\sigma} +
    \eta^{\mu\sigma}\eta^{\nu\rho} -  {1\over 4}\eta^{\mu\nu}\eta^{\rho\sigma}\right) \Delta(x, x^\prime)\ ,  
   \eea
where for the graviton we used the  standard de Donder gauge.  Putting everything together gives   after some straightforward algebra
   \bea\label{Epp}
   {E}_{\rm int}  \ =\   2 V_p\,  \kappa^2_{10}\,  {{(\rho_{{\rm D}p}^2 - T_{{\rm D}p}^2)}} \, \Delta_\perp( \vert \vec r - \vec r^{\ \prime}\vert)\ , 
   \eea
 where $V_p$ is the volume of the D$p$-branes which can be  made finite by wrapping
   them  on a    torus,  and 
   \bea
   \Delta_\perp (r) \ =\ 
    {\Gamma({1\over 2}d_\perp - 1)\over 4\pi^{d_\perp/2}}\,\, ({1\over  r})^{7-p}\,   \  
   \eea
   is the scalar propagator  in the $d_\perp = 9 -p$ transverse dimensions.   
 It is worth noting that  the graviton and
   dilaton exchanges  combine to  make the prefactor that multiplies $\Delta_\perp(r)$ the same for all 
   values of $p$.

 We will now repeat the calculation in string theory and compare. Since the bulk supergravity fields correspond to  massless closed strings,
 their  exchange  is given by the cylinder diagram shown in figure {\color{red}{5}}. To compute the  full  diagram one would in

\medskip \vskip 2mm


\begin{center}
 \hskip 1.4cm
 \begin{tikzpicture}[scale=1.0]
 \draw[pattern=north west lines, pattern color=ovgreen] (0,-1) -- (0,3) -- (2,2) -- (2,0) -- cycle; 
 \draw [thick] (1,1.35) arc (90 :270 :0.35) ;
  \draw[dashed,thick] (1,1.35) arc (90 :-90 :0.35) ;
\draw (4, 2.4 ) node{closed}; 
\draw (9.5,1.5) node{open}; 
\draw [->,thick] (9.5,1.2) arc (90 :270 :0.3 ) ;
\draw [thick] (1,1.35) -- (7,1.35) ;
\draw [thick]  (1,0.65) -- (7,0.65) ;
\draw[pattern=north west lines, pattern color=ovgreen] (6,0) -- (6,2) -- (8,3) -- (8,-1) -- cycle;
  \draw [dashed,thick] (7,1.35) arc (90 :270 :0.35) ;
  \draw [thick] (7,1.35) arc (90 :-90 :0.35) ;
 \draw[->, thick] (3.5,2) -- (4.5,2) ; 
 \node[cylinder, 
    draw = black, 
    cylinder uses custom fill, 
    cylinder body fill = mygray, 
    cylinder end fill = ashgrey,
    minimum width = 0.85cm,
    minimum height = 6.73cm] (c) at (3.9,1) {};
\end{tikzpicture}
\end{center}
\vskip 3mm
\centerline{\small {\bf Fig. 5}\ \ A closed-string exchange or an open-string loop are given by the same cylinder diagram.}

 \newpage

\noindent  principle need  to  
 know how each closed-string state couples to the D-branes. 
But there is another way to view the  cylinder,  as a loop of an open  string stretching between 
the D-branes. Since we know the boundary conditions on the open-string worldsheet,  the diagram can be readily computed.\footnote{In  the open-string channel  the cylinder diagram 
 resembles a Casimir force, i.e. a force between probes that alter the   fluctuations of the vacuum.  }

Let us recall the calculation of vacuum energy in ordinary Quantum Field Theory.
For a scalar field in $d$ flat spacetime dimensions the one-loop vacuum energy is
$$
{E}_0 
\,   {\rm T }  = -{1\over 2} {\rm log\,  det}  (-\partial^2 + M^2) = -{1\over 2} \int {d^dx \,d^dk\over (2\pi)^d}\,\, 
{\rm log}  (k^2 + M^2)  \    
$$\vskip -4.5mm
 \be
\Longrightarrow \ \ {{E}_0\over V}   =  -{1\over 2} \int_0^\infty {dt\over t} \int {d^dk\over (2\pi)^{d}}\, e^{-(k^2+ M^2)\,  t }\  =\ 
-  \int_0^\infty {dt\over 2 t}  (4\pi  t)^{-d/2} \, e^{- M^2 t } \ .  \, \, 
\ee 
Here \,$V$T=$\int d^dx$\,  is the volume  of spacetime, and in 
  the lower line  we have expressed the logarithm using Schwinger's proper
   time.
Repeating the calculation for particles  with spin gives the same  result  times  the number of spin states, while  for fermions
  the sign in front must be  flipped.

   Treating now the stretched open string  as a collection of point particles and using the above expression gives
\bea\label{87}
 {E}_{\rm int} = 
  - V_p \int_0^\infty {dt\over 2t}  (4\pi^2 \alpha^\prime t)^{-(p+1)/2} \, {\rm Str} \, (e^{- \pi  \alpha^\prime M^2    t} \,) \ , 
 \eea
where the supertrace (Str)   stands for  the sum over  bosonic minus fermionic states of the  string,  and  
the momentum integration is over  the $p+1$  
dimensions of the D-brane worldvolume. We have also rescaled the integration  variable $t\to \pi\alpha^\prime\, t$\,  and define
 for convenience $q= \exp(-\pi t)$.   Since all Neveu-Schwarz states are spacetime bosons while all Ramond states  are fermions 
the   supertrace reads
\bea\label{str1}
{\rm Str} \, (q^{\alpha^\prime M^2 }) \ =\  \,  2\times {1\over 2}\, {\rm tr}_{\rm NS} \,  \left(1 + (-)^F\right)  q^{\alpha^\prime M^2 }\,
 - \ 2\times{1\over 2}\, {\rm tr}_{\rm R}  \, \left(1+  (-)^F\right)  q^{\alpha^\prime M^2}\ . 
\eea
The insertions of  the  worldsheet-fermion parity operator $(-)^F$ implement the GSO projection and the factor 2 accounts for the two orientations of the open string. 

We have seen that  the spectrum of an open string between identical parallel D$p$ branes is  isomorphic to one chiral  sector of the  closed  string.
The mass formula, given in  eq.\,(\ref{massopenQ}) for the bosonic string, is readily extended to the superstring   with the result
\bea
 \alpha^\prime M^2 = \hat N \ + \ {\, \,\vert \vec r - \vec r^{\ \prime}\vert^2\over 4\pi^2 \alpha^\prime}\ - \ \begin{cases}  1\ \  &{\rm for \ NS} \\   0  \ \   &{\rm for \ R} \end{cases}
\eea
where the level $\hat N$  is the sum of oscillator frequency.  
  The sum over states   becomes 

 \noindent  then the canonical 
partition function of  eight  free bosons $X^j$  and eight  free  fermions $\psi^j$.  
Partition functions for independent species factorize and  one can compute them separately. 
Each  boson  contributes a factor  $\eta(q)^{-1}$ where 
\bea
   \eta (q)\ =\ q^{1/24} \prod_{n=1}^\infty (1- q^n)\   
\eea
is  the Dedekind eta function encountered   
  in eq.\,(\ref{Dede}). 
For the  fermions the result  depends  on the sector and on whether  $(-)^F$ is inserted or not in the trace.  
 The four terms  in eq.\,(\ref{str1})  are  
  \vskip 3mm

\begin{center}
 \hspace*{0.1\linewidth}
 \begin{tikzpicture}
\draw[fill=mygray] (1,0) -- (2.3,0) rectangle (1,2) -- (2.3,2);
\draw (1.7,-0.4) node{NS}; 
\draw (0.7,0.9) node{1}; 
\draw(5.8,0.9) node{$ = \ q^{-1/6} \prod_{n=0}^\infty (1+ q^{n +1/2})^8 \ \equiv\  ({\theta_3/\eta})^4\ , $  };
\draw(10.6,0.9) node{ (91a) };
\draw [very thick,color=ovgreen] (1,0) -- (1,2.) ;
\draw [very thick,color=ovgreen] (2.3,0) -- (2.3,2.) ;
\end{tikzpicture}
\end{center}


\begin{center}
 \hspace*{0.044\linewidth}
 \begin{tikzpicture}
\draw[fill=mygray] (1,0) -- (2.3,0) rectangle (1,2) -- (2.3,2);
\draw (1.7,-0.4) node{NS}; 
\draw (0.3,0.9) node{$(-)^F$}; 
\draw(6.0,0.9) node{$ = \ - q^{-1/6} \prod_{n=0}^\infty (1- q^{n + 1/2})^8 \ \equiv\  ({\theta_4/\eta})^4\ , $};
\draw(10.6,0.9) node{ (91b) };
\draw [very thick,color=ovgreen] (1,0) -- (1,2.) ;
\draw [very thick,color=ovgreen] (2.3,0) -- (2.3,2.) ;
\end{tikzpicture}
\end{center}


\begin{center}
 \hspace*{0.107\linewidth}
 \begin{tikzpicture}
\draw[fill=mygray] (1,0) -- (2.3,0) rectangle (1,2) -- (2.3,2);
\draw (1.7,-0.4) node{R}; 
\draw (0.7,0.9) node{1}; 
\draw(5.8,0.9) node{$ = \ 2^4\, q^{1/3} \prod_{n=1}^\infty (1+ q^n)^8 \ \equiv\  ({\theta_2/\eta})^4\ , $};
\draw(10.55,0.9) node{ (91c) };
\draw [very thick,color=ovgreen] (1,0) -- (1,2.) ;
\draw [very thick,color=ovgreen] (2.3,0) -- (2.3,2.) ;
\end{tikzpicture}
\end{center}


\begin{center}
 \hspace*{0.043\linewidth}
 \begin{tikzpicture}
\draw[fill=mygray] (1,0) -- (2.3,0) rectangle (1,2) -- (2.3,2);
\draw (1.7,-0.4) node{R}; 
\draw (0.3,0.9) node{$(-)^F$}; 
\draw(6.3,0.9) node{$ = \  {  (1-1)^4} \, q^{1/3} \prod_{n=1}^\infty (1- q^n)^8 \ \equiv\  0 \times ({\theta_1/\eta})^4\ . $};
\draw(10.7,0.9) node{ (91d) };
\draw [very thick,color=ovgreen] (1,0) -- (1,2.) ;
\draw [very thick,color=ovgreen] (2.3,0) -- (2.3,2.) ;
\end{tikzpicture}
\end{center}

\setcounter{equation}{91}

\noindent We have  expressed the products in terms of the Jacobi theta functions $\theta_i(0\vert q)$. 
Note that ({{\color{red} 91d}) is zero because
the unprojected Ramond sector has
 equal numbers of states with even and odd fermion parity. 
   We kept this term  for completeness since it   contributes to other amplitudes. 
 Collecting everything 
   we arrive at the following  expression  for the   interaction energy   eq.(\ref{87}),  
$$
 {E}_{\rm int} = 
  - { 2}\,V_p \times  \int_0^\infty {dt\over 2t}  (4\pi^2 \alpha^\prime t)^{-{(p+1)/ 2}} \,  {{e^{-\vert \vec r -\vec r^\prime\vert^2 t/4\pi\alpha^\prime}
  }} \,  Z_{\rm open} (q = e^{-\pi t})
$$ \vskip -5.1mm
  \bea\label{E0Z}
  {\rm with} \qquad   Z_{\rm open}  =  {1\over 2\eta^{8}} 
  \left[ ({\theta_3\over \eta})^4 - ({\theta_4\over \eta})^4 - ({\theta_2\over \eta})^4 
  \right]\ . 
 \eea

 The first thing to note  about this expression  is that it vanishes by Jacobi's abstruse identity
 $ \theta_3^4 - \theta_4^4 - \theta_2^4  = 0$. Hence  there 
 is no  force between parallel  identical D-branes at any separation $\vert \vec r -\vec r^\prime\vert$. This is a consequence of unbroken supersymmetry, akin to
 the vanishing of the  force between $N=4$ SYM monopoles. 
Comparing in particular with the effective-theory result  (\ref{Epp})  we conclude that $\rho_{{\rm D}p} = T_{{\rm D}p}$, i.e.  the tension of a D-brane is  equal to its charge.
 
  To find the  actual value of the charge we must isolate the exchange of the  RR field  $C_{p+1}$, 
  in other words  separate the NS-NS from  the RR  contributions  in the closed-string channel of the cylinder diagram. 
  For this  it is useful to consider the  path-integral representation of the traces. As  familiar from  finite-temperature field theory, 
   fermions are   antiperiodic on  the (Euclidean) time circle, and periodic after  insertion 
  of $(-)^F$.\,\footnote{ \,This follows from the Feynman-Kac formula and the Grassmann-integral  identities
$$
{\rm tr} (A ) = \int d\bar\theta d\theta\, e^{-\bar\theta\theta}\,
      \langle - \bar\theta\vert A \vert \theta  \rangle\ , \quad  {\rm tr}\left( (-)^F A\right)  = \int d\bar\theta d\theta\, e^{-\bar\theta\theta}\,
      \langle\bar\theta\vert A \vert \theta  \rangle \  ,  
$$ 
where $\vert \theta \rangle  =  \vert 0\rangle - \theta \vert 1\rangle$ and $\langle \bar \theta \vert =  \langle 0 \vert   -   \langle 1 \vert  \bar \theta$
are  coherent  fermionic bra and ket states and $A$ any  2$\times$2-matrix operator acting
 on   $ \vert 0\rangle$ and $ \vert 1\rangle$.
  }
   In   eqs.\,({{\color{red} 91}) the open-string time runs upwards  whereas the closed string propagates
  from left to right. We conclude that  ({{\color{red} 91a,c})  describe the  exchange of a NS-NS closed string, whereas
  ({{\color{red} 91b,d}) that  of  a closed string in the RR sector.

        In order to compare with  effective field theory the D-branes must be sufficiently distant,     $\vert \vec r - \vec r^{\,\prime}\vert^2  \gg \alpha^\prime$,   so that the contribution of
        massive closed-string states is  exponentially small. 
      In this limit the integral in eq.\,(\ref{E0Z}) is dominated by the  region $t \ll 1$\,.  This is the high-energy limit for  the  open string dominated by  
      highly-excited states.\,\footnote{Another way of stating the same thing is  that the longer the stretching of the open string the softer its excitations.      
 } To extract the leading behaviour of  the RR exchange ({{\color{red} 91b}) we use the modular properties of the Jacobi $\theta$-functions,
 \bea
{\theta_4^{\ 4} \over \eta^{12}}\Biggl\vert_{
q= e^{-\pi t}} =  \  \left({t\over 2}\right)^6\, {\theta_2^{\ 4} \over \eta^{12}}\Biggl\vert_{
q= e^{-4\pi/ t}} \  \simeq  \  \left ({t\over 2}\right)^6  2^4\ + O(e^{-4\pi /t})\  . 
\eea
Inserting  in  eq.\,(\ref{E0Z}) gives the  following contribution of the RR antisymmetric gauge  field to the interaction energy  per unit volume 
(${E}_{\rm int}/V_p$): 

$$
   \int_0^\infty {dt\over 2t}\, {t^6\over 4}  (4\pi^2 \alpha^\prime t)^{-{(p+1)/ 2}} \,   e^{-\vert \vec r -\vec r^\prime\vert^2 t/4\pi\alpha^\prime} 
  =\     2\pi (4\pi^2 \alpha^\prime)^{3-p} \, \Delta_\perp(\vert\vec r - \vec r^{\ \prime}\vert)\ . 
 $$
 Comparing with the supergravity result (\ref{Epp}) we finally arrive at
 \bea\label{Dcharge}
  {\boxed{
 T_{{\rm D}p}^2 = \rho_{{\rm D}p}^2 = \ {\pi\over \kappa_{10}^2}\, (4\pi^2 \alpha^\prime)^{3-p}\ . 
 }}
 \eea
\vskip 1mm


This result is  unusual   when compared to  ordinary solitons, i.e. smooth $p$-brane solutions of the non-linear field equations.
 The tension of such solutions scales like
$m^{p+1}/g^2$, where $m$ is the typical mass of  perturbative excitations and $g$ is the strength of their interaction. For example, the mass of the  't Hooft-Polyakov monopole
is  $4\pi  M_e/g^2$,  with $M_e$ the mass of the charged gauge bosons and $g$ the Yang-Mills coupling constant, see  section 
\ref{sec:4.1}. In string perturbation theory the typical mass is $(\alpha^\prime)^{-1/2}$, while $\kappa_{10}$ is related to 
 the string coupling $g_s = \exp(\Phi_0)$   as follows \cite{Polchinski:1995mt}
\bea\label{kappa10}
\kappa_{10}^2  = {1\over 2} (2\pi)^7 \alpha^{\prime\ 4} g_s^2\ . 
\eea
Inserting in (\ref{Dcharge}) shows that $T_{{\rm D}p}$  scales like one inverse power of $g_s$.
Thus D-branes are  heavier at weak coupling than  fundamental strings, but lighter than  conventional solitons such as  the NS 5-brane that we will
encounter  in the coming  section. This unusual $1/g_s$ scaling can be  also related to the fact that the string-loop series expansion
  diverges much faster than in 
 quantum field theory \cite{Shenker:1990uf}.

\smallskip 
   A non-trivial check of formula (\ref{Dcharge}) is that it obeys Dirac's celebrated   condition, or rather its generalization to extended objects in 
   higher dimension  \cite{Nepomechie:1984wu,Teitelboim:1985yc}. Recall that
 in \,$d=4$\,  electric and magnetic charges  must obey $Q_eQ_m= 2\pi n$ for  integer $n$. \\
   \setcounter{figure}{5} 
        \begin{wrapfigure}{r}{0.48\linewidth}
\centering
\vskip -0.65cm
\includegraphics[width=0.85\textwidth]{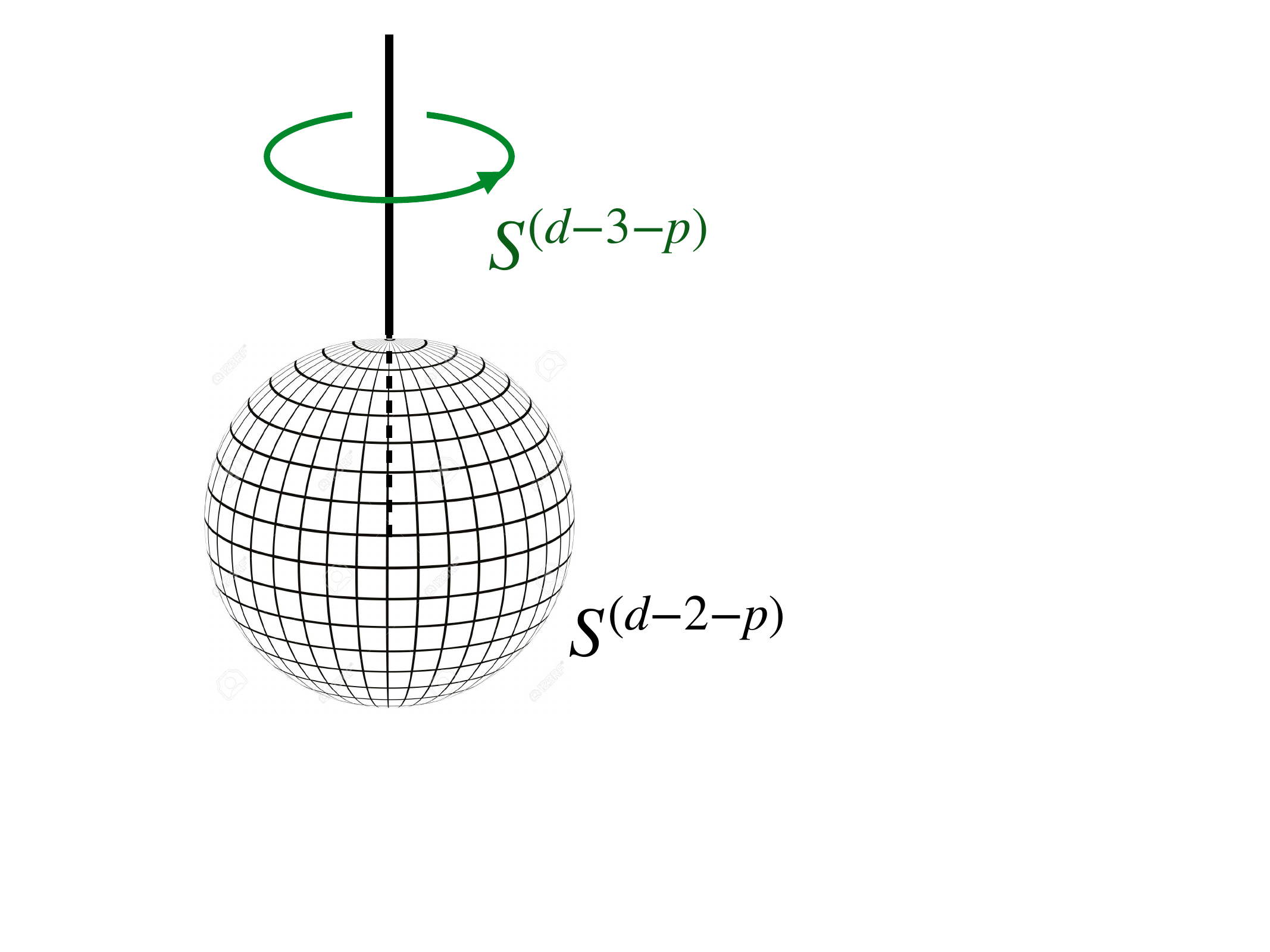}
\vskip -1.2cm
\caption{Dirac's thought experiment  for  extended  electric and magnetic objects in $d$ dimensions.  }
\label{fig:ws}
\end{wrapfigure} 

\vskip -8.5mm

 \noindent This is the condition that 
 the Dirac-string singularity of the monopole field cannot be detected  by  
 quantum interference of electric charges. Figure {\color{red} 6} illustrates the
extension of the argument  to higher dimensions. A $p$-dimensional magnetic object 
produces a field with a Dirac singularity that intersects the surrounding $(d-2-p)$-sphere
 at the north pole.  Electric charges are $(d-4-p)$-dimensional branes which would pick a Bohm-Aharonov phase
by  wrapping  a $(d-3-p)$-sphere around the Dirac string. For $d=4$ both the electric and magnetic charges are  point particles
and the $(d-3-p)$-sphere is a circle.
 For $d=10$ the dual charges  are   $p$-branes and $(6-p)$-branes. 
 In the 
  conventions of the supergravity action  (\ref{IIASe}), the  canonically\\   
   normalized  gauge fields are
\noindent     $\sqrt{2} \kappa_{10}\, C_{p+1}$, and   the quantization condition  for the D-brane charges  reads\,\footnote{\,The interpretation of this condition in the case  $p=-1$ is somewhat  different. 
It  guarantees that the complex
D-instanton  action is single-valued in the backgroud of a D7 brane.
}
 \bea\label{96}
2 \kappa_{10}^2 \,  \rho_{{\rm D}p}\, \rho_{{\rm D}(6-p)} = 2\pi n\  
 \eea
with $n$ integer. The charges in   (\ref{Dcharge})  show that the cylinder calculation is consistent with this quantization condition.

 D-brane charges actually  obey the {minimal} quantization, i.e. with  $n=1$. 
This should be contrasted to $N=4$ SYM   where the elementary dual charges $(1,0)$ and $(0,1)$ satisfy  $Q_eQ_m =  2\pi n$ with $n=2$,
see eq.\,(\ref{dyoncharges}). The explanation of  this fact  is simple:  Fields transforming as doublets of the gauge group $SU(2)$ have 1/2  the charge of the adjoint triplets in the
Coulomb phase. Even though such doublets  are not part of the $N=4$ supersymmetric theory, they can  be coupled consistently and should thus  be able to coexist with
the 't Hooft Polyakov monopoles. In string theory, on the other hand, we just showed that fractional D-brane charges  are not allowed.\,\footnote{Except when localized on orbifolds, 
in which case Dirac's argument does not apply \cite{Douglas:1996sw}.
}
This is in line with the  lore  that string theory does not admit  couplings  to  other forms of matter or to external probes.


\section{Dualities}\label{sec:5}

Part of the initial excitement around D-branes came from the fact that they provided the charged excitations 
predicted by string dualities. This section gives a lightning  account of dualities focussed  on the role of  D-branes.

 \subsection{The power of T-duality}\label{sec:5.1} 

 T-duality is a worldsheet symmetry and hence valid  order by order in the string-loop expansion.
It exchanges  a free  field $X$  with another free  field  $X^\prime$ defined by   $\partial_\alpha X = \epsilon^{\alpha\beta}\partial_\beta  X^\prime$. 
Let $X$ be a compact coordinate, $X = X+ 2\pi$\scalebox{1.2}{$r$}  with \scalebox{1.2}{$r$}  the    radius of the circle.
To simplify the formulae we  set  $\alpha^\prime = 1$, units will be restored when needed.  The mode expansion of $X$ reads
  \bea
  X =  \scalebox{1.3}{${1\over 2}$} ({n\over\scalebox{1.1}{$r$}} - \tilde n \scalebox{1.1}{$r$}) \sigma^- 
  + \scalebox{1.3}{${1\over 2}$} ({n\over\scalebox{1.1}{$r$}} +  \tilde n \scalebox{1.1}{$r$}) \sigma^+\, +
   \, \sum_{k\not= 0}  {i\over k\sqrt{2}} ( a_k^\mu e^{- ik\sigma^-} + \tilde a_k^\mu e^{- ik\sigma^+})\ . 
    \eea
  The   difference with the  earlier expression (\ref{modesclosed}), besides the dummy summation index $k$, is in the zero-mode part of the above expansion. 
 Because $X$ is now compact the centre-of-mass momentum  must be an integer multiple of  1/\scalebox{1.1}{$r$}
and there is  also an integer winding number  $\tilde n$.

 From  $\partial_0 X =  \partial_1 X^\prime$ one sees that 
T-duality interchanges momentum and  winding. Hence  it is an exact equivalence between theories with radii \scalebox{1.1}{$r$} and  \scalebox{1.1}{$r$}$^\prime$= 1/\scalebox{1.1}{$r$},\   
provided one shifts also the    dilaton by a constant
  \bea\label{Td3}
 \Phi^\prime =  \Phi - \log  \scalebox{1.1}{$r$}\  . 
 \eea
 The reason for this dilaton shift  is that the interactions of string states with neither momentum nor winding  in the compact dimension are
 controlled by the effective  coupling in one lower dimension  $ \exp({\Phi_0})/\sqrt{ 2\pi  \scalebox{1.1}{$r$}}$. 
 This is invariant provided the dilaton transforms as above. 
 
   In the bosonic string theory T-duality becomes an exact symmetry at the self-dual radius $r=1$. The story for the superstrings   is slightly twisted. To understand  how,  note
that T-duality flips the sign of the right-moving fields 
  \bea\label{Tdaa}
  (X_R^\prime,  \psi_{R}^{\prime})  =    ( - X_R ,  - \psi_{R}) \ , \quad   (X_L^\prime,  \psi_{L}^{\prime})  =    (X_L ,  \psi_{L} )
  \ . 
         \eea        
For $X$ this is a rewriting of $\partial_\alpha X = \epsilon^{\alpha\beta}\partial_\beta  X^\prime$, while its fermionic partner
  `goes along for the ride' in order  to preserve 
 worldsheet supersymmetry.  Now as was explained  in section \ref{susyDbr}, the operator   that changes the sign of $\psi^j_R$  
is represented on  the right-moving  Ramond ground states   by the matrix  $\Gamma\Gamma^j$.  
 Since this flips the   chirality of the spinor,  
  T-duality  maps type-IIA theory to type-IIB,  and vice versa. It is thus an equivalence but  not a   symmetry even at  $r=1$.

    This can be also understood from the  spacetime perspective.  String theory compactified   on a circle has a $U(1)_L\times U(1)_R$ gauge symmetry 
    whose gauge fields are   Kaluza-Klein components of the metric $G_{\mu\nu}$ and the Kalb-Ramond  field $B_{\mu\nu}$. 
      In the bosonic theory at the self-dual radius $r=1$ this  symmetry is enhanced to $SU(2)_L\times SU(2)_R$  thanks to the appearance 
      of four extra  massless vector  bosons \  
      $$a^\mu_{-1}\vert  n=\tilde n=\pm1\rangle\quad   {\rm and}  \quad \tilde a^\mu_{-1}\vert  n= - \tilde n=\pm1\rangle\ .  
      $$ 
  T-duality   acts as a Weyl reflection of  $SU(2)_R$  
which is   spontaneously-broken at
   generic $r$  by the expectation value of an (adjoint, adjoint) scalar. 
 In the superstring theories  the above vector  bosons are always massive and the $U(1)_L\times U(1)_R$ gauge  symmetry is never enhanced.
\smallskip

  Let us next try to understand how T-duality acts on a  D-brane. 
  The first thing to note is that the transformation  (\ref{Tdaa}) swaps
    Neumann with   Dirichlet boundary conditions. Thus a D$(p+1)$ brane wrapping the dimension $x$  is transformed to a
    localized D$p$ brane in the T-dual theory,  and vice versa. 
     This ties in nicely
    with the fact that the  type-IIA and   type-IIB theories have BPS  D$p$ branes  with $p$ even,  respectively odd.
 Both T-dual objects  look  $p$-dimensional   in the  non compact 9$d$ spacetime. Asking that  their tensions  match gives
      \bea\label{100}
      2\pi r \,  T_{{\rm D}(p+1)}  \, =\,  T_{{\rm D}p}^\prime\ = \ r \,  T_{{\rm D}p}\  , 
    \eea
where in the second step we have used  
    eq.\,(\ref{Td3}) together with  the fact that the D-brane tensions scale like $\exp(-\Phi)$. 
 The result of  the cylinder calculation, 
   eqs.\,(\ref{Dcharge}),    obeys  this relation (after restoring the units $\alpha^\prime =1$)  and 
   is therefore  consistent with T-duality,  as  expected. 
    
   Conversely,  T-duality  determines the 
    tension of  all D$p$ branes up to a common normalization. This can be  fixed by imposing  Dirac's minimal quantization, i.e.  eq.\,(\ref{96}) with  $n=1$. 
  The non-trivial content
   of  the cylinder calculation is  therefore  to show  that the  D-branes form a complete set of RR charges.
\smallskip

Let us take the discussion  one  step further and study  the implications of T-duality for the D-brane action.\footnote{\,A   review of duality constraints on  D-brane actions is
 \cite{Garousi:2017fbe}. 
} 
First we extend the  action of  the fundamental string, eq.\,(\ref{NG}), to account for non-trivial backgrounds of
$G_{\mu\nu}, B_{\mu\nu}$ and of the gauge field  $A_\mu$ that lives on the  D-brane worldvolume,
\bea\label{F1back}
S_F =  - T_F \int_\Sigma   d^2\sigma \sqrt{- {\rm det}(\widehat G_{\alpha\beta})} \ +\,  T_F \int_\Sigma \widehat B_2 \,+\,  T_F \oint_{\partial \Sigma} \widehat A\ . 
\eea
The first two terms are analogous  to those of the  D-brane action, eq.\,(\ref{SDpE}). We have used   the fact that the  fundamental string (alias F1 brane) is a BPS charge
for the Kalb-Ramond  2-form $B_2$, with
hats indicating  the pullback on the worldsheet, 
 e.g. $\widehat B_2 = {1\over 2} B_{\mu\nu}\partial_\alpha X^\mu \partial_\beta X^\nu d\sigma^\alpha\wedge  d\sigma^\beta$ etc. 
The extra  term $\oint_{\partial \Sigma} \widehat A$ shows that  the  endpoint of the string  is charged under  the Maxwell field of  the corresponding D-brane.
We  normalized for convenience  $A_\mu$ so that this charge is $\pm T_F$. 

Note that the last two terms in   (\ref{F1back}) are `topological' terms, i.e. they do  not couple to the metric of a curved worldsheet.  
 They can therefore be  added to  the Polyakov action eq.\,(\ref{Polyakov}) without introducing the Liouville field.

 An immediate consequence of (\ref{F1back}) is that the  gauge invariance $\delta B_2 = d\omega_1$ of closed strings
must be modified  in the presence of D-branes as follows 
\bea\label{gaugeB}
\delta B_2 = d \omega_1\quad {\rm and} \quad \delta A = -\omega_1\bigl\vert_{{\rm D} p}\  , 
\eea
where $\omega_1\vert_{{\rm D} p}$ is the pullback of the 1-form gauge parameter to the D-brane.\footnote{\,We don't use here the hat notation
to avoid confusion with  pullbacks on the F-string worldsheet.}
If the D-branes at the  string endpoints are different, both worldvolume gauge fields  transform as above  modulo  a sign   for orientation.

Consider now a D$p$ brane transverse to a compact dimension, say  $x^9= x^9+2\pi r$. 
We use the static gauge,  eq.\,(\ref{staticgauge}), in which the embedding of the D-brane in  the 9th dimension is given by the coordinate  function $Y^9(s^\alpha)$ 
with $\alpha=0, \cdots, p$. For simplicity  we set all other transverse fluctuations   to zero. 
The pullback of  the RR gauge field  on  the D$p$-brane worldvolume then reads
 \bea\label{103}
 \widehat C_{p+1} =  C_{\mu_0 \cdots \mu_p}  dY^{\mu_0}\wedge \cdots \wedge dY^{\mu_p} \quad {\rm with}\quad \  dY^\mu = (\delta^\mu_{\,\alpha} + \delta^\mu_{\,9}\partial_\alpha Y^9) ds^\alpha\ . 
 \eea
 Let us try to understand how  this  transforms  if   the  coordinate $X^9$ is T-dualized. 
First, as explained above,   the  D$p$ brane becomes  a  D$(p+1)$ brane wrapped around the $x^9$ circle.  
Second, 
   $Y^9(s^\alpha)$ is traded  for the  component $A^9(s^\alpha)$ of the  gauge field on the worldvolume of the D$(p+1)$ brane, see section \ref{sec:4.2}.  
Finally  the  RR bispinor field 
  ${\bm C}$  transforms to $\Gamma \Gamma^9 {\bm C}$. Combining it all and working out the algebra  
   leads to the following transformation of the RR form eq.\,(\ref{103}), 
 \bea
  \widehat C_{p+1} \to  \widehat C_{p+2} +  \widehat C_{p}\wedge F
 \eea
 where  $F=dA$.  Since T-duality is an equivalence of theories,   we have just learned that 
 in addition to $C_{p+2}$ a D$(p+1)$ brane  couples also  to $ C_{p}$
 when its worldvolume gauge  field is switched on  \cite{Li:1995pq,Douglas:1995bn}.

 Extending this reasoning to more  dimensions  and applying it also to  the pullback of the metric in the Nambu-Goto part, 
 leads to the following generalization of the action 
 of an isolated D$p$ brane\,\footnote{\,Our T-duality argument  assumed that   background fields  do not depend on  $x^9$.  
 This is because  T-duality  would trade this for dependence on a dual coordinate $\tilde x^{\,9}$     conjugate to winding rather than momentum, 
 and geometric intuition would be lost. But after discovering the  couplings required by T-duality,  one can  extend them   to $x^9$-dependent backgrounds by locality
 in the $r\gg 1$ limit.
   }

  \bea
  \hskip -3mm  S_{{\rm D}p} \, =\,   -T_{{\rm D}p} \hskip -0.4mm  \int [ds] \, e^{-\Phi}
 \sqrt{ -{\rm det }( \hat G_{\alpha\beta} +\hat B_{\alpha\beta}+  F_{\alpha\beta}) }
  \,+\,
  \rho_{{\rm D}p} \hskip -0.4mm \int    \left[ e^{\hat B+ F}\wedge \hat {\rm \bf C} \right]_{p+1} \ \ \,  
  \label{DpF}
    \eea
where  here ${\rm \bf C} = \sum_n C_n$ is the formal sum of all RR forms, the exponential is likewise
the  formal finite sum $e^{\cal F}  = 1 + {\cal F} + {1\over 2} {\cal F}\wedge {\cal F} + \cdots$, and $[{\cal A}]_{p+1}$ projects   the 
$(p+1)$-form out of the  sum of forms ${\cal A}$.  The coupling of $B_{\mu\nu}$ in this action has been
fixed by   invariance under the gauge transformations (\ref{gaugeB}). The two terms in  eq.\,(\ref{DpF})  are known as the Dirac-Born-Infeld (DBI) \cite{Leigh:1989jq}
and Wess-Zumino (WZ) terms of the D-brane action.

The Wess-Zumino term raises a puzzle about the proper definition of RR charge \cite{Bachas:2000ik}\cite{Marolf:2000cb}. 
To illustrate the problem  consider the D-particle  charge induced on   a spherical D2-brane.  The gauge-invariant charge  \, 
$q=  \rho_{{\rm D}2}\int_{S^2} (\hat B_2 + F)$, the one that couples to the potential $C_1$,  is neither  conserved nor quantized. Indeed, the charge of a  D2 brane
tracing  a worldvolume ${W}$ changes  by an amount $\delta q= \int_W \hat H_3$ which
 does not vanish in general unless $H_3 = dB_2  = 0$. To remedy the situation  one can define the  alternative `Page charge' 
\bea\label{Pagech}
q_{\rm Page} = \rho_{{\rm D}2}\int_{S^2} F\ . 
\eea
This  is indeed quantized,  as  follows from Dirac's quantization of the worldvolume flux $\int_{S^2} F=  2\pi  n / T_F $\,\footnote{\, Recall that $F$ has been 
normalized so that string endpoints carry charge $T_F$.}
and  from the  tension formula eq.\, (\ref{Dcharge}) 
which implies that  $\rho_{{\rm D}2}\int_{S^2} F = n \rho_{{\rm D}0}$. Something however has to give, and this  is the invariance of the
 charge  under large gauge transformations of $B_2$, i.e.
transformations $\delta B_2$ such that $d (\delta B_2)  =0$ but $\delta B_2 \not= d \Lambda_1$. Since $\delta F = - \delta B_2$ such large gauge transformations change $q_{\rm Page}$.  
The non-existence of an invariant quantized RR charge is a general phenomenon  
that  can be traced to the anomalous Bianchi identities and  the Chern-Simons terms of  supergravity actions. 

We  note for completeness, before moving on,   that  it is possible to extend the D-brane action (\ref{DpF}) so as  to include the worldvolume gaugini \cite{Bergshoeff:1996tu}, and
that both the WZ and the DBI terms receive  curvature corrections  \cite{Green:1996dd,Bachas:1999um}.  
These are useful in many applications but they are beyond our scope here.


 \subsection{SL(2, $\mathbb{Z}$) duality of type IIB} \label{sec:5.2}

    In section \ref{sec:4.1}  we gave a rudimentary account of 
     the  duality of  $N=4$ SYM theory. The effective  Maxwell theory
     on the Coulomb branch 
 has a continuous symmetry under   SL(2, $\mathbb{R}$)   
     rotations of $F_{\mu\nu}$ and  $\,-^*F_{\mu\nu}$.  The discrete subgroup SL(2, $\mathbb{Z}$) that respects the quantization of electric and magnetic charges
   is  conjectured to generate exact equivalences of the full quantum theory
        \cite{Montonen:1977sn}. 
 It includes a non-trivial element called  S-duality that maps strong    to
weak coupling,  $\tau \to -1/\tau$. 
   The role of supersymmetry  is  to protect both $\tau$  and the dyon spectrum from quantum corrections,  making it possible to test the   
   conjecture.  
 
      The  U-dualities of string theory \cite{Hull:1994ys,Witten:1995ex,Sen:1994fa} are  similar in spirit but much  richer.
 They are discrete remnants of the  
  exceptional symmetry groups  \cite{BJ} of the maximal  supergravities in various   dimensions, that  are believed to relate 
  consistent  quantum-gravity theories. U-dualities    include both T-dualities and strong-weak equivalences.
 Tests of the  latter use  various protected quantities,   in particular the spectra of BPS 
 excitations. For a comprehensive review the reader may consult  ref.\,\cite{Obers:1998fb}, here we only present a few salient features.

\smallskip 
 
 In  $d=9$ dimensions the U-duality group is SL(2, $\mathbb{Z}$)$\times \mathbb{Z}_2$. The   $\mathbb{Z}_2$ factor  is the    T-duality  of the previous subsection
 that identifies the  type-IIA and  type-IIB string theories. These two theories are  perturbatively equivalent but 
   the action of  SL(2, $\mathbb{Z}$)  on them  is strikingly different. 

 
 We start  with  type-IIB  which  is conceptually simpler. 
 The effective   supergravity has in addition to   the metric $g_{\mu\nu}$,  a complex scalar field $\tau =  C + i e^{-\Phi}$, two 2-form fields $B_2$ and $C_2$, 
  the  self-dual 4-form $C_4$, and their  fermionic partners.   The action
   has a 
 continuous SL(2, $\mathbb{R}$) symmetry  that leaves $g_{\mu\nu}$ and $C_4$ invariant and acts on the other fields as follows
  \bea
  \left(\begin{matrix}
  B_2^{\,\prime}  \cr -C_2^{\,\prime} \end{matrix}\right) = \  
   \left(\begin{matrix}  a & b  \cr c & d   \end{matrix}\right)
    \left(\begin{matrix}  B_2 \cr -C_2 \end{matrix}\right)   \quad {\rm and}\quad 
    \tau^\prime =  {a\tau + b\over c\tau + d} \ . 
 \eea
 It is  conjectured  that  SL(2, $\mathbb{Z}$)$\subset  {\rm SL}(2, \mathbb{R}$)   is  an exact duality of the quantum theory. Let us see how D-branes 
  fit  with this conjecture.

   Begin  with the D-instanton whose tension   reads  $T_{{\rm D}(-1)}= 2\pi/g_s$, see 
   eqs.\,(\ref{Dcharge}) and (\ref{kappa10}). Plugging in the action (\ref{SDp}),   rotating  to imaginary time and 
    reinserting  $g_s= e^{\Phi_0}$ inside the  dilaton   gives  the Euclidean action
   \bea
  S_{\rm D(-1)} = 2\pi i \tau\ .
 \eea
 This makes manifest that  $\tau\to \tau+1$ is a symmetry of the instanton weight $\exp(2\pi i \tau)$
 in the Euclidean path integral. 
The full  SL(2, $\mathbb{Z}$)  duality implies that the  sum  of  all  multi-instanton contributions to any observable quantity
must be  a modular-invariant function of $\tau$. 
Evidence for  this   comes  from  the protected  $R^4$ corrections to the supergravity action \cite{Green:1997as}.

   Consider next the   D-string  whose tension is  $T_{{\rm D}1}= T_{\rm F}/g_s$,  with
    \,$T_{\rm F} = 1/(2\pi\alpha^\prime)$ 
  the  fundamental-string tension.  The S-duality $\tau \to -1/\tau$ 
  exchanges the NS-NS and RR 2-form fields and inverts
  the coupling  $g_s$ when  $C=0$.  We see that the  tension formula 
    is consistent with  the swapping of  F-strings and  D-strings, thus  validating  the S-duality conjecture.
   
  More generally,  the duality predicts 
   an  SL(2, $\mathbb{Z}$) orbit of $(p,q)$ strings which  is isomorphic to  that  of dyons in  $N=4$ SYM. The string tension
     is given by the same  invariant  mass formula 
    eq.\,(\ref{dyoncharges}) 
   with  the replacements  $(n_e, n_m) \to (p,q)$  and with
   $v \to T_F /\sqrt{4\pi g_s}$\,.
      Let us see how the D-brane action (\ref{DpF}) leads to the correct spectrum  of $(p, 1)$ strings. 
      Start with  a D-string wrapping a  circle of unit radius, say  in the $x^1$ direction.
Assume for simplicity  a  constant  axion-dilaton  $\tau = i/g_s$\,. The D-string action in static gauge reads
  $$
 S_{{\rm D}1} =\, -T_{{\rm D}1}\int dx^0 dx^1 \sqrt{- ({\rm det} \,\widehat g + {\cal F}^2)} \, + \  T_{{\rm D}1} \int \widehat C_2 
 $$ 
 where $\hat g_{\alpha\beta}$ is the induced metric, $\widehat C_2$ is the pullback of the RR 2-form and 
 ${\cal F}$ is the $(01)$ component of   the 2-form  $\widehat B_2 + F$. 
 The worldsheet gauge field   has only one degree of freedom, the Wilson line $T_{\rm F} \oint dx^1 A_1 \in [0, 2\pi]$. 
 The  momentum  conjugate to $A_1$ is   constant and  quantized, 
 \bea
\pi_{A_1} =  {  T_{{\rm D}1} {\cal F} \over    \sqrt{- ({\rm det} \hat g + {\cal F}^2)}}  =    p\,  T_{\rm F} \ \ \ 
 \Longrightarrow\ \ \ 
 {{\cal F} \over \sqrt{- {\rm det} \hat g}}  =    { p\,  T_{\rm F} \over \sqrt{T_{{\rm D}1}^2 + p^2 T_{\rm F}^2}}\   
 \eea
 for  integer $p$.  Performing the partial  Legendre transformation
   $S_{{\rm D}1}^\prime  = S_{{\rm D}1} - p\,  T_{\rm F}\int F$\, to  eliminate $A_1$ from the 
action\,\footnote{\,Such a partial Legendre transform is known in classical mechanics as a
 Routhian. It is useful 
for  eliminating  cyclic variables from the action.}
   gives after some  algebra
 \bea
 S_{{\rm D}1}^\prime  = T_{(p,1)} \int dx^0 dx^1 \sqrt{- {\rm det} \,\widehat g } \, + \  T_{{\rm D}1} \int \widehat C_2  +  p\,  T_{\rm F} \int \widehat B_2\ , 
 \eea
 with $T_{(p,1)} = (T_{{\rm D}1}^2 + p^2 T_{\rm F}^2)^{1/2}$\,. This is   the action of a $(p,1)$ string 
 with the  tension  predicted by SL(2, $\mathbb{Z}$) duality. Repeating the calculation with $C= {\rm Re}\,\tau\not=0$ gives the
 more general  formula     (\ref{dyoncharges}). To obtain  arbitrary $(p,q)$ strings one must start with $q$ D-strings and  endow each of them  with  $p/q$ units of  $\pi_{A_1}$. 
 We will see why such fractional momenta are allowed in the coming section.

     The  analogy with the BPS spectrum of $N=4$ SYM is   not a sheer coincidence.  
The  SYM dyons are actually  realized  by  $(p,q)$ strings stretching between parallel D3 branes, so that the duality of $N=4$ SYM follows
from the SL(2, $\mathbb{Z}$) duality of type-IIB string theory (but not the other way around).  
Note that a D3-brane in the  ground state   couples only  to the self-dual 4-form   
and is invariant under SL(2, $\mathbb{Z}$) transformations.

    Consider next  the D5 brane which is a magnetic source for  $C_2$. Since  $(B_2, -C_2)$ transforms  as a doublet of SL(2, $\mathbb{Z}$),   duality also predicts  that there exists
    an  orbit of  supersymmetric  $(p,q)$   five-branes.  Set again  for simplicity the RR scalar $C=0$.
    Dirac's minimal quantization, eq.\,(\ref{Dcharge})  and the BPS relation between tension and charge  determine  the tension of  
   the  (1,0) five-brane which is a  magnetic source  for the NS-NS field   $B_2$\,,   
\bea
 T_{\rm NS5} = {2\pi^2 \alpha^\prime  \over \kappa_{10}^2} \ . 
\eea
 As explained below eq.\,(\ref{Dcharge}),  the  $\kappa_{10}^{-2}$ scaling  is characteristic of  ordinary solitons, 
 i.e. smooth localized solutions of the classical supergravity equations. The
classical  solutions  of superstring field theory correspond to superconformal  $\sigma$-models  on the string worldsheet.  
Such a model does exist for the solitonic NS5 brane 
  \cite{Callan:1991at},  although our understanding of it  
   remains incomplete.  
We will revisit this NS5-brane solution   in section \ref{sec:6.3}.

 Seven-branes must  also form  SL(2, $\mathbb{Z}$) orbits. To understand why recall  that the D7 brane  is a conical point defect in the transverse 
 complex plane with  axion  monodromy
 $\tau \to \tau +1$ that leaves invariant the D-instanton weight $\exp(2\pi i\tau)$. Since  the choice of  SL(2, $\mathbb{Z}$) frame is arbitrary,  we may  trade the fundamental string for 
 a $(p,q)$ string which has the effect of  conjugating the monodromy matrix, 
\bea
 \hskip -4mm   \left(\begin{matrix}
  p  \cr q  \end{matrix}\right) = 
   \left(\begin{matrix}  p  & r  \cr q & s   \end{matrix}\right)
    \left(\begin{matrix}  1 \cr 0 \end{matrix}\right)   \ \ 
    {\rm \longrightarrow }\ \ 
 \left(\begin{matrix}  p  & r  \cr q & s   \end{matrix}\right) 
   \left(\begin{matrix}  1  & 1 \cr 0 & 1   \end{matrix}\right) 
   \left(\begin{matrix}  p  & r  \cr q & s   \end{matrix}\right)^{-1}\hskip -2mm 
   =\left(\begin{matrix}  1-pq   & p^2   \cr -q^2 &  1+pq  \end{matrix}\right)
   \,. 
\eea
So    $[p,q]$ seven-branes must exist  for each pair of relative primes, but only the D7 brane is amenable to a perturbative  treatment. 
It is  sometimes convenient  to 
think of SL(2, $\mathbb{Z}$) as large diffeomorphisms of a 2-torus in a  twelve-dimensional theory  dubbed `F-theory'
\cite{Vafa:1996xn}.
There is however no local   supersymmetric  action  in twelve  dimensions, so this  is no more than  a  mathematical device.

  \subsection{Type IIA and M theory}\label{sec:5.3}  
  
By contrast a local  action  does exist   in eleven  dimensions,  the 11$d$  supergravity of Cremmer, Julia and Scherk \cite{Cremmer:1978km}. 
As we saw
in section \ref{sec:4.2},     type-IIA supergravity is obtained from it by dimensional reduction. 
The Kaluza-Klein  relation
\bea\label{k10}
 {1\over \kappa_{10}^2}\, =\,  {2\pi r_{10}\over  \kappa_{11}^2} \   
\eea 
\vskip -1mm
\noindent   suggests  at first sight
 that  the strongly-coupled type-IIA  sting theory   corresponds to the small-radius limit, but  as we will see   the dimensionless ratio
 $r_{10} \kappa_{11}^{-2/9}$ is actually large. So the limit looks like a theory in eleven dimensions which has been 
  given the provisional name `M theory.' 
 Contrary however  to the case of type IIB   whose 
  strong-coupling limit was another weakly-coupled string theory,  here quantum M theory cannot be independently defined.

  Nevertheless, much   is  known about the  half-BPS excitations of this putative theory \cite{Witten:1995ex,Townsend:1995kk}. 
 To begin with, the
 3-form of 11-dimensional supergravity can couple consistently  to a
supersymmetric membrane \cite{Bergshoeff:1987cm}. Wrapped around the eleventh dimension this looks  like a 10$d$  string.
Second,   there exists a  five-brane soliton that is charged under the (Hodge dual)  magnetic 6-form
  \cite{Gueven:1992hh,Gibbons:1994vm}. 
Wrapped around the circle it looks like a four-brane in 10$d$.
Finally one has 
  the  standard
Kaluza-Klein (KK) modes as well as  the   Kaluza-Klein monopole whose three-dimensional transverse space is described  by 
the   Taub-NUT solution of Einstein's equations \cite{Sorkin:1983ns,Gross:1983hb}.
 The one-to-one correspondence between these  excitations and the branes of
 type-IIA  string theory  is shown in table {\color{red} 3}.

 Missing from the table is the D8 brane that sources  the constant  RR  field strength $F_{10}$. This    was identified  in section \ref{sec:4.2}
 with the  Romans mass   of type-IIA supergravity which  has no
  {{local}} lift to eleven dimensions.\footnote{\,But can be described by a non-local extension of supergravity known as  generalized
 exceptional field theory  \cite{Ciceri:2016dmd}.}

\begin{table}[htp]
\begin{center}
\begin{tabular}{|c|c||c|c|}
 \hline
& & &  \\
{\bf tension}&  {\bf type-IIA}  & {\bf  M theory  on $\bm S^1$}
  &  {\bf tension } \\
&  & &  \\
\hline \hline
 & & & \\
$\displaystyle{\sqrt{\pi}\over  \kappa_{10}} 
\textstyle (2\pi\sqrt{\alpha^\prime})^3 $
 &\  \ D-particle \ \ &  KK modes & $\displaystyle {1\over  r_{10}}$  \\
 & &  &\\
 \hline
 & &  & \\
 \scalebox{1.25}{$ T_{\rm F}= {1\over 2\pi\alpha^\prime}$} & string  &\ wrapped membrane \  &
$2\pi r_{10} \displaystyle \left( {2 \pi^2 \over \kappa_{11}^{\
      2}}\right)^{1/3} $ 
  \\
 & & & \\ 
 \hline
 & & & \\
$\displaystyle{\sqrt{\pi}\over  \kappa_{10}}\textstyle
 (2\pi\sqrt{\alpha^\prime})$ & D2 brane  & membrane \  &
\ $ T_2^M= \displaystyle \left( {2 \pi^2 \over \kappa_{11}^{\
      2}}\right)^{1/3} $  \\
& &  &\\
  \hline
 & & & \\
$\displaystyle{\sqrt{\pi}\over  \kappa_{10}}\textstyle 
 (2\pi\sqrt{\alpha^\prime})^{-1}$
 & D4 brane   & wrapped five-brane &
$r_{10} \displaystyle \left( {2 \pi^2 \over \kappa_{11}^{\
      2}}\right)^{2/3} $
 \\ 
& & & \\ 
 \hline
 & & & \\
$\displaystyle {2\pi ^2 \alpha^\prime\over \kappa_{10}^{\ 2}}\textstyle
 $ &\  NS5 brane & five-brane \  & 
$  \ T_5^M = \displaystyle {1\over 2\pi} \left( {2 \pi^2 \over \kappa_{11}^{\
      2}}\right)^{2/3} $\ 
 \\
& &  &\\
  \hline
 & & & \\
\ $\displaystyle{\sqrt{\pi}\over  \kappa_{10}}\textstyle
 (2\pi\sqrt{\alpha^\prime})^{-3}$\ 
 & D6-brane   & KK monopole & 
$ \displaystyle  {2 \pi^2 r_{10}^{\ 2}  \over \kappa_{11}^{\  2}} $ \\ 
& & & \\ 
 \hline
\end{tabular}
\end{center}
\vskip 0.3cm
\caption{
Correspondence between  half-BPS excitations of type-IIA string theory and of
M
theory compactified on a  circle. As explained in the main text the two sides can be computed independently, 
but  only one of the five relations is real  evidence for the existence
of the 11th  dimension. }
\end{table}


The  tensions in  table {\color{red} 3} are those coupling to the ten-dimensional Einstein-frame metric,   with the quadratic   supergravity action
multiplied by
  $1/2\kappa_{10}^2$. 
The type-IIA entries have been  computed in  section \ref{sec:4}.
  To compare with M theory  we need to express $r_{10}$ and $\kappa_{11}$ in terms of the  string-theory parameters
 $\alpha^\prime$ and $\kappa_{10}$. 
This can be done by using eq.\,(\ref{k10}) and by identifying 
 the D-particle mass with the mass of the first KK excitation.
Then the  remaining M-theory entries are  predictions of the conjectured  IIA/M-theory duality.

Although  we have no  independent formulation of M-theory, we can still compute   the right column  of table {\color{red} 3} from first principles as I now explain. 
 Thanks to BPS saturation one can think of   tension or charge interchangeably. 
The M2 brane and the M5 brane are electric/magnetic sources of the 3-form in eleven dimensions, so they must obey  Dirac  quantization 
\bea\label{Dir11}
 2\kappa_{11}^2\, T_2^{\rm M}T_5^{\rm M} \,=\, 2\pi n\ . 
\eea
String theory predicts that $n=1$. Alternatively this follows   from  the `Completeness Hypothesis'  \cite{Polchinski:2003bq,Banks:2010zn} which states  that all  gauge charges
 obeying  Dirac quantization must be  in the spectrum.\footnote{\,The Completeness Hypothesis is  related to the absence of global symmetries including $n$-form and non-invertible symmetries,  
 see  \cite{Harlow:2018tng,Heidenreich:2021xpr}. } 
 This is one of the best-motivated `principles' of  quantum gravity,  
 so we will  admit  it. 
  
     The condition (\ref{Dir11}) is not sufficient to determine each tension  separately. 
   But a second relation verified by the entries in  table {\color{red} 3}, 
   \bea\label{CS11}
   2\pi\, T_5^{\rm M}\,=\, (T_2^{\rm M} )^2 \ , 
   \eea
   also follows from topological considerations in eleven dimensions \cite{Duff:1995wd}. 
  The main character of the argument   is  the Chern-Simons term in the action (\ref{Mthry}),  which can be written
  as a twelve-dimensional  integral 
  \bea\label{CSM}
  -{1\over 12\kappa_{11}^2} \int_{\cal M} F_4\wedge F_4\wedge F_4
 \eea
 with ${\cal M}$ is an open manifold whose boundary is eleven-dimensional spacetime. 
Now Dirac quantization implies that the  minimum flux through any closed 4-cycle is  \,$T_2^{\rm M}  \int F_4 =  2\pi  $. 
The Euclidean action (\ref{CSM}), on the other hand, should not depend on the choice of ${\cal M}$, which   is the case provided that
\bea
 {1\over 2\kappa_{11}^2} ({2\pi\over T_2^{\rm M} })^3 = 2\pi m\  
\eea   
   for some  $m\in \mathbb{Z}$. Choosing $m=1$   gives  precisely  eq.\,(\ref{CS11}), as can be seen by using Dirac quantization to eliminate from this  equation  $T_5^{\rm M}$\,.   

 The choice $m=1$ is in the same spirit,  but does not follow immediately from  the Completeness Hypothesis. 
One can argue that if  $m> 1$ Euclidean configurations with fractional flux  have
 a well-defined action and should be included. It would be interesting to relate this to global symmetries.

   Let us now take stock of what we have learned. Dirac quantization and the single-valuedness of  the Chern-Simons  action  give four  consistency conditions
  that must hold independently on each side of the  type-IIA/M-theory duality.  
 Combined  with the Completeness hypothesis, which is verified in string theory and is assumed in M-theory, they provide four relations among the six entries of table 
{\color{red} 3}. Since we used the top entry  to establish the $(r_{10}, \kappa_{11})\leftrightarrow (\alpha^\prime, \kappa_{10})$ dictionary, only one relation 
is left as an independent test of the duality conjecture. We can choose it to be 
  \bea\label{118}
       T_{\rm D0}\, T_{\rm F} \ = \ 2\pi T_{\rm D2} \ . 
      \eea
This is a topological relation in M theory: it follows from the fact that the wrapped M2 brane is the  fundamental string 
and   the KK excitation is the D-particle. 
 From  the type-IIA perspective,  on the other hand, this relation  follows from T-duality or   from the cylinder calculation of section \ref{sec:4.3}, 
 but it has no geometric meaning. 
This is the  evidence from  table {\color{red} 3} for the existence of the eleventh dimension.

 
 \section{Interactions between D-branes}
 
  The massless degrees of freedom  of an isolated D-brane are  a supersymmetric
   \begin{wrapfigure}{r}{0.4\textwidth}
\centering
\vskip -0.70cm
\includegraphics[width=0.95\textwidth]{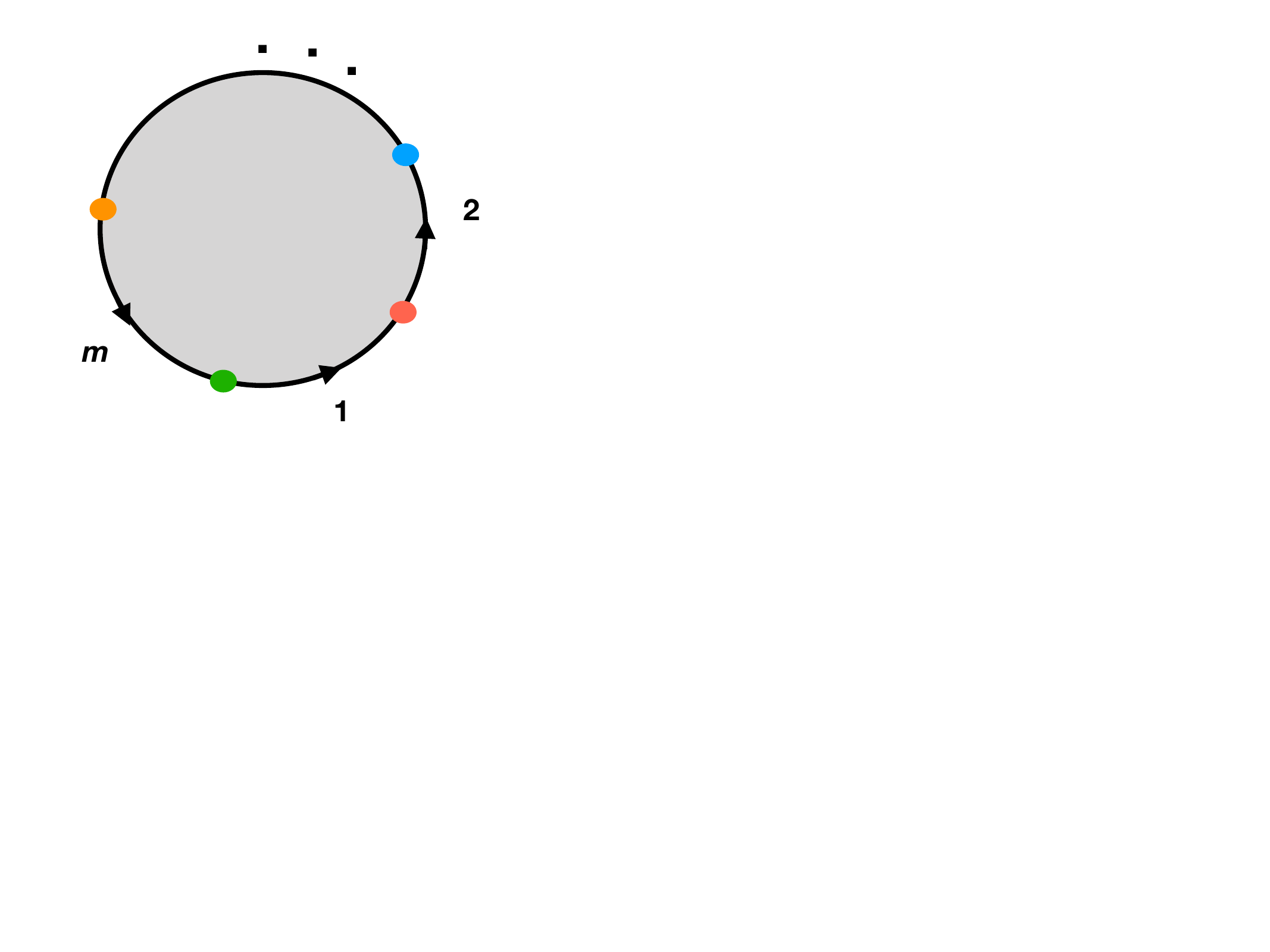}
\vskip -4.cm
\caption{Disk diagram for the  interaction of  $m$ open strings.  
 The coloured dots represent D-branes. The amplitude is
  proportional to tr$(\lambda^{1}\lambda^{2}\cdots \lambda^{m})$.}
\label{fig:vertex}
\end{wrapfigure}
 Maxwell multiplet.
 When many D-branes are present one must associate 
to the open strings a matrix-valued wavefunction $\lambda_{i j}$
 such that $\vert \lambda_{i j}\vert^2$ is the probability that the  string stretches from the $i$th to the $j$th D-brane.
 The indices $i,j$ 
are called for historical reasons  Chan-Paton factors.  
 Classical interactions of open strings are proportional to the trace of the product of their matrix wavefunctions,
 as illustrated in  figure \ref{fig:vertex}. General amplitudes have a    trace for each boundary of the Riemann surface.
 When  $N$ D-branes are  identical and coincident  the amplitudes must be  
 invariant under $U(N)$ rotations of the Chan-Paton factors. The worldvolume gauge theory
     \clearpage \eject

  \noindent  thus becomes   non-abelian   SYM
  with  the  D-branes coordinates matrix-valued. 
  In contrast  to  
    ordinary  gravitational 
  solitons  like black holes which have little  or no internal  structure, 
  D-brane solitons  hide a rich SYM theory in their interior. It is this unique feature that   
opens a new window  into the microscopic structure of  spacetime. 
We will now introduce some  features of composite D-brane systems that will be   developed in much greater  detail elsewhere
in  this  volume.


 \subsection{Scattering   D-particles}\label{sec:6.1}
 
Scattering strings at high energy cannot probe distances below the minimal size of a fundamental string\, 
 $\sim \sqrt{  \alpha^\prime}$\,. Scattering  strings  off  D-branes does not help  either,  when one smashes  a nail with a  hammer it is the hammer that
 limits the resolution. One can however  probe substringy distances  by scattering  slow D-branes off one another,  as I  now   explain.
 
  We focus on  D-particles, the extension to other compact D$p$ branes  is easy. 
At weak  coupling (\,$g_s\ll 1$)  the  D-particles are   heavy and their  Compton wavelength  $\sim g_s \sqrt{  \alpha^\prime}$
  is much smaller than the  fundamental-string scale.
But  probing it  brings in strong-gravity effects
since the Schwarzschild radius of the D-particle is of the same order as its Compton wavelength, 
  \,$ r_{\rm S}\sim \kappa_{10}^2 T_{\rm D0} \sim g_s \sqrt{  \alpha^\prime}$.
The best one can do  
  \cite{Douglas:1996yp}
 is probe the  11$d$  Planck length,  which is larger than the Schwarzschild radius  but still much below the  fundamental-string length 
($\ell_{11} = \kappa_{11}^{9/2}  \sim g_s^{3/4} \sqrt{  \alpha^\prime}$).

\vskip 1mm

To show this consider the thought  experiment illustrated in  figure \ref{fig:scatter}. 
A D-paricle moving on a linear trajectory in the direction $x^1$ passes near
 a  second D-particle at  
 \begin{wrapfigure}{r}{0.45\linewidth}
\centering
\vskip -0.8cm
\includegraphics[width=0.8\textwidth]{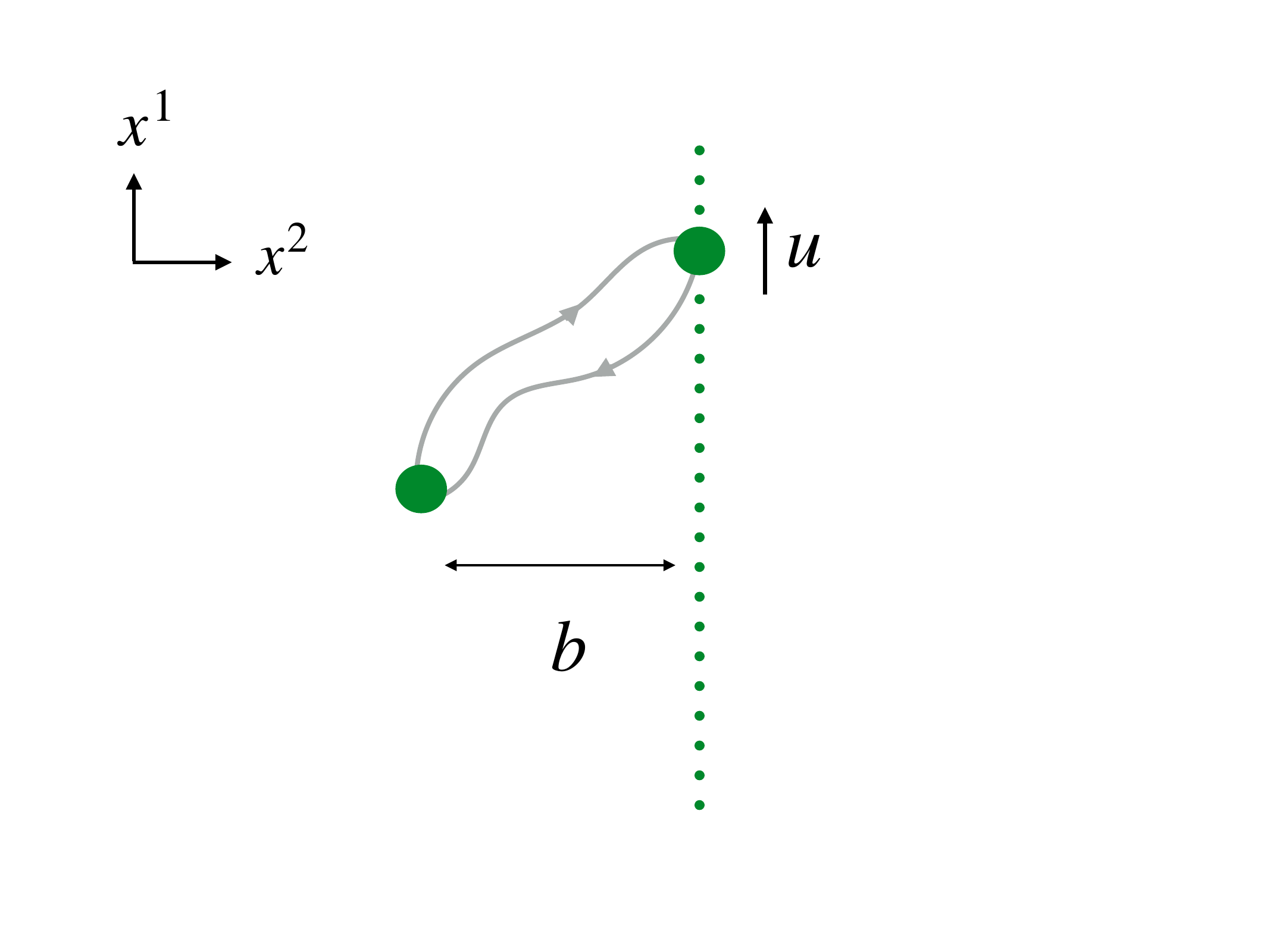}
\vskip -6.mm
\caption{Scattering of  two D-particles (green dots).  A pair of virtual open strings   (in  light grey color) can materialize
from the vacuum and slow  down the motion.
}
\label{fig:scatter}
\end{wrapfigure}
rest with impact parameter $b$ along the  $x^2$ direction.   
We are interested in the phase shift,  $\delta$, 
as function of $b$ and the  relative velocity $u>0$. 
The leading contribution comes from  the annulus  diagram
with (DD) boundary conditions for  $X^{2, \cdots , 9}$
and mixed boundary conditions for $X^{0,1}$\,:   
  \bea
  \partial_1 X^0 = X^1  = 0\,  \ \  &{\rm at} \ \sigma^1=0\ ;  \nonumber
\\
  \partial_1 X^{\prime\,0} = X^{\prime\,1} = 0\, \ \ \ &{\rm at} \ \sigma^1=\pi \,. \ \  \ \ 
\eea
Here  $X^{\prime\,\pm} = e^{\mp\pi \varepsilon} X^\pm$ are the coordinates in the rest frame
of the moving D-paricle expressed in terms of  the rapidity
\bea
 \pi\epsilon =  {\rm arctanh} (u) \,\, .
 \eea
 The reader will   notice  the similarity with the analysis of  D-strings making an angle,
  \clearpage \eject
 
\noindent see
section \ref{sec:2.2}. The two problems are related by rotating to  imaginary time and
   identifying $\pi\varepsilon$ with an imaginary angle  $\vartheta$. 
The mode expansion of the  coordinates  $(X^\pm, \psi^\pm)$ and their contribution 
to  the annulus diagram can be worked out easily
and are left as an exercise.  The final result for the phase shift is \cite{Bachas:1995kx}
$$
\delta (b,\varepsilon)  =
  -  \int_0^\infty
{dt\over t}\; 
 e^{-b^2 t/4\pi\alpha^\prime}\ Z_{\rm open}(\epsilon, q= e^{-\pi t}) \ , 
\label{eq:phaseshift}
$$ \vskip -2.5mm  
\begin{equation}
 {\rm where} \qquad Z_{\rm open}   = -{1\over 2} \sum_{j=2,3,4} (-)^j \;
 {\theta_j\left({1\over 2}\epsilon t \left|\right. q\right)
\over \theta_1\left({1\over 2} \epsilon t \left|\right. q\right)}
\;
{\theta^{3}_j\left(0
  \left.\right|
q\right) \over 
 \eta^{9}\left(q\right)}
\   
\label{eq:Z}
\end{equation}
and $\theta_j$ are the  Jacobi  theta functions. 
This  generalizes  Polchinski's  calculation of the static force, eq.\,(\ref{E0Z}),   to  moving D0 branes.
\smallskip

 Using  standard Jacobi identities 
one can write $Z_{\rm open}$ in the equivalent   form
\begin{equation}
Z_{\rm open}(\epsilon, t)  = 
 {\theta_1^4\left({1\over 4}\epsilon t \left|\right. q\right)
\over \theta_1\left({1\over 2}\epsilon t \left|\right. q\right)
 \eta^{ 9}\left(q\right) }
\ \   
\label{eq:Jacobi}
\end{equation}
from which   the small-velocity  limit is easier to extract. 
The product formula
$$
{\theta_1(z\vert q) \over \eta(q)} = 2  \sin (\pi z)\, \,  q^{1\over 12} \prod_{n=1}^\infty (1- 2 q^{n}\cos (2\pi z)   + q^{2n}) 
$$
leads indeed  after a little algebra  to the    expansion
\bea
Z_{\rm open}  =    {(\pi\varepsilon t)^3  \over 16} + O(\varepsilon^7)\, 
\qquad {\rm where}\quad 
(\pi\epsilon)^3 =  u^3 + u^5 + O(u^7)\ . 
\eea
Inserting in  eq.\,(\ref{eq:Z}) and performing the integral gives
\bea\label{124}
\delta (b, \varepsilon) =  \bigl( {2\pi\alpha^\prime \over b^2}\bigr)^3  (\pi\varepsilon)^3  \, + \, O(\varepsilon^7)\ . 
\eea
  In order to interpret the result we  define the   interaction energy $E_{\rm int}(u, r)$,  where
$r$ is the instantaneous distance between the  D-particles, via an integral transform
\bea
\delta(b, u) =   \int_{-\infty}^{\infty} d\tau\; { 
  E}_{\rm int}(\sqrt{u^2\tau^2+b^2},\, \, u)\   =\ 
{2\over u}  \int_{b}^{\infty} dr \;  {r\over \sqrt{r^2 - b^2}}\,
\, E_{\rm int}(r, u)\  . \ \ \ 
\eea
The phase shift  (\ref{124}) gives
\bea
E_{\rm int}(r, u) \ = \,  {15  \over 16 }   {  (2\pi\alpha^\prime)^3 \over   r^7}    (u^4+ u^6) + O(u^8)\ . 
\eea
One learns from this expansion (i) that not only the static but also the $O(u^2)$ force between  D-particles vanishes, and 
(ii) that the leading $O(u^4)$ interaction does not depend on the string scale $\alpha^\prime$. 
Both results follow from spacetime supersymmetry, or more precisely from the fact that only half-BPS states of the open string  contribute
 at this order to the supertrace \cite{Bachas:1996bp}. Since only the ground states of the open superstring are BPS, we understand why $\alpha^\prime$  dropped out.

\smallskip

 We will return to these remarks in a minute, but first let us point out that $\delta (b, \varepsilon)$ also has an imaginary, i.e. absorptive part.
This arises from  the zeroes of $\theta_1(z\vert q)$ at $z= k\in \mathbb{Z}$ which produce poles in $Z_{\rm open}$ at $t= 2k/\varepsilon$ for $k$ odd, 
see eq.\,(\ref{eq:Jacobi}). The integration contour must leave all the poles on the same side in order not to obstruct the rotation to the imaginary-$t$ half-axis
which corresponds to imaginary rapidity. The absorptive part is the sum of the residues at these poles, explicitly
 \begin{equation}
{\rm Im}\;\delta\; = \;  {1 \over 2}  \sum_S\, 
 \sum_{k\ {\rm odd}} {1\over k} \, {\rm exp}
\left[ - {  k\over \epsilon}
 \left(   {b^2\over 2\pi\alpha^\prime }
   +  2\pi\alpha^\prime \, M_S^2 \right)  \right]\ ,  
\end{equation}
where the sum
runs over all open-string states $S$
 with mass
 $M_S$, and the even powers of $k$ cancel between bosons and fermions.   
 
   This absorptive part expresses the probability $P = e^{-2\,{\rm Im} \delta}$  that a pair of open strings  
 materializes from the vacuum and
   slows  down  the relative motion, see figure \ref{fig:scatter}. The phenomenon  is T-dual to  Schwinger's pair creation in an electric field, generalized 
  to the case where the charge is  carried by open strings
 \cite{Bachas:1992bh}. T-duality maps the speed of light to the limiting Born-Infeld electric field. 
 As $u\to 1 \Leftrightarrow \varepsilon \to \infty$ the process is unsuppressed even if the impact parameter is large, 
 $b\gg \sqrt{\alpha^\prime}$. In this ultrarelativistic  limit $ 2\pi \varepsilon \simeq -\log (1-u)$, 
 so the critical impact parameter below which 
  the scattering becomes  inelastic  reads
 \bea
  b_{\rm cr} \simeq 
  \sqrt{  - \alpha^\prime \log(1-u)  } \  \simeq \ \   \sqrt{  2  \alpha^\prime   \log(s/T_{\rm D0}^2)   }
 \eea
where $s$  is the Mandelstam variable of the collision. This  $\sim  \sqrt{  2  \alpha^\prime   \log(s)}$ 
growth of the critical impact parameter
  with  $s$ is universal, 
 it  was  also  found in  string-string and string-brane  collisions at sufficiently high energies \cite{DiVecchia:2023frv}. 
  It is a stringy effect, to be distinguished from gravitational tidal effects,
    see e.g.  ref.\,\cite{DiVecchia:2023frv,Giddings:2006vu}.\footnote{I thank Rodolfo Russo for a discussion of this point.}
 \smallskip 
 
   Let us go back now  to substringy  $b$ and  low velocity $u\simeq \pi\varepsilon \ll 1$. To suppress the inelastic process we need 
   $b \gg \sqrt{2\alpha^\prime u}$\,. The quantum  uncertainty of  the velocity on the other hand, $\delta u \simeq (T_{\rm D0} \,  b)^{-1} $,  must be
   much smaller than $u$ so we need  $T_{\rm D0} \, b \, u \gg 1$. 
 The smallest $b$  that is consistent with  both bounds is given by  
 \bea
 b^3_{\rm min} \,\sim \, {\alpha^\prime \over T_{\rm D0}} \, \sim \, {1\over  T_2^M}\,  \sim\,  l_{11}^3 \ , 
 \eea
where we have used eq.\,(\ref{118}) and the fact that in 11$d$ Planck units the membrane tension 
  is an $O(1)$  number, see table {\color{red} 3}. 
 This shows that D-particle scattering can probe the 11$d$ Planck scale as announced.
 
       At  energies comparable to those of a stretched open string, $E \sim T_F\ell_{11}  \sim g_s^{1/3}$ in string units,
        string excitations and higher-order terms of the DBI action  can be safely  ignored and the system is described by an
         effective quantum-mechanical model which is
        the   reduction of  SYM from $d=10$  dimensions to only  time. The Hamiltonian  of this Matrix Quantum Mechanics  reads
      \bea\label{MQM}
     H_{\rm MQM} =  r_{10} \, {\rm Tr} \Bigl(  8\pi^2 \ell_{11}^6  \sum_i \Pi^i \Pi^i  - \sum_{i,j}\,\, [Y^i, Y^j]^2 -  \sum_i  \bar \lambda \Gamma^i [Y^i, \lambda]\ 
     \Bigr)
     \eea   
   where $\Pi^i$ are the   momenta conjugate to the (appropriately rescaled)  coordinates $Y^i$,   the gauge group
   for $n$ D-particles  is $U(n)$ and its role is to project 
   non-singlet states out of  the  spectrum.  This  Hamiltonian  controls  the D0 brane interactions at substringy 
   scales and   reproduces, as we saw,  their  leading $O(u^4)$ long-range force.    
 Other remarkable properties of $H_{\rm MQM}$ are  that it provides a discretization of the supersymmetric membrane \cite{deWit:1988wri}, 
 and   that its spectrum  (most likely)  includes the threshold bound states that are dual  to  the higher
    KK modes  of eleven-dimensional supergravity  \cite{Yi:1997eg,Sethi:1997pa}. 
   
      Ordinary  solitons have size comparable to the Compton wavelength of the fundamental quanta.  For example
      the size of the 't Hooft-Polyakov monopole  of section \ref{sec:4.1}  is $\sim 1/gv$, the Compton wavelength of the charged gauge bosons. 
      Since weakly-coupled D-branes  are smaller than  
      quantum strings one may suspect that the latter are not the   fundamental quanta of   gravity. 
      A  bold proposal  by Banks et al  \cite{Banks:1996vh}  was that  Matrix Quantum Mechanics in the $n\to\infty$ limit 
    may  serve as the  non-perturbative definition of M theory.   
    There is,  however, no   evidence  that at higher energies
      the Hamiltonian (\ref{MQM})  reproduces  the  rich D-particle dynamics of  string theory.  Soon after this BFFS proposal
            a similar but sharper conjecture, 
     the duality between $d=4$ SYM and quantum gravity in a Anti-de Sitter box, 
     alias AdS/CFT correspondence  \cite{Maldacena:1997re,Gubser:1998bc,Witten:1998qj},   
     revolutionized the subject.
     

\subsection{Bound states and moduli spaces}\label{sec:6.2}

   In section \ref{sec:5} we have seen  that the worldvolume gauge field $F=dA$ can endow a D-brane with extra charges.
   For example,  the Wess-Zumino term of the action (\ref{DpF}) shows that  a D$p$ brane   with a magnetic field 
  couples both to $C_{p+1}$ and to $C_{p-1}$.  And as   explained in section \ref{sec:5.1}, 
  a  D-string with a  worldvolume electric field is 
   a $(m,1)$ string, i.e.  it carries $m$  units of  
      F-string charge.  We will  now discuss  new effects that  arise because of the non-abelian 
      nature of the worldvolume fields, and more generally  in  composite   D-brane systems.
      Many of these effects are  easy to understand  in  one duality frame but look highly non-trivial in others. 

We start with the description of  two effects, 
 long strings and D1/D5 bound states,  entering the construction of the  supersymmetric three-charge black hole
which lead to the first microscopic derivation of the Bekenstein-Hawking entropy \cite{Strominger:1996sh}.

\medskip

\scalebox{1.1}{\bf{Long strings}.} Consider   $n$ identical D2 branes wrapped around  a  torus in the (12) plane. 
Take  for simplicity the torus to be  orthogonal   with  radii   $r_1, r_2$. We work in   units  $\alpha^\prime =1$. 
 On  the Coulomb branch  the gauge group 
is $U(1)^n$,  and we may  switch on  magnetic fields   on each D2 brane separately. 
To minimize  energy these fields have to  be constant. A convenient gauge is  
 $({\bm A}^1, {\bm A}^2) =({\bm a}, {\bm b}+ {\bm B}\, x^1)$,  where bold-face symbols denote $n$-dimensional vectors with  one 
component for each $U(1)$ factor  of the gauge group. 
 The   Wilson lines  $a_j$ and $b_j$ (for $j=1, \cdots , n$) are  periodic with periods $2\pi/r_1$ and $2\pi/r_2$, 
   and each  magnetic field $B_j$ must be a multiple  
of  $1/(r_1r_2)$ in accordance with  Dirac quantization.\footnote{\,We folllow the  
convention  of  section \ref{sec:5.1} so that  T-duality trades the gauge field for a D-brane coordinate with no multiplicative factor.
In this convention  string endpoints have charge  
  $q= 1/2\pi\alpha^\prime$.}
The total energy  and  D0-brane  charge of this composite system can be  readily  computed  from the action (\ref{DpF}) with the result
\bea\label{131}
 T_{{\rm D}2} \, (4\pi^2 r_1r_2)  \,  \sum_{j=1}^n  \sqrt{1+{m_j^2\over (r_1r_2)^2}} \qquad {\rm and} \qquad  \ \rho_{{\rm D}0}\sum_{j=1}^n m_j \equiv \rho_{{\rm D}0}\,m
\eea
where  $m_j = r_1r_2 B_j  \in \mathbb{Z}$. The energy of $n$  D2 branes and   $m$ 
D-particles, all widely-separated in the transverse space,  
  is \ $n T_{{\rm D}2} \,(4\pi^2 r_1r_2)   + m T_{{\rm D}0}$.   Comparing 
with eq.\,(\ref{131}) and recalling that   $T_{{\rm D}0} = (2\pi)^2 T_{{\rm D}2}$ 
 shows  that our  configuration describes 
  $n$ separate D2/D0 bound states.

 \vskip 10mm

 \hspace*{0.01\linewidth}
 \begin{tikzpicture}
\draw (0,1) node{$x^2$}; 
\draw (-1,2) node{$x^1$}; 
\draw[fill=mygray] (1,0) rectangle (4,2.2);
\draw[fill=mygray]  (5.5,0) rectangle (8.5,2.2);
\draw[->] (-1,1) -- (-1,1.7);
\draw[->] (-1,1) -- (-0.3,1);
\draw [ultra thick,color=ovgreen] (1.4,0) -- (1.4,2.2) ;
\draw [ultra thick,color=ovgreen] (2.1,0) -- (2.1,2.2) ;
\draw [ultra thick,color=ovgreen] (1,0) -- (4,2.2) ;
\draw [ultra thick,color=ovgreen] (5.5,0) -- (6.5,2.2) ;
\draw [ultra thick,color=ovgreen] (6.5,0) -- (7.5,2.2) ;
\draw [ultra thick,color=ovgreen] (7.5,0) -- (8.5,2.2) ;
\end{tikzpicture}  
\vskip 5mm
{\small {\bf Fig. 9}\ \ 
Three   D-strings, one of which winds around  the $x^2$ circle (left), can lower  their  energy
 
  by recombining  into one long D-string
(right).}

\vskip 7mm

For general $\bm B$ this is not the configuration  of lowest energy. This becomes clear after  T-dualizing
the coordinate $x^2$ which has the effect of converting 
  the magnetized D2-branes to  D-strings
with diagonal  embeddings 
 ${\bm Y}^2 = {\bm b}+ {\bm B}\, x^1$, and   inverting also the radius ($\tilde r_2 = 1/r_2$). 
For given  D2 and D0 charges, $n$ and $m$,  the T-dual configuration of minimal energy is one  long D-string winding $(n,m)$ times around the two circles,
 as illustrated  in figure {\color{red} 9}. This  is the unique ground state whenever $(m,n)$ are relative primes.
  More generally the $n$   D-strings recombine into several
     longer ones,  unless  $m$ is a multiple of $n$ in which case they are all free to separate. 

In  the original D2-D0 duality frame the long string corresponds to the following $U(n)$ gauge field
\bea
{\bf A}^1 = a \, {\bm 1}_{n\times n}\ , \quad 
{\bf A}^2 =      \bigl( {2\pi\over n r_2}\bigr) \, {\rm diag}( 0, 1, \cdots , n-1) +   \bigl(b+ {m\, x^1 \over n\, r_1r_2} \bigr)\, {\bm 1}_{n\times n}\ , \ \ 
\eea
where ${\bm 1}_{n\times n}$ is the identity $n\times n$ matrix. For    fractional  $m/n$ this is not an admissible $U(1)^n$ gauge bundle. But  it becomes
admissible if the gauge group is $U(n)$,  since 
\bea
q\, {\bf A}_\mu(x^1+ 2\pi r_1) = {\cal U}^{-1} [q\,{\bf A}_\mu(x^1) -   i\partial_\mu] {\cal U}
\eea
with
\bea
{\cal U}(x)  =  {\rm diag}(e^{\, i(1-n) x^2/ nr_2}, \, e^{ \,ix^2/ nr_2}, \cdots ,\, e^{\, ix^2/nr_2} ) \,  {\bm P}_{1\to 2 \cdots \to n}
\eea
and ${\bm P}_{1\to 2 \cdots \to n}$  the cyclic permutation matrix. We let the reader verify that ${\cal U}(x)$ is indeed  a single-valued 
unitary matrix on the torus. It is actually  single-valued also  in $SU(n)/\mathbb{Z}_n$, but fractional $m/n$ obstructs its lifting to a good $SU(n)$ bundle.
This  twisting of the boundary conditions of the gauge field
is familiar in the study of gauge theories  where it is known as a   't Hooft flux \cite{tHooft:1979rtg}.  
 
    A similar story holds for the $(p,q)$ strings of section \ref{sec:5.1}.  After T-dualizing  the circle wrapped by the string,  this latter 
    becomes a collection of $q$ D-particles sharing $p$ units of momentum in the dual dimension. When  $p$ and $q$ are relative primes 
    the ground state is unique and  corresponds to an eleven-dimensional  KK graviton with
     incommensurate momenta in the  compact dimensions.

  One should  note
   that the  non-abelian worldvolume theories  are strongly-coupled  in these low-dimensional  systems. 
   Our  classical arguments are nevertheless reliable for  the above (half-BPS) ground states
  thanks to  supersymmetry.

\medskip\medskip

\scalebox{0.95}{$\bm{1/4}$}\scalebox{1.1}{{\bf -BPS systems}}.  Such systems offer less supersymmetry protection but also  richer  possibilities.  
They contain  open-string sectors with four Neumann-Dirichlet 
boundary conditions, as 
 explained  in  section  \ref{susyDbr}. 
 A prototype
  consists of $m$ parallel D9 branes and  $n$ D5 branes, but  our discussion will be equally valid  for systems obtained from D5/D9  by   duality transformations.
Some  systems  that enter in important  applications are listed  in table \ref{table:4}.

\begin{table}[htp]
\begin{center}
\noindent\begin{tabular}{|c|c|c|c|c|c|c|c|c|c|c|}
  \cline{2-11}
  \multicolumn{1}{c}{}  &   \multicolumn{1}{|c|}{\ 0\ \ }   &  \ 1 \  & \  2  \ & \  3 \ & \  4 \  & \  5  \ &  \ 6 \ &  \  7  \ & \  8 \ & \  9 \  \\    \hline
  \  {\bf D(-1)} \ &         &     &      &  &  &  &   &    &  &  \\  \hline
 \   {\bf D3} \ &   &    &    &   & &    &    \cellcolor{green}   &  \cellcolor{green}     &  \cellcolor{green}     & \cellcolor{green}   \\  \hline\hline
   \  {\bf D3} \ &  \cellcolor{ovgreen}       &    \cellcolor{ovgreen}    &   \cellcolor{ovgreen}      &  \cellcolor{ovgreen}   &  &  &   &    &  &  \\  \hline
 \   {\bf D7} \ &    \cellcolor{ovgreen}   &  \cellcolor{ovgreen}     &  \cellcolor{ovgreen}     & \cellcolor{ovgreen}   &    &  & \cellcolor{ovgreen}   &   \cellcolor{ovgreen}  & 
  \cellcolor{ovgreen}    &   \cellcolor{ovgreen}     \\  \hline\hline
    \  {\bf D3} \ &  \cellcolor{green}       &    \cellcolor{green}    &   \cellcolor{green}    &  &    &  &  \cellcolor{green}    &    &  &  \\  \hline
 \   {\bf D5} \ &    \cellcolor{green}   &  \cellcolor{green}     &  \cellcolor{green}     &   &    &  &  & \cellcolor{green}   &   \cellcolor{green}  & 
  \cellcolor{green}     \\  \hline\hline
   \  {\bf D1} \ &  \cellcolor{ovgreen}       &   \cellcolor{ovgreen}   &      &  &  &  &   &    &  &  \\  \hline
 \   {\bf D5} \ &    \cellcolor{ovgreen}   &  \cellcolor{ovgreen}     &   &   &  &    &  \cellcolor{ovgreen}     & \cellcolor{ovgreen}   & \cellcolor{ovgreen}   &   \cellcolor{ovgreen}   \\  \hline\hline
    \  {\bf F1} \ &  \cellcolor{cyan}       &   \cellcolor{cyan}    &      &  &  &  &   &    &  &  \\  \hline
 \   {\bf NS5} \ &   \cellcolor{cyan}     &  \cellcolor{cyan}     &   & &    &  &  \cellcolor{cyan}      & \cellcolor{cyan}    & \cellcolor{cyan}    &  \cellcolor{cyan}     \\  \hline
\end{tabular}
\end{center}
\caption{
Brane pairs that are dual to the D5/D9 system. Coloured boxes show the dimensions  along which each  brane extends. 
The first four configurations are  T-dual to D5/D9,  whereas D1/D5 and F1/NS5 are related by S-duality. To underline their common 6$d$ origin
we chose the 
 non-common dimensions to be always  the last four.}
\label{table:4}
\end{table}

  The   low-energy theory of a composite D$p$\scalebox{1.2}{/}D$(p+4)$ system  is the dimensional reduction  of 
  the  6$d$ N=1 SYM that lives on the worldvolume of the parent D5/D9. The bosonic field content is as follows:  
  \begin{itemize}
  \item The strings on the  D$p$  branes 
  give the  usual    $(A^{\mu=0, \cdots , p},\, Y^{p+1, \cdots , 9})$ in the adjoint of the gauge group $U(n)$. 
 The coordinates along the    D$(p+4)$ dimensions   transform as a real vector of $SO(4)$, or as a complex doublet    of  $SU(2)_R\subset SO(4)$.
  They  are part of a   hypermultiplet of half-maximal supersymmetry with  $SU(2)_R$ the 6$d$ R-symmetry. 
  We  define  $Z= Y^6+iY^7$ and $\tilde Z^\dagger  =Y^8 + i Y^9$ (the  components of the doublet)  and 
    reserve  the symbol 
    $Y^j$ for  the  coordinates in  the remaining $5-p$  directions transverse to all  D-branes.
 \smallskip    \item The  fields on the D$(p+4)$  branes are similar except that they are  in the adjoint of $U(m)$. 
The four real coordinates packaged previously in  $Z, \tilde Z$ are now components  of the gauge field in the extra  dimensions. 
 \smallskip    \item Finally  we have the open  strings stretching between a D$p$ and a D$(p+4)$ brane. 
 Their  (ND) coordinates  have  integer modes  in the Neveu-Schwarz sector  
(and half-integer   in the Ramond sector). It follows from the analysis  in sections \ref{sec:2} and  \ref{sustrings}
 that  $\vert 0\rangle_{\rm NS}$ is a  $SO(1,p)$  scalar transforming as  a chiral spinor 
  of  the internal $SO(4)$, i.e.  as a doublet  of $SU(2)_{R}$. We call  $H$ and $\tilde H^\dagger$   the    complex components 
 of the doublet,  they  are  in the 
  $(n, \bar m)$ representation of the gauge groups.
 \end{itemize}

  The classical low-energy  action for all these fields is  fixed by supersymmetry and gauge invariance. 
  To simplify the discussion we place  the
   D$(p+4)$  branes  at the origin in transverse space,  and freeze all fields on their  worldvolume.\footnote{If  
    the four extra coordinates are non-compact
   the D$(p+4)$ worldvolume fields decouple from the D$p$-brane action and $U(m)$ becomes a global symmetry.
   }
  The action includes the  standard kinetic terms and a  scalar potential given by
$$
\scalebox{1.17}{${2\over g^2}$}\,V =   \,     \sum_{i>j}\, \bigr\vert \,  [Y^i, Y^j] \, \bigr\vert^2 +
 \sum_{i}\  \Bigl( 
  \bigl\vert \, [Y^i, Z] \, \bigr\vert^2  
+ \,   \bigl\vert \, [Y^i, \tilde Z^\dagger] \, \bigr\vert^2 
+  \,  \bigl\vert  Y^i H \bigr\vert^2 +  \,  \bigl\vert  Y^i \tilde H^\dagger \bigr\vert^2 
\Bigr)
$$ \vskip -7mm
\bea
+ \, \, {\rm tr}\,  \Bigl( H H^\dagger - \tilde H^\dagger  \tilde H
+  [Z, Z^\dagger] -  [\tilde Z^\dagger , \tilde Z ] \Bigr)^2
+ \,  {\rm tr}\,  \Bigl\vert \, H \tilde H  + [Z , \tilde Z ]\, \Bigr\vert^2\ . 
\eea
We have pulled out a factor of $g^2$ to make it clear  that all terms originate from gauge interactions.
In the first line  $\vert M\vert^2$ stands for tr$(M^\dagger M)$, and all entries come from gauge-invariant kinetic terms in six dimensions. 
The lower line is the contribution of an $SU(2)_R$ triplet of D-terms. Here $H$ and $\tilde H^\dagger$  are $n\times m$ matrices, so that
all matrices inside the traces are  $n\times n$. 

  Since $V$ is a sum of positive contributions, they must all  vanish in vacuum. 
One option is to have all  hypermultiplets vanish and allow commuting    $\langle Y^i\rangle$ of  the scalars in the vector multiplet.
Another is to set all scalars in vector multiplets to zero and solve the D-term equations for the hypermultiplets. 
These two branches of vacua are called respectively the Coulomb and the Higgs branch. The hypermultiplets contain $4(nm+n^2)$ real scalar fields. Since there
 are  3$n^2$  D-term conditions and there is a redundancy of $n^2$ gauge transformations,  
  the dimension of the Higgs branch is  $4mn$. On  the Higgs branch the transverse 
coordinates $Y^i$ are massive and the D$p$ branes are bound to
the   D$(p+4)$ branes.

 From the perspective of the D$(p+4)$ branes we can understand this  bound state as follows.
The $U(m)$ non-abelian generalization of the action (\ref{DpF}) contains the  Wess-Zumino term
\bea\label{136}
 \rho_{{\rm D}(p+4)} \int \Bigl[  C_{p+5} + {1\over 2}{\rm tr} (F\wedge F) C_{p+1} \Bigr] \ . 
\eea
The integral of  ${1\over 2}{\rm tr} (F\wedge F)$ over the four extra directions of the D$(p+4)$ brane is a topological invariant 
 equal to\,   $n\, \rho_{{\rm D}p}/\rho_{{\rm D}(p+4)}$, where
$n\in \mathbb{Z}$ is the  instanton number (second Chern class) of the gauge-field background. 
 By inserting in (\ref{136}) we learn  that 
    the D$(p+4)$ brane has been endowed with $n$ units of D$p$-brane charge.
 This suggests  that  the Higgs branch of the  D$p$\scalebox{1.2}{/}D$(p+4)$ system
  is the   moduli space of $n$-instanton solutions of  theYang-Mills theory with gauge group $U(m)$.  
   Note that in the absence of a first Chern class, i.e. if tr($F) =0$, the D$p$ branes  are only marginally bound to the   D$(p+4)$ branes.

  The above statements can be made more precise.\footnote{As first noted in refs.\,\cite{Witten:1994tz, Douglas:1996uz}, for a nice review see  \cite{Tong:2005un}.}
The D-term equations (also called moment-map conditions) 
are  the starting point of the celebrated Atiyah-Drinfeld-Hitchin-Manin  (ADHM)  construction of multi-instanton solutions \cite{Atiyah:1978ri}. 
And the metric on this moduli space is precisely the one induced by the flat  $\mathbb{R}^{4n(m+n)}$ metric
 on the   $V=0$ hypersurface  quotiented by the action of   $U(n)$.

 The metric of the Coulomb branch  is also curved, though this is not a classical effect but rather  the result of quantum corrections.
 The annulus calculation of section 
\ref{sec:6.1} for a D-particle scattering off a D4 brane gives indeed a   $O(u^2)$ force which also receives non-perturbative corrections.  
For pure $N=2$ gauge theory the Coulomb-branch metric  is  the celebrated Seiberg-Witten solution \cite{Seiberg:1994rs}. More 
 generally its calculation is  hard. The one certainty comes from supersymmetry  which
restricts the Coulomb-branch moduli space  to be  a special K{\"a}hler manifold,  and the Higgs-branch moduli space  to be hyper-K{\"a}hler. 


\subsection{Special effects}\label{sec:6.3}

    In an effort  to learn more about  quantum gravity, 
     we  considered up to now gauge theories that describe the dynamics of  simple D-brane systems. 
     Inverting the logic one can  try to engineer brane systems  whose worldvolume theory is a quantum field  theory of 
    interest. This can  offer
  several advantages:   (i)   geometrize  moduli spaces and  parameters; (ii)   bring to light  surprising  connections between      
previously unrelated theories (one such example is 3$d$ mirror symmetry \cite{Intriligator:1996ex,Hanany:1996ie});   and (iii) provide  seeds of holographic dualities.

      Brane engineering is presented elsewhere in this volume (see also \cite{Giveon:1998sr} for an early review). 
        I   close  the present chapter with  some special effects that play a role  in these  lego constructions. 
        The first one  is straightforward:

\medskip

\scalebox{1.1}{\bf{Branes ending on branes}.} 
        As explained earlier,  D-branes can dissolve as   worldvolume fluxes in higher-dimensional D-branes.
But they can also end on other branes on whose worldvolumes   they appear as   gauge charges.  This is obvious if we apply  various U-dualities
to  a fundamental string  suspended between two  type-IIB D-branes. 
Consider for  example the F-string stretching between two parallel    D3 branes. Denote this configuration by
$[{\rm D}3 \rightarrow {\rm F}1 \rightarrow {\rm D}3]$  and  
apply  the following chain of dualities

 \vskip 6mm

 \hspace*{0.1\linewidth}
 \begin{tikzpicture}
 \draw [color=black] (1.4,0) rectangle (4,0.8);
 \draw (2.7,0.4) node{D3\,$\rightarrow$\,F1\,$\rightarrow$\,D3}; 
\draw[<->, very thick, color=red]  (4.8, 0.4) -- (6.2,0.4); 
 \draw (5.5,0.70) node{\color{red} S}; 
  \draw [color=black] (7.,0) rectangle (9.6,0.8);
 \draw (8.3,0.4) node{D3\,$\rightarrow$\,D1\,$\rightarrow$\,D3}; 
   \draw [color=black] (7.,-2) rectangle (9.6,-1.2);
 \draw (8.3,-1.6) node{D5\,$\rightarrow$\,D3\,$\rightarrow$\,D5}; 
 \draw[<->, very thick, color=red]  (4.8, -1.6) -- (6.2,-1.6); 
 \draw (5.5,-1.3) node{\color{red} S}; 
  \draw [color=black] (1.4,-2) rectangle (4,-1.2);
 \draw (2.7,-1.6) node{NS5$\rightarrow$D3$\rightarrow$NS5}; 
  \draw[<->, very thick, color=blue]  (8.4, -0.2) -- (8.4,-1); 
   \draw (8,-0.6) node{\color{blue} ${\rm T}^{\,2}$};
 \end{tikzpicture}  

 \vskip 6mm

 From the  S-duality on the top we learn  that  D-strings can end on D3 branes. Such D-strings  are  the S-duals of charged gauge bosons, so they must 
 correspond to  
  the 't Hooft-Polyakov monopoles of the worldvolume $N=4$ SYM. 
By  T-dualizing two of the common transverse dimensions we also learn  that a D3 brane can  end on D5 branes with which it  shares
two  spatial  dimensions. Finally, another  S-duality shows that the D3 branes may also end on  NS5 branes.

  Continuing like this one can discover all other possibilities.  Note that while U-dualities are a quick way to discover these rules,  the 
 fact that a brane X may end on a
  brane Y  could  be deduced ab initio  from the possible  gauge charges of  the worldvolume theory of Y.  As one example of creative engineering
  one can start from the lower-left end of the above duality chain, perform one more T duality to convert the D3 branes to D4 branes and 
  the NS5 branes of type IIB to NS5 branes of type IIA, and add $N_f$ D6 branes oriented so as not break extra supersymmetries.
  The resulting configuration can be lifted to M theory and realizes N=2 supersymmetric QCD with $N_f$ flavours in four dimensions
\cite{Witten:1997sc}. 
 
 \medskip
 
 \scalebox{1.1}{\bf{Brane creation and the s-rule}.}  Which brane can end on which other brane is a local rule. Each junction leaves 
 1/4 of the supersymmetries unbroken,  but together the two suspending branes  will break  all   supersymmetries unless  they are 
 carefully oriented in space, as in section  \ref{susyDbr}. There is however a robust phenomenon  that happens when two  D-branes  
 share only one  transverse dimension or only a  longitudinal one, but not both. This is the generic situation for two planar D4 branes in 
  $\mathbb{R}^9$,  
 similar  to two   straight  lines in $\mathbb{R}^3$.  
 
     When one   D4 brane moves past the other there could appear a tachyon indicating that the D-branes want to
     split  and reconnect. Such instabilities are absent if
       the rotation from one D-brane to the other is sufficiently `large', for instance if the D4 branes span orthogonal subspaces of $\mathbb{R}^9$.
      In this case the  stretched open strings have  
   eight coordinates with mixed (ND) boundary conditions and   1/4-maximal  supersymmetry is preserved. A T-dual  configuration is
     the D8/D0 brane  system which I will use to describe the phenomena at hand.

  Let the D8 brane extend  in the directions $x^{j=2,\cdots ,9}$ and consider the spectrum of the stretched  string.
     In the NS sector the   coordinates $x^{j=2,\cdots ,9}$  have half-integer modes and their fermionic partners integer modes.
The ground state subtraction is thus minus that in eq.\,(\ref{DN=8}),  all the states are massive and we are safely inside the stability regime. 
 The  Ramond ground state, on the other hand,  is always massless. Since 
 there are only  two fermionic zero modes,  $\psi_0^{\mu=0,1}$, 
   it is a two-dimensional Weyl fermion with space momentum   replaced by the  separation $\Delta x^1$ between the D-particle and the 
 D8 brane.\footnote{A standard 
 2$d$ fermion lives on the T-dual 
D9/D1 intersection. Note that the absence of a
 bosonic partner for the fermion  is due to the fact that the unbroken  supersymmetries act in the sector of opposite chirality, similarly to
 what happens on the worldsheet of  the heterotic string. }

  In the effective  D8/D0 worldvolume  theory
   this open-string sector contributes therefore just a qubit:  $\vert 0\rangle$ if there is no such string,  or  $\vert 1\rangle$
 if there is a string   in its unique (supersymmetric)  ground  state. Any extra (D8-D0) strings must occupy excited states with 
 masses  $O(\sqrt{T_F})$   and break supersymmetry.

\begin{wrapfigure}{l}{0pt}
\begin{tikzpicture} 
\draw[fill=ovgreen] (1,0) circle (1ex);
\draw[-, ultra thick, color=olive]  (3, -0.5 ) -- (3,2.5); 
\draw[-, ultra thick, color=olive]  (3.05, -0.5 ) -- (3.05,2.5); 
\draw[->, dashed, thick] (1.3,0.15) -- (4.5,1.7);
%
\draw[very thick, color=red] (3.05,2) -- (5,2);
 \draw (6,0.4) node{\,}; 
 \draw[fill=ovgreen] (5,2) circle (1ex);
  \draw (1, -1.7) node{{\small\bf Fig. 10}}; 
    \draw (3.6, -1.7) node{{\small A D-particle crossing a D8 brane}}; 
        \draw (2.3, -2.05) node{{\small and creating a string (in red).}}; 
\end{tikzpicture}
\end{wrapfigure}

Let us now see what happens when the D-branes cross. The 
   Dirac-Weyl equation for the open-string  fermion  is   $E =   \pm T_F \Delta x^1$. 
   Suppose that for $\Delta x^1 <0$    the negative-energy solution of the Dirac sea is filled
   so there is no string.
  As the D-particle crosses the D8 brane   the  hole    becomes positive-energy 
  and a string pops out of the vacuum, see fig.\,{\color{red}10}. 
  This is the familiar story of the 2$d$ chiral anomaly. 
At first sight it looks as if $\Delta x^1 >0$ is not a flat direction of the D-particle potential anymore since the 
 energy of the created  string grows linearly with separation.  This  conclusion is  however wrong 
 because the  D8 brane  generates a piecewise-linear background for the dilaton, 
   $\Phi = 2\kappa^2_{10} T_{\rm D8}\,\Theta(\Delta x^1)\Delta x^1$
   with $\Theta$ the Heaviside step function.\footnote{We assume $\Phi=0$ on the left where the observer of this gedanken experiment measures energy.}
 This  is  computed  using   the effective actions of section \ref{sec:4.2}.  
 The mass of the D-particle, proportional in the string frame to $e^{-\Phi}$,   is therefore reduced at   leading order   by  
 \bea
  T_{\rm D0} \Phi  =  T_F \Delta x^1 \ , 
 \eea
 where we have used the tension formula (\ref{Dcharge}). 
  This  compensates precisely  the  energy of the stretched string, so 
   $\Delta x^1$ is a modulus for both signs, as should be expected from the unbroken supersymmetry.

    These simple  facts about the quantum mechanics of the D8/D0 system become highly non-trivial
  in other,  non-perturbative duality frames. Consider for example the duality chain T$(56)$\,S\,T$(789)$ where the T-dualized dimensions  are shown 
 in parentheses. 
 It  converts  the D8/D0/F1 system to a D3 brane suspended between a D5  and a NS5 brane which have only one transverse dimension in common.
If S-duality is valid 
 we conclude (i)  that  more than  one   D3 branes  suspended between the two 
 five-branes  break all supersymmetries, and  (ii) that whenever a NS5 brane  crosses a D5 brane  a D3 brane is created or
destroyed depending on orientation.  

These rules were first conjectured by Hanany and Witten \cite{Hanany:1996ie} using plausibility arguments about  gauge theories 
on  D3 branes. Their connection with the Pauli exclusion principle  and the $2d$ anomaly 
was shown in \cite{Bachas:1997ui, Bachas:1997kn}. Another interesting duality frame is the D6/D2/F1 system which lifts in M theory to a membrane
  in the background of the KK monopole, see section \ref{sec:5.3}. The Fermi-Dirac statistics of the  open F-string translates in M theory  to  the 
  fact that the  classical  membrane equations admit no 
    holomorphic multi-membrane solutions 
 \cite{Bachas:1997sc}. This is a rather surprising  check  of the type-IIA/M-theory duality conjecture.

\medskip

\scalebox{1.1}{\bf{Myers effect}.}\, We have till now considered  vanishing  supergravity backgrounds. 
The study of D-branes in non-trivial geometries  such as Calabi-Yau manifolds\,,\footnote{For a review see e.g. \cite{Aspinwall:2004jr}.}
 or in the context of AdS/CFT,  is a  rich subject in its own right.  
  As an appetizer I will conclude this chapter with the description of a  general phenomenon
 with multiple  manifestations and names: Myers or dielectric effect, giant gravitons, supertubes 
  \cite{Emparan:1997rt, Myers:1999ps,Bachas:2000ik, McGreevy:2000cw, Mateos:2001qs} (see also \cite{Myers:2003bw, Bena:2007kg} for reviews). 
 In a nutshell, this is the puffing up of D-branes into higher-dimensional ones  that wrap topologically-trivial cycles, but
    are   prevented from collapsing by  background   fields. 
 \smallskip
  
   To illustrate the phenomenon consider a spherical D2 brane in the background of a non-vanishing but weak electric 4-form flux. 
   In  locally-flat polar coordinates
 $F_4 \simeq  f  r^2   dr\wedge d ({\rm cos}  \vartheta) \wedge d\varphi \wedge  dt $\, with $f$ constant. 
We also switch on a  worldvolume gauge field that endows the D2 brane with $n$ units of D-particle charge.
 Choosing a convenient gauge for $C_3$ and in static coordinates for the D2 brane we have
     \bea 
     C_3\,  \simeq\, { f  r^3\over 3}    {\rm sin}  \vartheta\,  d \vartheta \wedge d\varphi  \wedge dt\,   \quad {\rm and} \quad
    F\bigl\vert_{D2}  =  { n\over 2 T_F}  \,  {\rm sin}  \vartheta\,  d \vartheta \wedge d\varphi\ . 
    \eea
Inserting these fields in the D2-brane action  eq.\,(\ref{DpF}) leads to the energy 
\bea
 E =  4\pi  \, T_{\rm D2} \left[  \sqrt{r^4 + ({n\over 2T_F})^2 } 
 \, -   \, {1\over 3} \, f  r^3 \right] \ . 
\eea
We assume $f>0$, otherwise we change the D2 brane for an anti-D2 brane. The minimum of the energy is  at the radius
\bea\label{140}
 r_{\rm min}  = {nf\over 4T_F} +O(f^5) 
 \quad {\rm with} \quad
E_{\rm min}  = n T_{\rm D0} \left[ 1 - n^2 f^4 { (\pi\alpha^\prime)^2\over 96} + O(f^8)\right] \  
   \ ,    \ \ 
\eea
where we  used   $2\pi n T_{\rm D2}/T_F = n T_{\rm D0}$\,.   One  sees that for $f=0$ the 
  D2 brane shrinks to a point  and its energy is   that of $n$ D-particles.  But the non-vanishing   $F_4$  exerts an outward pressure 
  which combines with the pressure of the  magnetic field to puff up the D2 brane to a sphere. 
    Note that both $f$ and $n$  are 
  necessary to stabilize the brane at non-zero radius.
  

    It is instructive to  consider  the D0-brane  perspective. In the absence of the RR background  
  the low-energy Hamiltonian of the $n$   D-particles   is the reduction  to 0+1 dimensions of $N=4$   SYM with gauge group $U(n)$. 
  We choose now  Cartesian local coordinates in which $ C_3 \simeq   f\, x^1\wedge dx^2\wedge dx^3 \wedge dt$.
  This background adds a new term  to the  potential  \cite{Myers:1999ps}  which 
  restores  the  invariance  of the non-abelian extension
  of the D2-brane coupling $\ T_{\rm D2}\int (C_3 + F\wedge C_1)$ under the  T-duality of $X^2, X^3$ that exchanges 
   the D2 branes with D-particles, the  RR
  3-form  with  the  1-form,   and the D2-brane  gauge field with the  D-particle position.
  The potential to lowest order reads
    \bea\label{fuzzy}
    &V  = \, T_{\rm D0}\,     {\rm tr}\bigl( -\scalebox{1.1}{${1 \over 4 }$}     {T_F^2 } \, [Y^i, Y^j][Y^i, Y^j] + 
 \scalebox{1.1}{${i  \over 6 }$}   { T_Ff   }\epsilon_{ijk}\, Y^i [Y^j, Y^k]\bigr) \  
   \eea
where repeated indices  are implicitly summed,  and the coefficients are  fixed by the appropriately normalized open-string 
coupling.\footnote{We 
did not   bother with this normalization up to now,  but it  is needed for the  
 non-abelian terms in the  D-brane actions. One quick way to determine it
  is by relating the instanton number on a D4 brane to the  number of D-particles, see section
 \ref{sec:6.2}. }

 Varying the above  potential gives   the   equations
 \bea
  \Bigl[[ Y^i , Y^j] -  i {f\over 2T_F} \epsilon_{ijk} Y^k\, , 
  \, Y^j \Bigr]  = 0\ , 
 \eea
 which can be solved by  matrices  that obey the $su(2)$ algebra  $[Y^i, Y^j] = i (f/2T_F) \epsilon_{ijk} Y^k$. 
 The solution that minimizes the energy is the one that maximizes the $su(2)$ Casimir, so it corresponds to the irreducible
 $n$-dimensional representation of the Lie algebra. 
  Inserting back in (\ref{fuzzy}) we find
 \bea
  \sum_{i=1}^3 Y^i Y^i    = 
  {f^2(n^2-1)\over  16\,T_F^2}\,{\bm 1}_{n\times n}  
  \ ,   \quad
  V_{\rm min}  = - T_{D0} \, {f^2\over 24}   \, {\rm tr}\bigl( \sum_{i=1}^3 Y^i Y^i \bigr) 
 \ . 
 \eea
The D-particles have thus puffed up to  the fuzzy (alias non-commutative) sphere
with radius $\simeq fn/ 4T_F$ for
 $n$ large. These results are in perfect agreement 
with  eq.\,(\ref{140}). Note that computing subleading terms in the $f$-expansion would require the full
non-abelian generalization of 
 the DBI and WZ actions which is  plagued by ordering ambiguities, see e.g. \cite{Myers:2003bw,Hatefi:2012zh}.

  \medskip


\scalebox{1.1}{\bf{NS5 brane   synthesis}.}\,  
We   conclude with an example that
  illustrates  the   effects of this subsection all at once \cite{Bachas:2000ik,Elitzur:2000pq}. It has also the advantage of being exact
 to all orders in $\alpha^\prime$  which means that it can be handled with the methods of Boundary Conformal Field Theory (BCFT).
  
   The background is the near-horizon region of $k$ type-IIB 
NS5 branes that we have briefly encountered in the duality section \ref{sec:5}. This is
 an  exact worldsheet CFT made of three parts: (i)   free string coordinates $X^{\mu=0, \cdots , 5}$
along the worldvolume of the five-branes; (ii) a Wess-Zumino-Witten (WZW) model whose target space is the round 3-sphere that surrounds the branes; and (iii) 
a   linear-dilaton CFT for the radial direction  $\log \vert X^\perp\vert$  \cite{Callan:1991at}.   
 The   metric and 2-form field  of the WZW model read
  \bea\label{NS5WZW}
  ds^2 =   k   \alpha^\prime  (d\psi^2 + \sin^2 \psi \, d\Omega^2)\   \quad {\rm and} \quad
  B_2 = k  \alpha^\prime \Bigl(\psi - {1\over 2}\sin \, 2\psi \Bigr)\, \omega_2 \,\, ,\ \ 
                 \eea       
where $d\Omega^2$ and $\omega_2$ are the metric and the area form of the 2-sphere of unit radius.
The field strength $H_3 = dB_2$ is the volume form of the 3-sphere;  
 the gauge potential $B_2$ has a Dirac 
  singularity at the south pole,  $\psi =\pi$, 
but this cannot be detected by  fundamental strings as long as  $k$ is an  integer. 

Consider now $n$ coincident  D3 branes ending  on the NS5 branes and preserving 

 \begin{wrapfigure}{l}{0pt}
\begin{tikzpicture}
\draw (0.1,2.5) node{$\psi=0$}; 
\draw (0.1,-2.5) node{$\psi=\pi$}; 
\draw[color=blue]  (2.3,1.2) node{\scalebox{1.2}{$S^2$}}; 
\draw (-2.7,1.25) node{$\psi =$\scalebox{1.2}{${\pi \,n\over k}$}}; 
  \shade[ball color = gray!40, opacity = 0.4] (0,0) circle (2cm);
  \draw  (0,0) circle (2cm);
  %
  \draw[ultra thick,color=blue] (-1.7,1.) arc (180:360:1.7 and 0.28);
    \draw[dashed,ultra thick,color=blue] (1.7,1.) arc (0:180:1.7 and 0.28);
  \fill[fill=black] (0,2) circle (1pt); 
    \fill[fill=black] (0,-2) circle (1pt);   
\draw[thick,black][->] (1.2,2) arc (40:20:2.7 and 1.5);
\draw[black] (0,0) -- (0,2);
\draw[black] (0,0) -- (1.7,0.9);
  \draw (-2.8, -3.4) node{{\small\bf Fig. 11}}; 
    \draw (0.08, -3.4) node{{\small $n$ D3 branes puffing up to a  D5 brane  }}; 
        \draw (-1.02, -3.75) node{{\small in the near-horizon of $k$ NS5 branes.}}; 
\end{tikzpicture}
\end{wrapfigure}

  \noindent   1/4 supersymmetry. 
 These must be oriented as  described in the beginning of this subsection, 
 i.e. they  span  the radial direction as well as
two of the NS5-brane directions, say $x^1, x^2$. 
  They   intersect   the 3-sphere at a point 
(which we  choose without loss of generality to be  the north pole $\psi=0$).
Nothing  prevents, however,  the D3 branes from puffing  up into a D5-brane if this can  lower  the energy.
In addition to the spectator dimensions $x^1, x^2, \vert x^\perp\vert$,   the D5 brane must now wrap  a
 2-sphere at some polar angle $\psi$ of the 3-sphere, as depicted in   figure {\color{red} 11} on the left. 
Furthermore a  worldvolume magnetic field  must endow  the D5 brane with  $n$ units  of D3-brane charge. 
 Since there are no RR backgrounds in the problem the energy is given entirely by the DBI part of the D5-brane action. It  is proportional to
  \bea\label{145}
  E \  \propto  \  \,  k \alpha^ \prime 
   \left[\sin^4\psi + (\psi - {1\over 2} \sin\,2\psi - 
  {\pi n\over k} )^2\,\right]^{1/2}\ . 
  \eea
We have used here the appropriately normalized worldvolume flux $F= - n   \omega_2/ 2 T_F$.
The minimum of  (\ref{145})  is  at   $\psi_0  = {\pi n/ k}$, so for $n$ small the D3 branes  puff up to a
 D5 brane of radius $\sim \psi_0 \sqrt{k\alpha^\prime}$. This is a variant of the Myers effect.
  
  \smallskip
  But   something else happens now as $n$ keeps  increasing.
 The D5 brane crosses the equator and then starts shrinking again untill, when $n=k$,
  it becomes a point  at the south pole.  To understand this  phenomenon note 
    that in the asymptotically-flat geometry the  3-sphere radius diverges and the D5 brane opens up to
an infinite  hyperplanar brane.  
This is  the [$k$\,NS5 $\to$ $n$\,D3 $\to$ D5] configuration of  Hanany and Witten,    
with the back-reaction of the  NS5 brane on the supergravity fields taken into account. 
 The fact that   $n$ takes values in  $[0, k]$ is our  s-rule. 
    Furthermore when the asymptotically-planar D5 brane crosses to the
 other side of the NS5 branes, its trajectory wraps the 3-sphere from north to south creating  $k$ D3 branes. 
 Finally,  $n$  transforms under large  gauge transformations of $B_2$ 
 (intuitively, these add  extra NS5 branes at infinity). It is   the Page charge defined in eq.\,(\ref{Pagech}). \footnote{A mathematically precise  
  definition of the Page charge in this coontext can be given   in terms of twisted K-theory \cite{Fredenhagen:2000ei}.
 }

The puffed-up D5 branes in the above example have  an exact CFT description as  Cardy   states of  the $\widehat{su}(2)_k$ WZW model
  \cite{Alekseev:1998mc}. From the perspective of the D3 branes, this  is a reincarnation 
 of a celebrated phenomenon in condensed-matter physics, 
  the screening of a magnetic spin-\scalebox{0.92}{$(n-1)/2$} impurity in the $k$-channel Kondo model,   see  
 \cite{Affleck:1995ge,Bachas:2004sy}.  
 


\end{document}